\RequirePackage{fix-cm}
\documentclass[natbib,smallextended]{svjour3}       
\smartqed  
\usepackage{amsmath}
\usepackage{graphicx}
\usepackage[T1]{fontenc}
\usepackage[utf8,latin1]{inputenc}
\usepackage{epsfig}
\usepackage{txfonts}
\usepackage{amssymb}
\usepackage{siunitx}
\usepackage{natbib}    
\usepackage{hyperref} 

\hypersetup{
    colorlinks=true,citecolor=blue,urlcolor=blue,breaklinks
}

\usepackage{url}
\usepackage{color}
\usepackage{bm}
\usepackage{sgl-commands}

\newcommand{\KS}{\textcolor[rgb]{0.5,0,0.5}}

\begin{document}

\title{Strong Lensing by Galaxy Clusters}

\titlerunning{Cluster Lensing}

\author{P.~Natarajan$^{1,2,3}\,^*${\thanks{* \email: priyamvada.natarajan@yale.edu}}, L.~L.R. Williams$^4$, M.~{Brada{\v{c}}}$^{5,6}\,^*${\thanks{* \email: Marusa.Bradac@fmf.uni-lj.si}},C.~Grillo$^7$, A.~Ghosh$^8$, K.~Sharon$^9$, J.~Wagner$^{10}$}

\authorrunning{P.~Natarajan et al.}

\institute{
$^{1}$Department of Astronomy, Yale University, New Haven, CT 06511, USA\\
$^{2}$Department of Physics, Yale University, New Haven, CT 06511, USA\\
$^{3}$Yale Center for Astronomy \& Astrophysics, Yale University, New Haven, CT 06520, USA
$^{4}$Minnesota Institute for Astrophysics, School of Physics and Astronomy, University of Minnesota, Minneapolis MN 55455, USA\\
$^{5}$Department of Physics, University of California, Davis, CA 95616, USA\\
$^{6}$University of Ljubljana, 1000 Ljubljana, Slovenia\\
$^{7}$Department of Physics, University of Milan, 20122 Milano MI, Italy\\
$^{8}$Minnesota Institute for Astrophysics, School of Physics and Astronomy, University of Minnesota, Minneapolis MN 55455, USA\\
$^{9}$University of Michigan, Department of Astronomy, Ann Arbor, MI 48109, USA\\
$^{10}$ Bahamas Advanced Study Institute and Conferences, 4A Ocean Heights, Hill View Circle, Stella Maris, Long Island, The Bahamas\\
}

\date{Received: date / Accepted: date}


\maketitle

\begin{abstract}
   Galaxy clusters as gravitational lenses play a unique role in astrophysics and cosmology: they permit mapping the dark matter distribution on a range of scales;  they reveal the properties of high and intermediate redshift background galaxies that would otherwise be unreachable with telescopes; they constrain the particle nature of dark matter and are a powerful probe of global cosmological parameters, like the Hubble constant. In this review we summarize the current status of cluster lensing observations and the insights they provide, and offer a glimpse into the capabilities that ongoing, and the upcoming next generation of telescopes and surveys will deliver. While many open questions remain, cluster lensing promises to remain at the forefront of discoveries in astrophysics and cosmology.
   
\keywords{Gravitational lensing \and Strong gravitational lensing \and Galaxy clusters}
\end{abstract}

\section{Introduction}
\label{sec2:brief_introduction}

In hierarchical structure formation in the Universe, per the standard cold dark paradigm, galaxy clusters are now understood to be the most recently assembled and most massive structures to form. Clusters are therefore also expected to be the largest repositories of dark matter. As a result, clusters are some of the most efficient gravitational lenses known to date. \cite{Zwicky1937} was the first to point out that gravitational lensing by galaxy clusters might serve as a valuable tool to measure the amount of this unseen mass component and permit study of distant, magnified objects that lie behind clusters. Despite his bold prediction, the lack of appropriate resolution and imaging facilities at the time and the lack of theoretical understanding of how structure assembles deferred progress in gravitational lensing studies. Several decades later, the presence of a foreground cluster was proposed by \cite{Young1980} as the explanation for a larger-than-expected separation between the images of a double quasar in the first detected double quasar Q0957+561 \citep{wal79}. With this observational impetus and hence renewed interest, the lensing properties of clusters were revisited and explored in detail by \cite{Narayan1984}.

 \begin{figure*}
    \centering
    \includegraphics[trim={0cm 0cm 0cm 0cm},clip,width=0.95\textwidth]{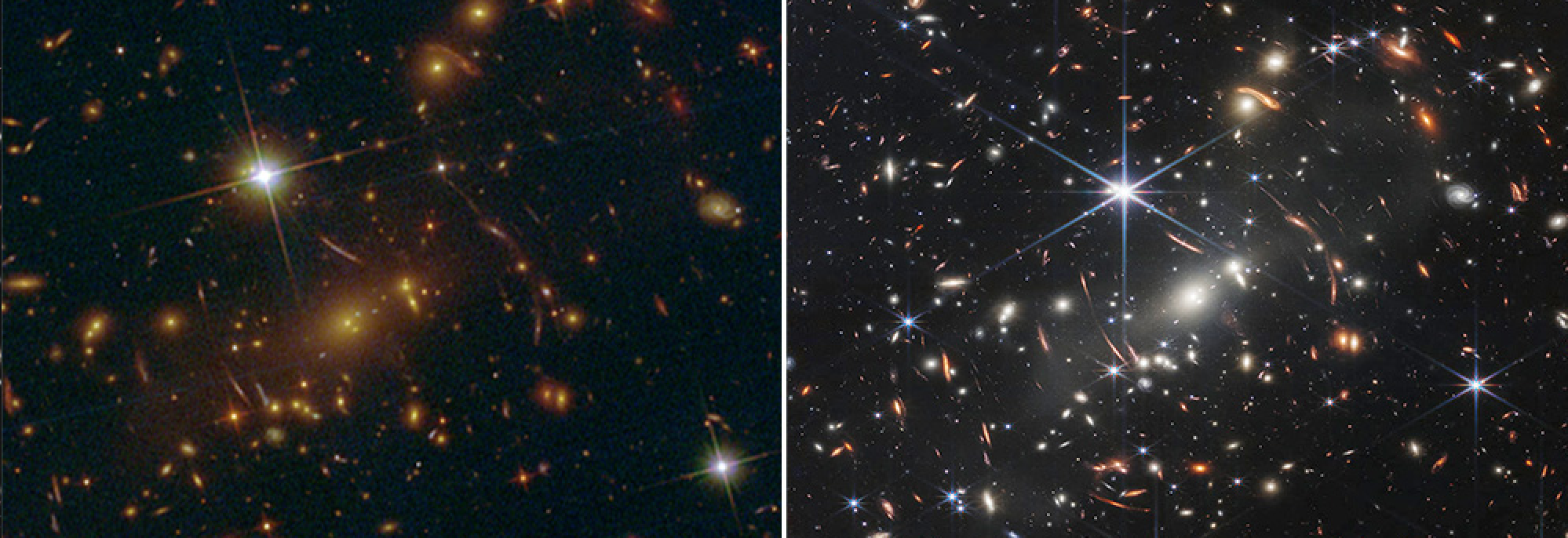}
    \includegraphics[trim={0cm 5.5cm 0cm 5.5cm},clip,width=0.5\textwidth]{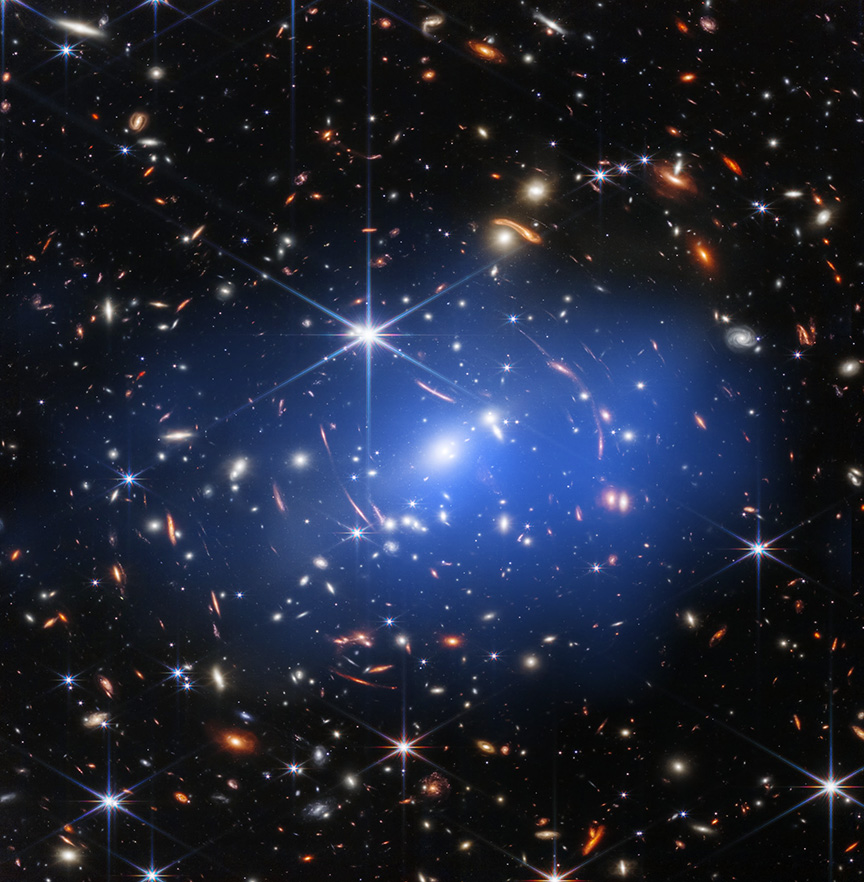}
    \caption{The galaxy cluster lens SMACS J0723.3-7327 at $z = 0.39$. {\it Top left:} HST image of the cluster showing multiple strongly lensed features visible by eye. {\it Top right:} JWST image of the same cluster that in turn reveals a significantly larger number of strongly lensed features. The improvement in the quality of the image is evident upon visual inspection. {\it Bottom:} Chandra X-ray image showing the spatial distribution of the hot X-ray gas (blue) overlaid over the JWST image. The images span nearly the same field of view and scale. The top right and bottom panels are about 2.35 arcmin wide, or 750 kpc at the redshift of the cluster, $z=0.388$.}
    \label{fig:smacs}
\end{figure*}

The first observational detection of strong lensing by a cluster---a giant arc in Abell 370---was reported in \cite{Lynds1989} and independently by \cite{sou87}. These extremely elongated distorted shapes seen in the core of the massive cluster Abell 370 were recognized by \cite{Paczynski1987} as merging images of a strongly lensed background galaxy, and were soon confirmed as such with the measurement of the redshift of the arc by \cite{Soucail1988}. With this discovery of giant arcs, the strong lensing regime was established as a property of the deflection produced by dense cluster cores. 

With the discovery of the weak, systematic alignment in the shapes of distant background galaxies produced in the outer regions of clusters as reported in \cite{Tyson1990}, observational evidence for both strong and weak lensing were confirmed. Several significant early papers \citep{Schneider1984,Blandford1986,Blandford1989,Kochanek1990,MiraldaEscude1991, Kaiser1992,Kaiser1993} helped establish the formalism for ``inverting'' observed lensing effects into mapping the mass distributions of galaxy clusters. Subsequent technological advances in the CCD imaging that allowed deeper and sharper images as well as the development and availability of spectrographs drove the rapid discovery of cluster lenses.

The truly transformative revolution in lensing, and strong lensing studies in particular, came with imaging from space with the deployment of the {\it Hubble Space Telescope}. After correction of its initially blurry images with the repair mission that installed the WFPC-2 camera in 1993, Hubble recovered its image sharpness, and one of the first post-repair image releases was the now iconic image of the massive cluster lens Abell 2218 \citep{Kneib:96}, replete with strong lensing features that are discernible by eye. High resolution images combined with a large field of view have been a game changer for studies of cluster lenses \citep[e.g.,][]{Smail1995}, for example, for tracing out the circularly averaged density profile of equilibrium clusters and comparing these to the predictions of cosmological models (Sect.~\ref{sec:clscale}), and for studying the most distant galaxies using the magnifying power of cluster lenses (Sect.~\ref{sec:SPmagnif}). Strong lensing features in clusters are now detected in the inner one arc-minute central region and on arc-second scales (see left panel of Figure~2 for a gallery of HST detected dramatic cluster arcs). On smaller scales, multiple images produced by individual cluster galaxies have also been detected in the deepest HST images \citep{Natarajan+1998,Natarajan:2017,Meneghetti+2020}.

The \textit{James Webb Space Telescope} (JWST), launched late 2021, looked at a galaxy cluster, SMACS J0723.3-7327, as one of its first targets, offering an unprecedented and detailed view of a massive lens in infra-red wavelengths. Figure~\ref{fig:smacs} compares the HST images to the newly obtained JWST images (top two panels), and the bottom panel shows the X-ray image superimposed on the JWST image. The improvement in the quality of the JWST image, compared to an already spectacular HST image is visually apparent and striking. 

\section{Strong lensing as a probe of cluster properties}
\label{sec2:clusters_CDM}

Clusters are powerful astrophysical laboratories as they are objects where the interplay of dark matter and baryons can be effectively probed on multiple physical scales \citep[see reviews by][]{JPK-PN2011,Umetsu2020}. On spatial scales greater than about 1-few kpc, clusters are dominated by dark matter, even at their centers. The baryonic component in clusters is mostly in the form of very hot plasma with $T \sim 10^7K$, the Intra-Cluster Medium (ICM) detected through its bremsstrahlung, or free-free, emission observed in X-rays. The stellar component is mostly confined to cluster member galaxies, including, in many cases, a massive central dominant (cD) galaxy or the brightest cluster galaxy (BCG) that has been known to have an extended stellar distribution \citep{Schombert1986,Uson+1991} and deeper recent observations have helped delineate this diffuse emission from a population of intra-cluster stars more clearly, referred to as the Intra-Cluster Light (ICL) component, in the form a diffuse cloud of light, which appears to trace the overall dark matter distribution with great fidelity \citep{Gonzalez+2005,mon19}. Notably, as discussed first by \cite{Uson+1991}, disentangling the BCG and the extended cD envelope is challenging even today. Even including all the baryonic components in the mass inventory in clusters, they still stand as dark matter dominated cosmic objects.

As noted in Sect.~\ref{sec2:brief_introduction}, the central $\lesssim 0.3$ Mpc of galaxy clusters are the regions that contain a high concentration of matter (dark + baryonic); these are the spatial scales on which the gravitational coupling between baryons and dark matter manifests. Currently the dark matter component of the Universe is believed to be cold and collision-less, rendering it inert in terms of interactions aside from gravity with the baryonic components. Cosmological simulations of structure formation over cosmic time find that dark matter settles into a near universal density profile across several decades in mass---the Navarro, Frenk and White profile (NFW)---with an inner slope of $\rho\sim r^{-1}$, that is expected in clusters of galaxies as well. However, if for example, dark matter interacts weakly with itself, this unusual, additional self-interaction, would preferentially produce discernible effects in this central dense region (see Sect.~\ref{sec2:constraints_warm}). While the mass distribution in the innermost regions is typically dominated by baryons, fortunately, these regions are also the ones that are best probed by strong lensing observations. 
 
One of the key measured properties for clusters is their mass and the detailed spatial distribution of the mass. The radial mass distribution in clusters can be probed and measured using 3 independent methods that all map the underlying gravitational potential: (i) the dynamics of cluster member galaxies assuming they are in equilibrium and robustly trace the gravitational potential, (ii) the X-ray emission from the hot plasma, also assuming it is in hydrostatic equilibrium with the cluster potential, and (iii) gravitational lensing, that makes no prior assumptions about the gravitational potential. The advantage of lensing as a technique is that light deflection is achromatic, and is also entirely independent of the dynamical state of the cluster. Lensing observations therefore permit study of merging and out-of-equilibrium clusters as well. Besides, lensing delivers not just the projected, {\it radially averaged} mass profiles for cluster, but also permits mapping of the {\it detailed} two-dimensional mass distribution. 
 
The power of cluster lensing derives from the fact that deeper imaging data not only permit probing the detailed spatial distribution of dark matter in the cluster but that it also simultaneously acts as a cosmic telescope allowing us to probe faint sources in the background Universe. Strong lensing by galaxy clusters brings into view distant, faint background galaxies that would otherwise remain inaccessible to even the most powerful telescopes on the ground or in space; the high magnification of intermediate and high redshift star-forming galaxies allows us to study the building blocks of these galaxies at an unprecedented spatial scale and at a fraction of the observing time otherwise necessary. Modeling and characterizing the mass distribution of cluster lenses therefore stands to open a new window into the early Universe by magnifying the most distant first sources that assemble, a prospect keenly awaited for with JWST data. However, as massive clusters are rare objects, defining systematic criteria to characterize cluster lens samples has proved to be challenging. Furthermore, given the dependence of the strength of lensing on the underlying geometry of the Universe, lensing has also been established as a powerful method to constrain cosmological parameters that characterize the underlying world model \citep{GilmorePN2009,Jullo:2010,jullo+2015,Magana+2018,Caminha+2022}. There are synergies in the general modeling approaches 
 for characterizing the lensing properties of individual galaxies and those adopted for clusters. However, the existence of a richer phenomenology and more extensive data available in both the strong and weak lensing regimes, warrants more sophisticated models tailored for lensing clusters. We outline the range of mass modeling techniques that have been developed and applied to cluster lenses in Sect.~\ref{sec2:modelling_methods}. 

\begin{figure*}
    \centering
    \includegraphics[trim={0cm -1.25cm 0cm 0cm},clip,width=0.44\textwidth]{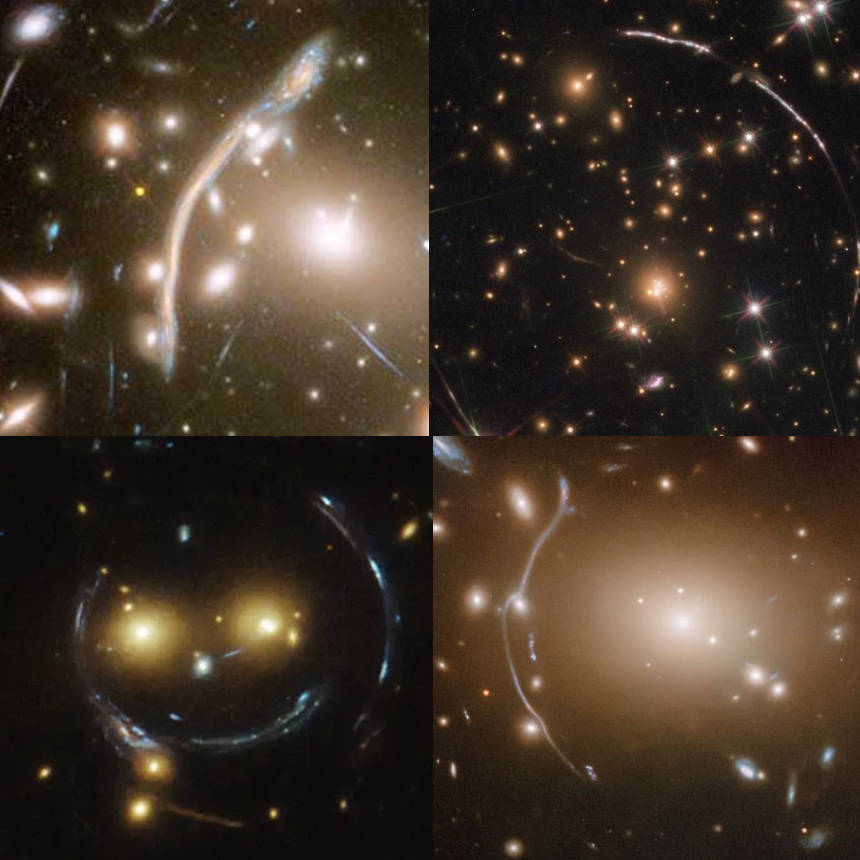}
    \includegraphics[width=0.49\textwidth]{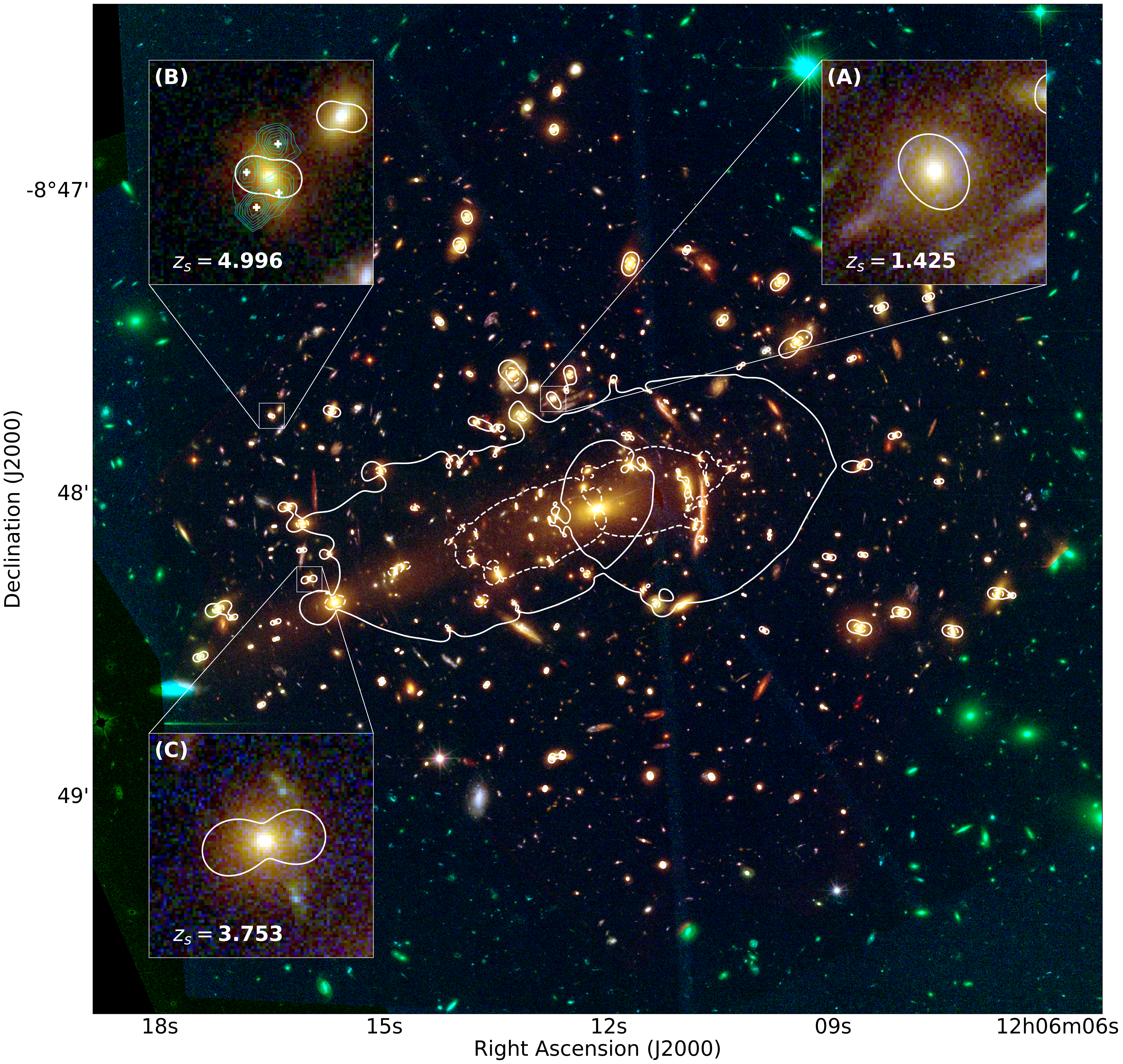}
    \caption{{\bf Left Panel:} Gallery of HST detected cluster arcs seen in the cluster Abell~370 (top-left); SDSS\,J1038$+$4849 (bottom-left); PSZ1\,G311.65$-$18.48 (top-right) and Abell~611 (bottom-right). {\bf Right Panel:} GGSL in the central region of the HFF galaxy cluster MACSJ1206: \citep{Meneghetti+2020}. The image is a composite and combines HST observations in multiple bands: F105W, F110W, F125W, F140W, F160W (red channel), F606W, F625W, F775W, F814W, F850LP (green channel), F435W and  F475W (blue channel). The dashed and solid lines in \KS{the right} panel show the lens critical curves for potential sources at redshifts of 1 and 7, respectively. Panels A, B, C are zoom-ins of three individual GGSL events produced by sources measured to be at redshifts $z=1.425$, $z=4.996$, and $z=3.753$. The GGSL sources in panel B  were not visible in the HST image and were discovered  in an observation with the Multi-Unit-Spectroscopic-Explorer (MUSE) spectrograph of the VLT.}
    \label{fig2:ggsl}
\end{figure*}

The Hubble Frontier Fields program (HFF thereafter), a key Director's Discretionary large program \citep{Lotz:2017} dedicated
840 HST orbits to ultra-deep strong lensing imaging of six selected massive cluster lenses. These fields were also observed across several wavelengths, adding to wealth of information for modeling and deeper understanding of clusters and high redshift Universe. These coordinated multi-wavelength surveys include ALMA \citep{gl2017}, the Spitzer Frontier Fields \footnote{\url{https://irsa.ipac.caltech.edu/data/SPITZER/Frontier/overview.html}} and Chandra, and JVLA Frontier Fields Campaigns \cite{vanWeeran+2017}. HFF has yet again transformed our understanding of the mass distribution and evolutionary states of clusters. Chosen as the most efficient lenses, these six massive cluster lenses permit probing  high mass end of the cluster mass function. Intriguingly, with deeper HFF imaging, in addition to the giant arcs and multiple image families observed in cluster lenses, additional strong lensing features produced by individual cluster members, referred to as Galaxy-Galaxy Strong Lensing (GGSL), have also been detected (see the right panel of Fig.~\ref{fig2:ggsl} for some notable GGSL examples). 

\section{Modeling the mass distribution in cluster lenses}
\label{sec2:modelling_methods}

The strong distortion of space-time by a massive cluster is so significant, that its effect on the shapes of background galaxies is unambiguously detectable, as evident in HST and even in ground-based images. Background sources, strongly lensed by foreground galaxy clusters result in the production of magnified, extremely distorted images referred to as giant arcs; and multiple images of the same source, that exhibit similar color and morphology, that albeit may have different levels of distortion and mirror symmetry. This wide range of phenomena---ranging from their geometry, brightness and observed spatial locations in the cluster---can and is used as input observational constraints to reconstruct the total mass of the cluster lens. Utilizing the observed distorted images to recover the original undistorted shapes enables constraining the mass distribution of the cluster lens, yielding ``lens models''. Lensing observations and the derived foreground cluster mass distributions have revealed that the bulk of the gravitational potential in clusters comprises the undetected "dark matter".

Several independent lens modeling algorithms to recover cluster mass distributions have been developed, widely tested and used by the strong lensing community. Regardless of the specific technique, all strong lensing modeling algorithms use as input constraints the positions of multiple images of the same background sources, sometimes referred to as ``families'', and ideally, when available the measured spectroscopic redshifts of the sources. Although the magnification in the strong lensing regime is conducive to follow-up spectroscopy to obtain redshifts, it is challenging, and often photometric redshifts are used instead.

The purpose of strong lens modeling algorithms is to find a representation of the projected mass distribution on the lens plane that offers the smallest scatter between the observed and model-predicted lensing observables. In the source plane (a.k.a. ``source plane minimization''), this is accomplished by minimizing the scatter of the model-predicted (unobserved) source position: for a set of $N$ images of the source, the source position $\vec\beta_i$ is calculated for each image $i$ using the lens equation:
\begin{equation}
\vec\beta_i = \vec\theta_i - \vec\alpha_i=\vec\theta_i-\frac{D_{\rm ls}}{D_{\rm os}}\hat{\vec\alpha_i},
\end{equation}
where $\alpha_i$ is the deflection angle at the observer, $\hat{\vec\alpha}$ is the deflection angle at the lens plane, and $D's$ are angular diameter distances connecting lens, source and observer; and $\theta$ is observed source position \citep[a schematic of the lensing geometry can be found in Figure~3 of the review by][]{JPK-PN2011}. In the image plane, the predicted positions of counter-images are calculated for each observed image, and compared to the observed ones. The latter, commonly referred to as ``image plane minimization'', is significantly more computationally intensive since the lensing equation cannot be inverted. The best-fit model is typically identified by maximizing a likelihood function, which usually means minimizing a $\chi^2$ function. Finally, missing information can be dealt with in different ways, such as setting priors on the source redshifts based on photometric information using what are now standard methods, or allowing the model to predict the unknown redshifts of sources.

The different approaches to lens modeling are generally divided into two broad categories, referred to as ``parametric'' and ``free-form'' (the latter is also often referred to as ``non parametric'') methods. Parametric approaches involve representing the mass distribution of the cluster lens with a set of physically-motivated profiles for the surface mass density, which are well described by analytic functional forms, and profile specific properties such as the ellipticity, and positional angles. Parametric models are conceptually motivated by cosmological simulations of structure formation in which cluster scale masses are observed to comprise of large and small scale dark matter halos that are gravitationally bound. Cluster mass distributions are modeled as a superposition of one or more large scale components, corresponding to the smooth cluster and/or group scale dark matter halo seen in simulations and several small scale sub-halos  that are associated with the positions of individual cluster member galaxies. The overall gravitational potential of a cluster is therefore modeled as a sum of contributions from large scale dark matter halos and smaller scale sub-halos: 
\begin{eqnarray}
{\phi_{\rm cluster} = \phi_{\rm smooth} + \Sigma_{i}\,\phi_{\rm sub}}.
\end{eqnarray}
As a result, at each position in the cluster, we can explicitly write the convergence, $\kappa(\vec x)$, and the shear, $\vec\gamma(\vec x)$, as a sum over these components \citep{Natarajan+1997},
\begin{equation}
\kappa(\vec x)=\kappa_{\rm smooth}(\vec x)+\sum_i \kappa_{{\rm sub},i}(\vec x)
\end{equation}
\begin{equation}
\vec\gamma(\vec x)=\vec\gamma_{\rm smooth}(\vec x)+\sum_i \vec\gamma_{{\rm sub},i}(\vec x)
\end{equation}
where $\kappa_{\rm smooth}$; $\vec\gamma_{\rm smooth}$ and $\kappa_{{\rm sub},i}$; $\vec\gamma_{{\rm sub},i}$ are the contributions to the convergence and shear from the smooth component and from the sub-halos respectively.

\begin{figure*}
    \centering
    \includegraphics[width=0.49\textwidth]{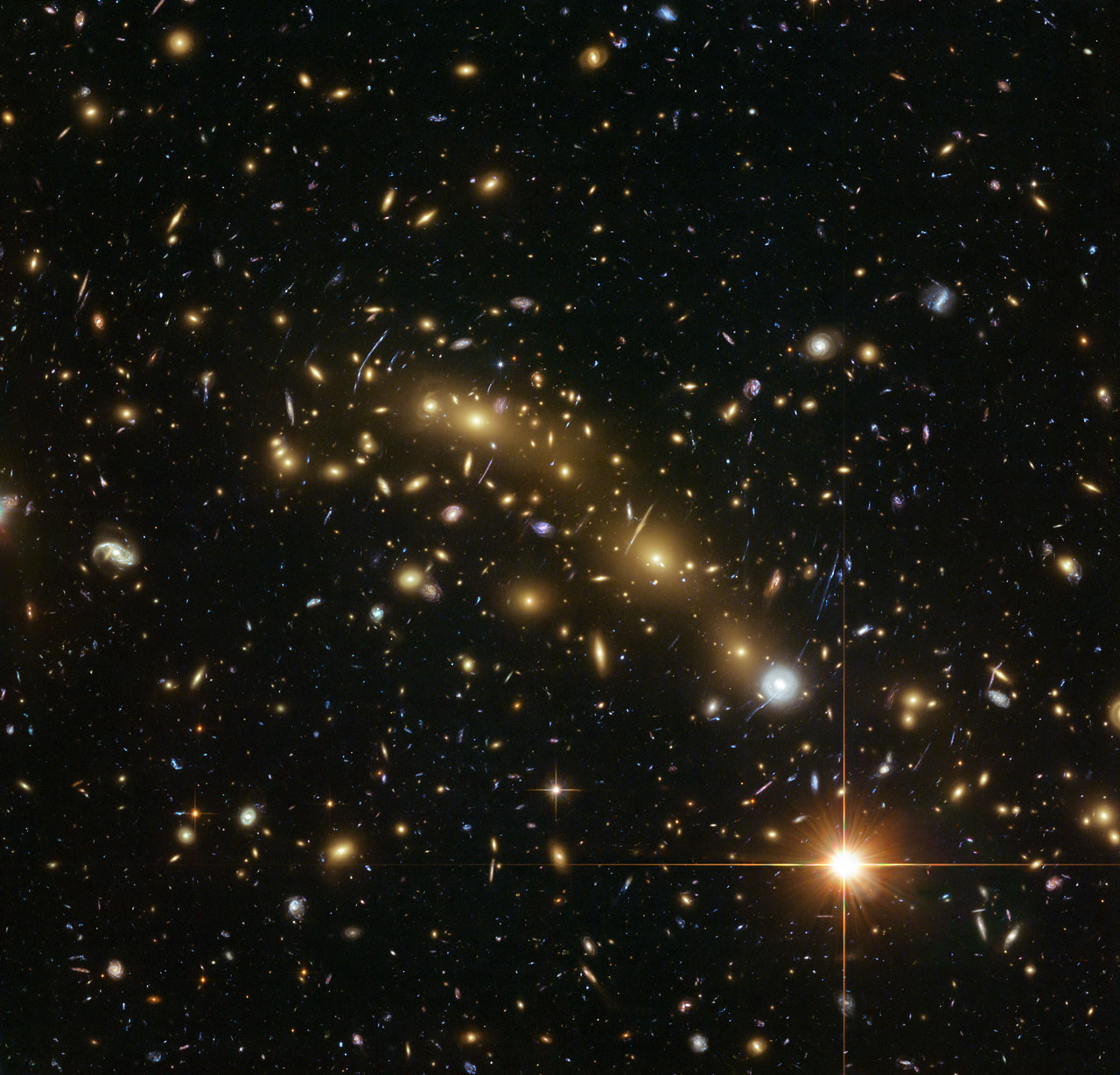}
    \includegraphics[width=0.49\textwidth]{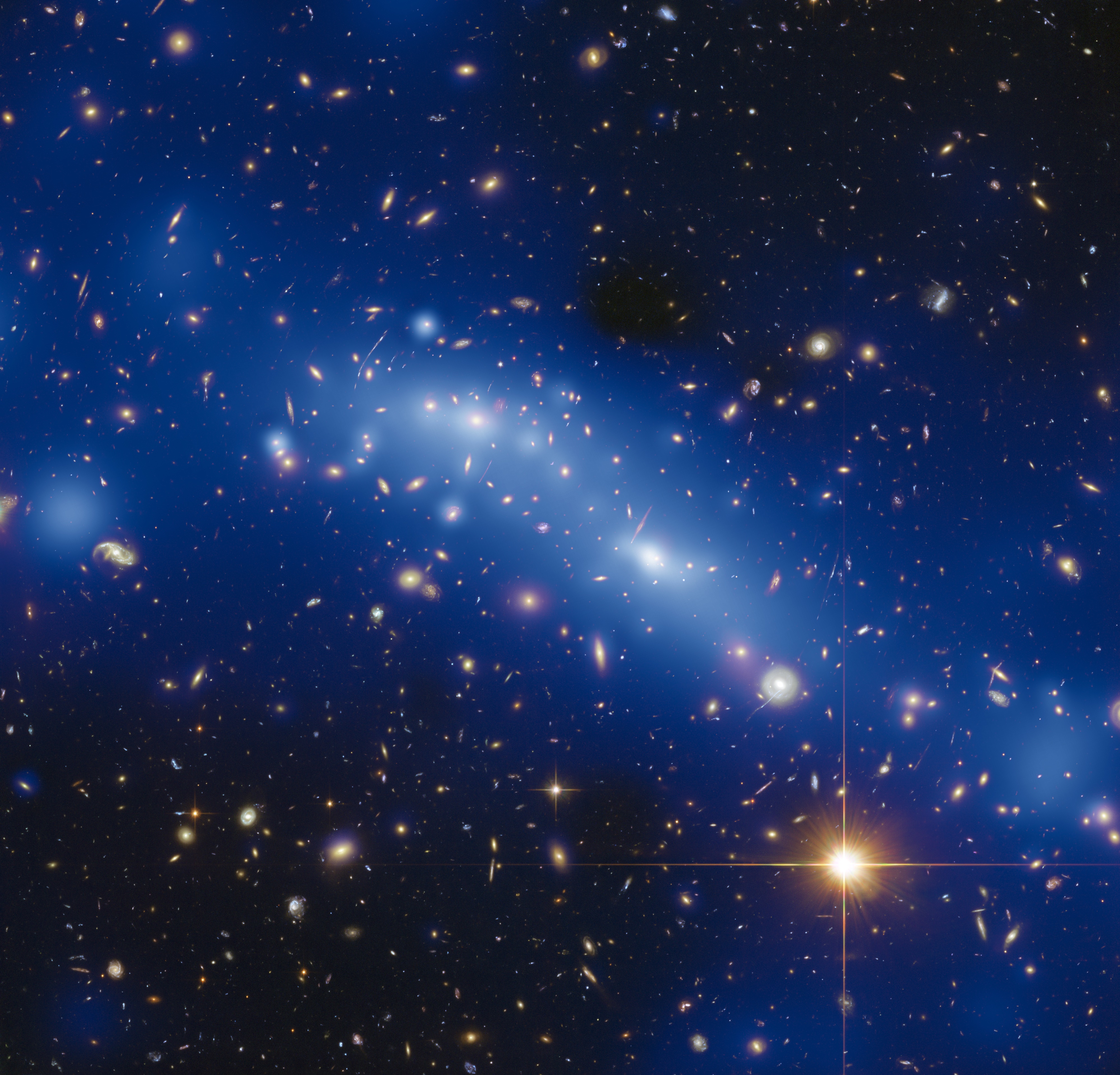}
    \caption{Left: HST Frontier Fields image of the cluster MACS 0416 at $z=0.397$. Right: Overlaid on HST image is reconstructed dark matter distribution of the cluster recovered using \lenstool shown as a blue haze. Image Credit: NASA/Hubble, ESA, HST Frontier Fields and J. Lotz, M. Mountain, A. Koekemoer, and the HFF Team (STScI). Acknowledgement for overlaid reconstructed dark matter map for Abell 2744: Mathilde Jauzac (Durham University, UK), Jean-Paul Kneib (Ecole Polytechnique Federale de Lausanne, Switzerland) \& Priyamvada Natarajan (Yale University, USA). Image taken by the Advanced Camera for Surveys (ACS) and WFC3 including B, V, I, Y, J, H \& W filters, with a FOV that is 2.08 x 2.32 arcminutes.}
    \label{fig2:macs0416}
\end{figure*}

Examples of commonly used analytic density profiles used to model the gravitational potentials are: a generalized \cite{Navarro:1997}, and truncated pseudo isothermal models (PIEMDs) that permit the potential to have finite mass and non-zero ellipticity or cored isothermal models. Parametric models have therefore been particularly powerful for comparing properties of observed cluster lenses with theoretical predictions from cold dark matter simulations. An example of parametric reconstruction of the mass distribution of the HFF cluster MACS 0416 with the publicly available software \lenstool\ is shown in Fig.~\ref{fig2:macs0416}.

On the other hand, free-form approaches describe the gravitational lens without assuming a predetermined mass density profile. For example, the lens plane is divided into a fixed or adaptive grid. The lens plane is then populated with some basis functions set, each grid cell is assigned one member of that set. Depending on the method, the number of grid cells can range between a few dozen and many thousands. The amplitude, i.e., mass of each basis function is determined directly by image properties. The motivation for free-form methods is that the mass distribution need not follow simple parametric forms, especially on cluster-wide scales and in merging clusters.

Modeling approaches vary by the degree to which they assume that mass and light distribution in clusters are correlated. Most, if not all, of the parametric approaches assign mass to observed individual cluster-member galaxies, motivated by empirical correlations like the Faber--Jackson relation derived from observations of early-type galaxies in clusters \citep{Natarajan+1997}. Some algorithms make no such assumption, and rely only on the lensing evidence to reveal the presence of mass. Finally, several algorithms adopt a hybrid approach: some mass components are inserted as physically-motivated parametric halos, while other components within the same model are set as free-form functions \citep[e.g.,][]{Niemiec+2020}.

As we note in the following section, when sufficient lensing constraints from rich observational data are available the different approaches outlined above perform equally well. However, they do vary in their strengths and weaknesses, and the decision of which approach to adopt depends primarily on the science goal and motivating question. The primary advantage of the parametric approach is that it can produce satisfactory results even with a small number of constraints that sparsely sample the image plane, as the assumed functional form that describes the lens plane fills in the gaps. And it is particularly well suited for comparison of the properties of observed cluster lenses with simulated clusters. Therefore, for tests of the CDM paradigm, where large cosmological simulations are the only available theoretical test-bed, parametric lensing models have been particularly useful and insightful \citep{Natarajan:2007,Natarajan:2017,Meneghetti+2020}. The high flexibility\footnote{It may be somewhat confusing that free-form models are also called ``non-parametric'', since they often have \textit{more} free parameters than parametric models.} of free-form models, on the other hand, require rich data sets as a high density of lensing constraints in all regions of the lens plane. However, given sufficient constraints, free-form models can be used to compare properties like the radial slope of the projected mass density against theoretical predictions, and have the flexibility to uncover substructure that may be potentially not be associated with light \citep{gho21}. Models that do not include a prior assumption about the correlation between mass and light are particularly useful for investigating this exact question---to what extent does mass follow light in the Universe as a function of physical scale \citep{seb16,gho22}.

Development of the formalism and first cluster-scale lens modeling attempts appeared in the literature nearly 25 years ago \citep{nar89,kov89,kne93,Kneib:96,Natarajan+1997,abd98}.  As noted above, at present, with the availability of richer data from deeper HST imaging surveys over the past two decades, a number of reliable cluster lens mapping methods are available as tabulated below. While the most commonly used methods tend to be parametric, recent years have seen the introduction of several new free-form methods. 
\begin{itemize}
\item \lenstool\ \citep[parametric;][]{Kneib:96, Jullo:07, Jullo:09, Niemiec+2020}. An extension of \lenstool\ includes free-form B-splines in addition to parametric functions \citep{bib2:beauchesne21,bib2:limousin22}.
\item Light-Traces-Mass \citep[LTM, parametric;][]{Broadhurst05b, Zitrin:09, Zitrin:15}, which closely ties the total mass to the observed light distribution.
\item \glafic\ \citep[parametric;][]{oguri10,Ishigaki:15,Kawamata:16}.
\item \gravlens\ \citep[parametric;][]{Keeton:10,McCully:14}.
\item {\sc Glee} \citep[parametric;][]{Suyu:2010, Suyu:2012}.
\item Weak \& Strong Lensing Analysis Package \citep[WSLAP+, hybrid;][]{Diego:05, Diego:07, Diego:16}, which combines parametrized galaxy-scale mass distributions with a free-form distribution on larger scales.
\item Strong and Weak Lensing United \citep[SWUnited, free-form;][]{Bradac:06,Bradac:09}, which, unlike most other methods reconstructs lensing potential, not projected mass.
\item PixeLens \citep[free-form;][]{wil04,sah06,koh14}, which is primarily designed for modeling systems with small number of images, but has been used for cluster reconstruction .
\item \grale\ \citep[free-form;][]{Liesenborgs:06,wil19,lie20,gho21}, which uses an adaptive grid and a genetic algorithm to find solutions in a high dimensional parameter space.
\item JPEG parametrization \citep[free-form;][]{lam19}, which takes inspiration from JPEG compression to do strong lensing inversion.
\item {\tt relensing} \citep[free-form;][]{tor22}, which reconstructs gravitational deflection potential on an adaptive irregular grid.
\item {\tt MARS} \citep[free-form;][]{cha22}, which uses a MaxEnt-regularized strong lensing inversion.
\end{itemize}

\subsection{Comparison of mass modeling methods using synthetic and real clusters} 
\label{sec2:comparison_different}

The independent cluster lens modeling methodologies outlined above have been carefully compared and contrasted in order to better understand their relative advantages and limitations. Below, we describe significant community-wide comparison exercises conducted to date and their conclusions. The first (Sect.~\ref{sec:simulated}) involved invitation to multiple teams to reconstruct the mass distribution of two simulated lensing clusters from their produced images wherein all teams were provided with the same input constraints on the identified families of multiple images and their redshifts. By construction, the details of the true underlying mass distributions in this case were known. This exercise was in preparation for the production of publicly available magnification maps from multiple teams for the HFF sample. 

For a second exercise (Sect.~\ref{sec:magnifications}), since it involved deep HFF imaging data sets, the mapping teams first arrived at a consensus set of input constraints, i.e., mutually agreed upon the identification of multiple lensed image families and either their inferred or measured redshifts to use uniformly for lens mass modeling. And from this cooperative and collaborative effort, magnification maps generated from various independent codes and teams were then made publicly available for the larger astronomical community to use. These maps have been successfully deployed for studies beyond lensing since. 

The third exercise (Sect.~\ref{sec:timedelays}) was an unanticipated continuation of the previous one, with the serendipitous discovery of a lensed supernovae in the field of two HFF clusters MACS J1149.6+2223 and Abell 2744. Multiple independent teams with their respective methodologies predicted magnification (for the cluster Abell 2744), and the arrival time and magnification of the final image (in the case of MACS J1149) as its observation was awaited.

\subsubsection{Comparison of lens modeling methods: reconstructing the mass and magnification of simulated cluster lenses}\label{sec:simulated}

\begin{figure*}
    \centering
    \includegraphics[width=\textwidth]{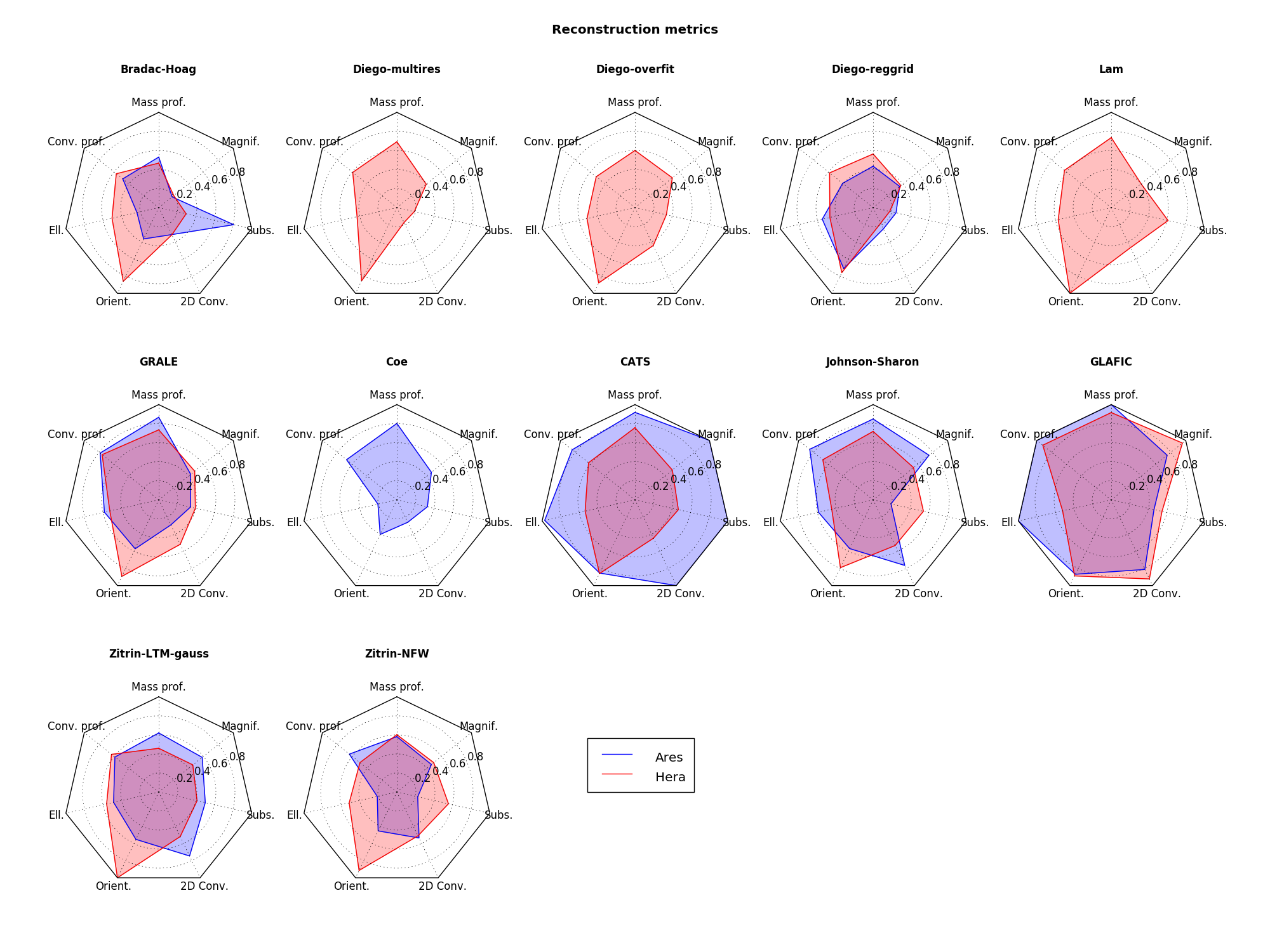}
    \caption{Reproduced from \cite{men17}, an HFF cluster lens mass model comparison project. Their caption reads: Radar plot showing the scores of each model for all metrics discussed in the paper. Larger polygons correspond to better overall performance. Each chart corresponds to a different lens model (see labels on the top) and shows results for both {\it Ares} (blue) and {\it Hera} (red), or whichever is available. The seven metrics are shown on the vertices of each chart. For each metric, the scores range from 0 (worst; plotted at the centre of the chart) to 1 (best; plotted at the vertex), normalized to the maximum value recorded by all models. A filled polygon is obtained by connecting the plotted scores of all metrics for each reconstruction.}
    \label{fig:FFmetrics}
\end{figure*}

In this detailed comparison exercise \citet{men17} compared mass distributions derived for the simulated clusters {\it ``Ares''} and {\it ``Hera''} from ten independent modeling groups. Simulated observations of these two simulated clusters that mocked the depth and resolution of HFF observations, were used as the reference sample. {\it Ares} was generated using the semi-analytical code {\sc moka}, as a superposition of analytical elliptical profiles representing the main cluster and its member galaxies consistent with cosmological initial conditions. {\it Hera}, on the other hand was more realistic, as it was directly taken from the output of a large volume, high resolution cosmological N-body simulation of the cold dark matter Universe. A concise summary of the comparison results are presented in a figure in \cite{men17}, which we reproduce here as Fig.~\ref{fig:FFmetrics}.  It is a so-called radar plot; each of the 7 spokes represents a particular cluster property, such as mass profile slope, ellipticity, magnification, etc.  If a property is reconstructed well (poorly) by a model the corresponding point is close to the vertex (center) of the black and white polygon. Reconstructed models of {\it Ares} and {\it Hera} are represented by blue and red colored polygons respectively. The larger the colored area the better that model reproduces the properties of the synthetic cluster. Upon detailed comparison it was found that several lensing properties that relied on integrated quantities were recovered equally well by the different lens modeling approaches using completely independent methodologies. For instance, the reconstructed mass profiles of the simulated clusters across methods matched at the level of $\sim 10 \%$ for both parametric and free-form methods; and the agreement on the mass enclosed within the Einstein radius, was found to be extremely robust for all methods. The results also showed that the parametric modeling methods tend to be generally better at capturing two-dimensional properties of the lens core. This was true in particular for {\it Ares}, because its true mass distribution resembled assumptions made by parametric methods. In addition, parametric models that permit comparison of additional quantities derivable directly from simulations like the subhalo mass function, and the radial distribution of subhalos, fared well. 

Predictably, the uncertainties in reconstructed magnification maps grew as a function of the magnification itself, resulting with largest uncertainties around the critical lines where the magnification diverges. The accuracy in the magnification estimates dropped by a factor of three when the magnification increased from $\mu_{\rm true}=3$ to $\mu_{\rm true}=10$. Another interesting result from this work was that even groups that used the same mass reconstruction algorithms with slightly different priors (e.g., CATS and Johnson-Sharon teams that both used the code \lenstool) and very similar inputs, obtained somewhat different maps, revealing the impact of the choice of priors and enabling quantifying their downstream effect.

\subsubsection{Comparison of lens modeling methods: magnification maps for the HFF sample}\label{sec:magnifications}

The six clusters of the HFF sample were selected to be amongst the most efficient lensing clusters on the sky and as they were legacy fields observed with Directors Discretionary time, the data was released immediately to the larger astronomical community. However, for exploitation of this unique data set, to interpret the properties of faint background lensed galaxies that these clusters bring into view for instance, reliable lensing maps for each cluster were required to be made available publicly as well. To do so, five independent lens modeling groups were selected to provide preliminary magnification maps for each of the HFF clusters prior to the observing campaign in order to facilitate rapid analysis of the HFF data by all members of the community. These models were initially based on a common set of input data, derived from shallower pre-HFF archival HST imaging, early spectroscopic campaigns, and a common shared catalog of lensed background galaxies. Once the data were taken, these models were continually collectively updated by each of the groups. As noted before the techniques adopted by groups spanned both parametric and free-form methods.

\begin{figure*}
    \centering
    \includegraphics[trim={0cm 1cm 0cm 0cm},clip,width=0.98\textwidth]{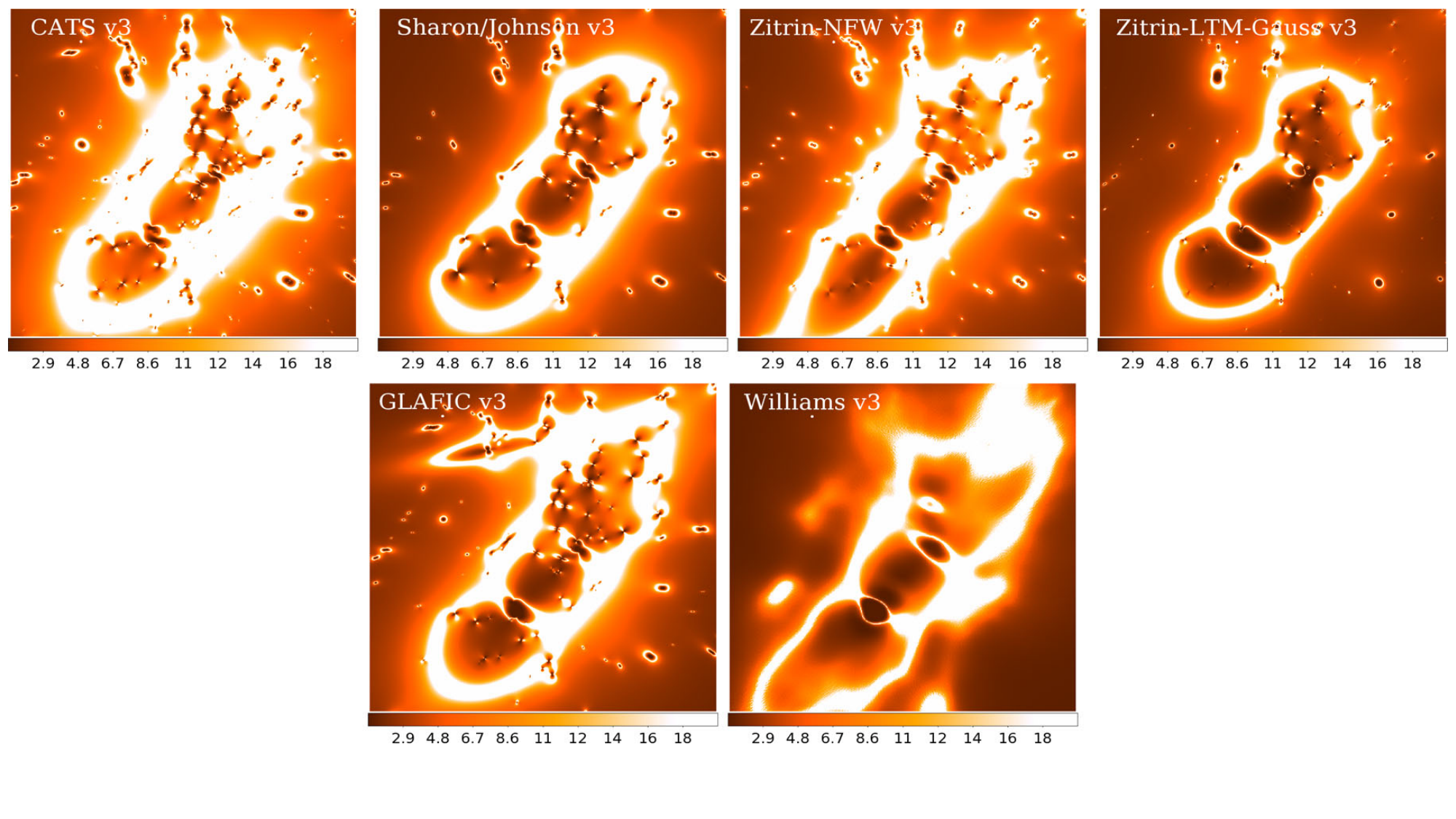}
    \caption{Magnification maps of 6 different reconstructions of galaxy cluster Abell 2744. The reconstruction team names appear in the top left of each panel. Magnification scale is the same for all panels, and goes from $\mu=1$ (darkest color) to $\mu=20$ (white). Each panel is a square 100 arcsec on the side. Taken from \cite{pri17}.}
    \label{fig2:A2744priewe}
\end{figure*}

The public HFF lens models include maps of mass ($\kappa$) and shear ($\gamma$) from which magnifications can be derived for sources at any redshift using scripts provided on the HST site. The models cover regions constrained by strongly lensed, multiply-imaged galaxies, within the HST ACS fields of view of the cluster cores. Some models extend to larger areas, including the HFF parallel fields, and also incorporate ground-based weak lensing data. All data, models and detailed description of individual methods and maps are available at the following website:\\
https://archive.stsci.edu/prepds/frontier/lensmodels/.

Since cluster magnification is of paramount importance for the HFF project, several papers compared magnification distributions as reconstructed by various independent groups. Figure~\ref{fig2:A2744priewe} shows magnification maps of Abell 2744 from 6 HFF teams, analyzed in \cite{pri17}. It was found that in the broadest terms the reconstructed magnification maps more or less agreed with each other. However, differences are also seen, and often exceed their quoted statistical errors.  The resulting dispersion ranged from $30\%$ at low magnifications, $\mu \sim 2$, to $70 \%$ at high magnifications, $\mu \sim 40$. This implies that the true uncertainties are probably currently underestimated in most of the modeling methods. These differences are mainly attributed to lensing degeneracies, suggesting that a small lens plane rms value is not a sufficient condition for a model to robustly and reliably predict the true magnifications. 

The more recent magnification comparison study by \cite{ran20} also arrives at similar conclusions, finding $\sim 40\%$ dispersion at low magnifications ({$\mu \sim 3$}) that increases to $\sim 65\%$ dispersion at high magnifications ({$\mu \sim 10$}). In contrast, excellent agreement exists for the derived integrated quantities like the cumulative mass distribution and it is significantly well constrained within $1'$  radius from the cluster core, with $< 5\%$ dispersion. Additionally, \cite{rem18} examined source plane scatter in 10 mass reconstruction models for the HFF cluster MACS J0416. Using source plane rms as a metric of diagnose lens model performance, they quantified the ability of different models to predict unknown multiple images. They found that while free-form reconstructions rely heavily on the quality and quantity of the lensing data, parametric models are not as susceptible to uncertainties in the data. However, parametric models do not necessarily benefit from adding more constraints with the inclusion of photometric redshift measurements when the existing amount of spectroscopic data available is large (of the order of measured redshifts for a hundred or so lensed images), as in the case of the HFF clusters.

\subsubsection{Comparison of lens modeling methods: using time delays and magnifications of quasars and supernovae}\label{sec:timedelays}

After the serendipitous discovery of the Supernova Refsdal in 2014, multiply imaged by a cluster member galaxy in the HFF cluster MACS J1149.6+2223 at $z = 0.544$, it was predicted that there would be the rare opportunity to see the supernova again in about one year, after the four images had faded away \cite{Kelly+2015}. This is because the initially observed four-image pattern was only one component of the lensing configuration. This supernova might have appeared as a single image some 10-20 years prior in the cluster field, and it reappeared at the predicted position between mid-November 2015 and December 11, 2015 (the approximately one month uncertainty is the interval between two consecutive Hubble observations). This was in good agreement with the blinded model predictions made by several independent groups deploying the range of lens reconstruction techniques prior to its observation \cite{SharonTraci2015,Oguri2015,Diego+2016,Treu+2016}. The time delay between the original quadruplet observed in 2014 and the latest appearance of the supernova in 2015 was used to infer the value of the Hubble constant. This is the first time this technique, originally proposed by Refsdal, was successfully applied to supernovae \cite{HubbleRefsdal2018}. 

\begin{figure}
    \centering
    \includegraphics[width=0.9\columnwidth]{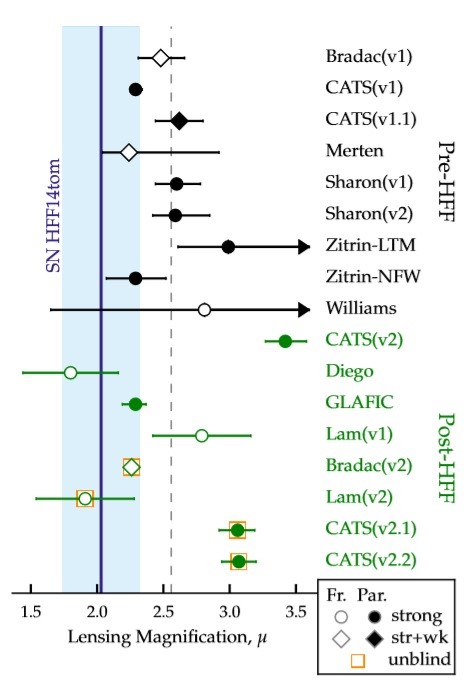}
    \caption{The observed lensing magnification is compared to predictions from multiple independent lens models. The vertical blue line shows the constraints from SN HFF14Tom, with the shaded region marking the total uncertainty. Markers with horizontal error bars show the median magnification and 68\% confidence region from each model. Circles indicate models that use only strong-lensing constraints, while diamonds denote those that also incorporate weak-lensing measurements. Free-form (parametric) models are shown as open (filled) markers. The top half, with points in black, shows the 9 models that were constructed using only data available before the start of the HFF observations. The lower 8 models in green use additional input constraints, including new multiply imaged systems and redshifts. The final four points, with square orange outlines, are the ``unblind'' models that were generated after the magnification of the SN was known. The black dashed line marks the unweighted mean for all 17 models, at $\mu=2.6$. Image reproduced with permission from \cite{rod15}, copyright by AAS.}
    \label{fig:SNuncertainties}
\end{figure}

In the case of another supernova detected behind the HFF lensing cluster Abell 2744 at $z = 0.308$, once again, specific predictions for the magnification derived from different lensing models have been compared in detail. \cite{rod15} computed the magnification predictions from various lens models for the location and redshift of the Type-Ia supernova HFF14Tom, leveraging the idea that the true luminosity of a SN-Ia can be inferred from its lightcurve (see Fig.~\ref{fig:SNuncertainties}). They report that the median of the lensing model magnification predictions is $25\%$ higher than the magnifications deduced from the observed brightness of the supernova, and that many model predictions disagreed by more than their stated uncertainties. A possible reason for the discrepancy is that there are not enough spectroscopically confirmed multiple image systems to constrain the lens models: accurate and precise image redshifts are as important for lens reconstructions as image positions.

\subsubsection{Comparison of lens modeling methods: detailed comparison of spatial regions within clusters}

\begin{figure*}
    \centering
    \includegraphics[width=0.9\textwidth]{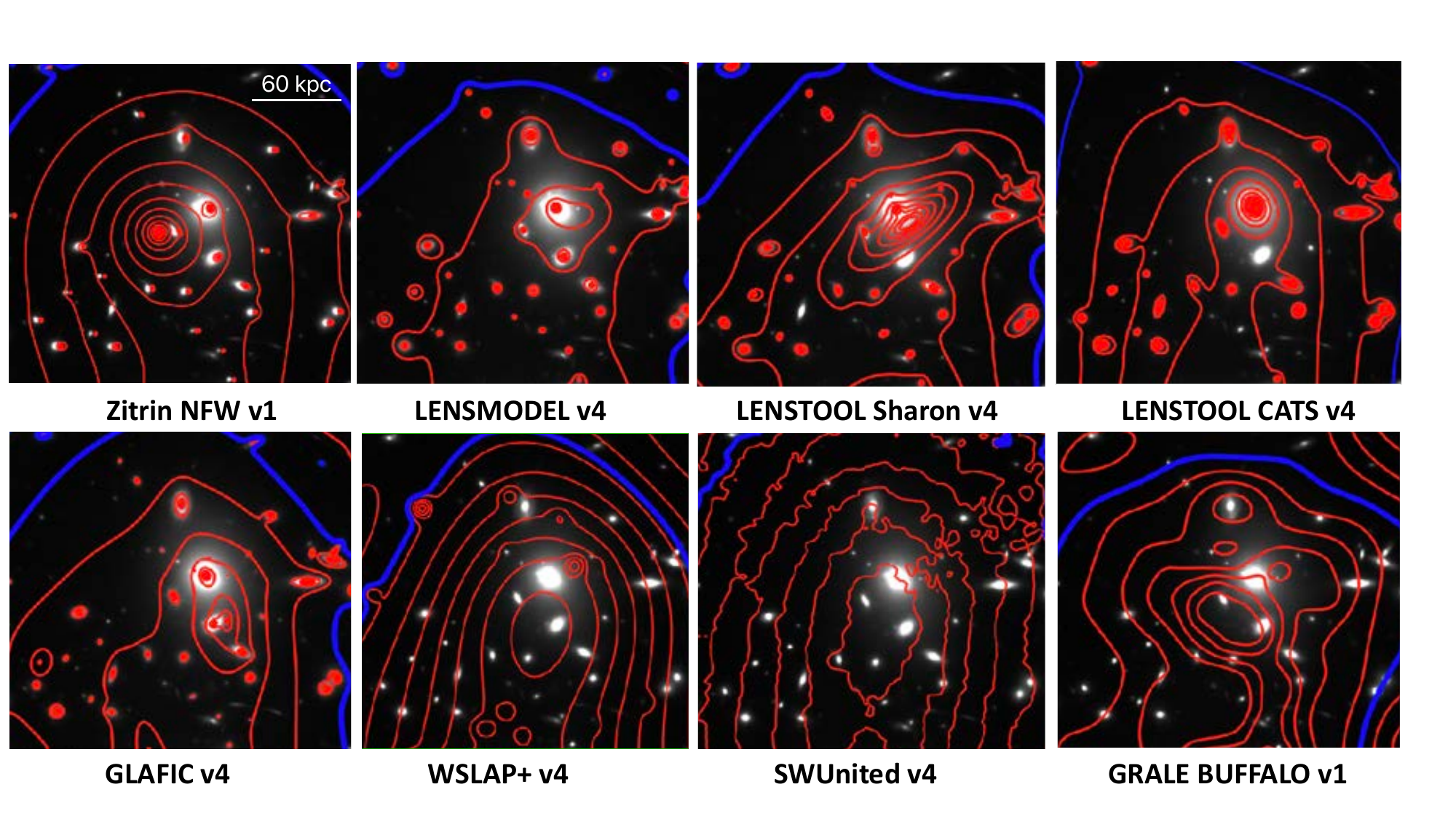}
    \caption{Comparison between different lens inversion methods mass maps of the same region in the HFF cluster Abell 370, centered on the northern of the 2 brightest cluster galaxies. Each panel is a square of 45 arcseconds on the side, or $\sim 250\,$kpc. The gray scale is the HST image in ACS/F814W filter. The panels show eight different reconstructions by various teams.  The contours are those of equal projected dimensionless surface mass density, $\kappa$, with the blue contour corresponding to $\kappa=1$. All but 3 (LENSMODEL HFFv4, LENSTOOL CATS HFFv4, Glafic HFFv4) show that the local center of mass and light do not coincide, with the center of mass offset the center of the BCG. For maps with offset centers, the offset is generally to the South of the BCG, though the morphology and the degree of displacement differs between models.}
    \label{fig2:A370BCG}
\end{figure*}

In addition to the exercises described above, it is instructive to compare regions of special interest in the reconstructed mass maps generated by various lens modeling methods. These could include regions around main cluster galaxies, or those that have excess reconstructed mass but are deficient in light. \cite{gho21} show that reconstructions with distinct modelling philosophies -- parametric vs. free-form -- often lead to similar results regarding specific mass features, while at the same time, models using the same algorithm can draw contrasting conclusions, just by using somewhat different model priors and data constraints. Figure~\ref{fig2:A370BCG} presents a region in an HFF cluster Abell 370, centered on the northern BCG. The red contours show the projected surface mass density, $\kappa$, with the blue contour indicating $\kappa=1$. Most models show that the center of mass appears to be displaced, generally towards the South, compared to the location of the center of light. This inferred feature of misalignment between mass and light is consistent with the merging, disturbed nature of this cluster. The fact that not all mass reconstructions agree on the detailed mass distribution in that region, including the misalignment between mass and light, is mostly due to lensing degeneracies, which are present not just on cluster scale, like the mass sheet degeneracy, but also on smaller scales \citep{lasko2023}. A larger number of multiply imaged sources than currently available will help to map out that region more accurately \citep{lin23} in the future. Other interesting examples are presented in \cite{gho21} and \cite{gho22}. The high spatial resolution with which strong lensing can measure the mass distribution can test predictions of dark matter properties \citep[e.g.,][]{bib2:harvey2019}. In summary, what these results imply is the importance of continued stress-testing of all lens mass modeling methods and continued comparison between them to delineate systematics that need to be understood in greater detail in cluster lensing studies.

\subsection{Connecting the Strong and Weak Lensing regimes}
\label{sec2:connecting_WL}

While strong-lensing driven mass modeling techniques have been used to obtain high signal-to-noise reconstructions in the central regions of clusters, the power of weak lensing is critical to extend the derived mass distributions out to the virial radius (e.g., \citealp{clowe06b, Umetsu:2014, Umetsu2020}). Connecting strong lensing and weak lensing regimes has therefore long been seen as a powerful way to recover the detailed and extended mass distribution of clusters from the central region to the outskirts. The method also has the advantage of breaking the so-called mass-sheet degeneracy (MSD), which says that the surface mass density $\kappa$ is invariant under the following global transformation $\kappa(\vec \theta)\to \lambda \kappa(\vec \theta) + (1-\lambda)$, where $\lambda$ is between $0$ and $1$. Non-zero $\lambda$ rescales the original surface mass density, and adds a constant density sheet to the lens. As a result, the overall density slope becomes less steep \citep{schneiderseitz95}. Weak lensing helps break MSD because it is reasonable to assume that beyond the weak lensing regime, i.e., far from the cluster center, cluster density is negligible.

Many of the lens modeling methods described above leverage the potency of this combination and therefore include both weak and strong lensing data, and some methods also include higher order information from flexion as well. Flexion parameters are similar to shear, but instead of the second derivatives of the lensing potential, they encode the third derivatives \citep[e.g.,][]{goldberg+2002,bacon2006,lanusse2016}. When both shear and flexion are important, image shapes deviate from purely elliptical distortion and can look `banana'-shaped. Flexion has been used to reconstruct mass distribution in some clusters \citep[e.g.,][]{okura08}.

The simplest adopted approach is to combine strong and weak lensing in a parametric fashion (e.g., \lenstool\ has included strong and weak lensing out to large cluster radii; \citealp{Natarajan:2009,Niemiec+2020,sch20}). These and other methods that were used to model cluster cores previously have an advantage that they are numerically stable, they are, however, inherently limited to the fitting a set of parametrized models \cite{Natarajan+1998}. Recently, free-form methods have also been used for combined analysis \citep{Diego:05, Bradac:06, merten16, lie20}. Modeling that include combining the two lensing regimes has been successful in predicting mass profiles on all scales and in determining other observables like the mass-concentration relation in clusters \citep{Merten:2015}. They have also proven useful in reconstructing magnification fields for high-redshift galaxies beyond the regime where strongly lensed images at lower redshifts are typically found as this pertains to the regime and region where magnification is poorly constrained \citep{pri17}. Inclusion of higher order constraints like flexion can also add useful information in this intermediate regime as demonstrated by \citet{cain16, leonard07}. 

\begin{figure*}[h!]
\centering
\begin{minipage}{\textwidth}
\begin{minipage}{0.5\textwidth}
\centerline{\includegraphics[width=0.9\textwidth]{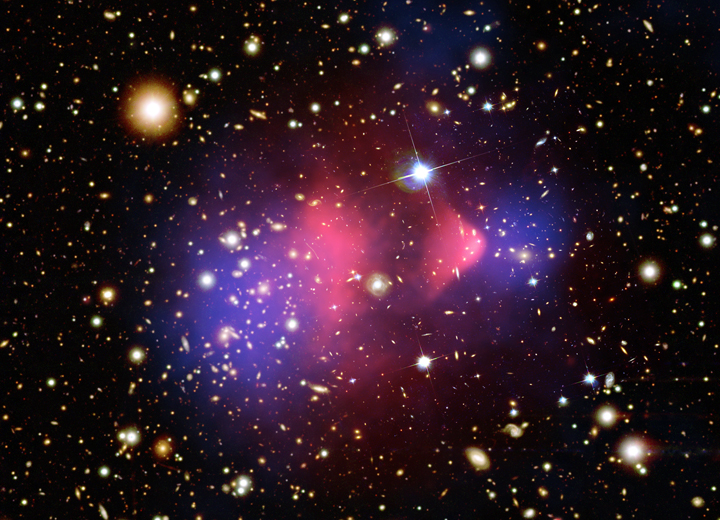}}%
\end{minipage}%
\begin{minipage}{0.55\textwidth}
\centerline{\includegraphics[width=0.89\textwidth]{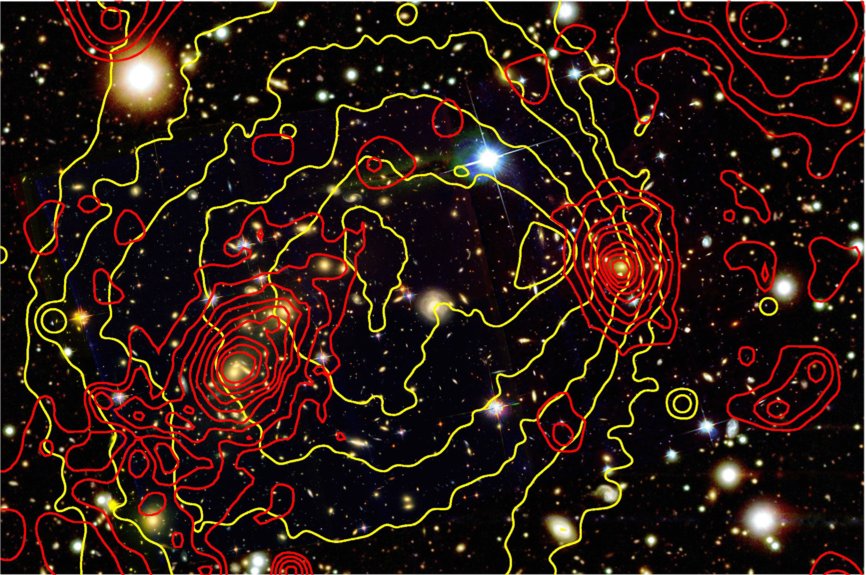}}%
\end{minipage}
\end{minipage}
\caption{{Left:} The color composite of the Bullet Cluster using only weak lensing measurement.  Overlaid in {\it
    blue} shade is the surface mass density map from the weak lensing
  mass reconstruction. The X-ray emitting plasma is shown in {\it
    red}.  Credit NASA/CXC/CfA/STScI/Magellan, \citealp{clowe06}. {Right:} Strong and weak lensing reconstruction for the Bullet Cluster. Overlaid in red contours is the surface mass density $\kappa$ from the combined weak and strong lensing mass reconstruction. The X-ray brightness contours from the 500 ks Chandra ACIS-I observations are overlaid in yellow. \citealp{Bradac:06,Bradac:09}. \label{fig:bullet}}
\end{figure*}

Another successful example of combining strong and weak lensing mass reconstruction has been the case of the Bullet Cluster (Fig.~\ref{fig:bullet}, \citealp{clowe06, Bradac:06}). The Bullet cluster comprises a pair of plane-of-the-sky merging sub-clusters, where the dominant baryonic mass in the form of X-ray emitting gas, and the the total mass, dominated by dark matter are not aligned post collision on the sky, providing strong evidence that dark matter is collisionless. In particular strong lensing provides useful constraints -- for instance, the offset between the peak of the baryonic mass, as traced by X-rays, and the peak of the total mass, as traced by strong and weak lensing, is over an arcminute, and was measured at $>10\sigma$ \citep{Bradac:06}, while the limits with only weak lensing were $8\sigma$ \citep{clowe06}. The Bullet Cluster configuration -- the location of the mass peaks, the offset between baryonic and dark matter distributions and the inferred relative velocity of the colliding sub-clusters was a challenge to reproduce for cold dark matter simulations \cite{SpringelFarrar2007}. The strong constraints on the inferred positions of the peaks of mass distribution provides a challenge for simulations that assume non-negligible self-interaction cross sections for dark matter \citep{randall08,lage15}.  In addition, results from combined analysis also provide useful and meaningful constraints on several alternative gravity models as they fail to predict large mass concentration needed to produce strong lensing constraints (e.g., \citealp{angus06}).

Many clusters now have successful measurements of mass and detailed mass distributions in combined regimes. HFF clusters have been reconstructed using combined strong and weak lensing analysis; see for example \citep{Jauzac+2016,merten11,wang15,hoag16,finney18,strait18}. The scientific results from these studies are discussed in Sect.~\ref{sec2:results_modelling}.

\section{Surveys for cluster lenses}

\subsection{HST cluster lens samples: CLASH, RELICS, HFF, BUFFALO Surveys}
\label{sec2:massive_clusters}

HST has devoted significant resources to cluster lensing observations via multiple programs including CLASH, RELICS, HFF and BUFFALO. Significant progress in the field of galaxy cluster physics has been made thanks to the data collected within the HST Multi-Cycle Treasury program Cluster Lensing And Supernova survey with Hubble (CLASH; \citealt{Postman:2012}), complemented with a comprehensive and coordinated spectroscopic campaign carried out with the ground-based telescopes (e.g., the CLASH-VLT Large Programme; \citealt{Rosati:2014}).

The CLASH survey was awarded 524 orbits of HST time (GO~12066; PI: M. Postman) to observe 25 massive (virial mass $M_{\rm vir} \approx$ 5-30\,$\times 10^{14}\,M_{\odot}$, and X-ray temperature $T_{\rm X}\ge$ 5 keV) galaxy clusters in 16 broadband filters, ranging from approximately 2000 to 17000 \AA, with the Wide Field Camera 3 (WFC3) and the Advanced Camera for Surveys (ACS). The sample, spanning a wide redshift range ($z\,=\,{0.18-0.90}$), was carefully chosen to be largely free of lensing bias and representative of relaxed clusters, on the basis of their symmetric and smooth X-ray emission profiles. CLASH had four main scientific goals: 1) to measure the total mass profiles of cluster over a wide radial range by using strong and weak lensing observations; 2) to detect new Type Ia supernovae out to redshift $z \sim$~2.5 to improve the constraints on the dark energy equation of state; 3) to discover and study the highly magnified first galaxies that formed after the Big Bang ($z$~>~7) and are brought into view by the cluster lens; and 4) to study galaxy evolution by comparing cluster member properties with those of and background galaxies (for a thorough overview, see \citealt{Postman:2012}).

A Large Programme (186.A-0798, PI: P. Rosati; \citealt{Rosati:2014}) of 225 hours with the VIMOS instrument at the VLT was also approved to perform a panoramic spectroscopic survey of the 14 CLASH clusters that are visible from ESO-Paranal. This observational campaign was aimed at measuring in each cluster the redshifts of 1) approximately 500 cluster members within a radius of 3 Mpc; 2) determine spectroscopic redshifts for 10-30 lensed multiple images inside the HST field of view, including possible highly-magnified candidates out to $z \approx$~7 (e.g., \citealt{Balestra:2013}); 3) possible supernova hosts. In the first CLASH cluster that was followed-up (i.e., MACS J1206.2$-$0847), the spectroscopic redshifts were exploited to build robust strong lensing models (\citealt{Zitrin:2012, Umetsu:2012, Eichner:2013, Grillo:2014}); to obtain an independent total mass estimate from the spatial distribution and kinematics of the cluster members (\citealt{Biviano:2013, Sartoris:2014}); to study the properties of the intracluster light (\citealt{Presotto:2014}), and to characterize the galaxy stellar masses as a function of environment and the stellar mass density profiles (\citealt{Annunziatella:2014}).

Using the unprecedented data set---panchromatic HST imaging and VLT spectroscopic data, astronomers have measured the gravitational potential in the inner regions of the cluster (i.e., the total mass profile in the core) for some of the CLASH clusters, by modeling the multiple images of several newly discovered strong lensing systems (e.g., \citealt{Zitrin:2015, Coe:2012, Monna:2014, Grillo:2015}). Moreover, the combination of the HST observations with deep, multi-band, wide-field imaging from Subaru has allowed  the weak lensing detection and therefore extension of the mapping of the total mass profile of several clusters out to their virial radii (\citealt{Medezinski:2013, Umetsu:2014, Merten:2015}).

The advent of the MUSE instrument at the VLT has revolutionized strong lensing studies in galaxy clusters. MUSE capabilities permit simultaneous identification and study of cluster members and multiply imaged background sources out to $z$~=~6.6. The exceptional suitability of MUSE to reach these goals has been demonstrated by several successful programs (e.g., \citealt{Karman:2015, Richard:2015, Grillo:2016, Caminha:2019, Bergamini:2021}), which have enabled the construction of detailed mass models of the cores of cluster lenses.

The results from systematic studies of lensing clusters with HST, including those from CLASH, led to the deeper Hubble Frontier Fields Initiative (HFF; \citealt{Lotz:2017}) that targeted six massive galaxy clusters, for a total of 140 HST orbits per cluster in 7 broadband filters, achieving in all of them unprecedented depth of $\sim$29 mag (AB) using Director Discretionary Time. This program has detected some of the highest redshift galaxies and enabled the first characterization of this sample of star-forming galaxies in a statistically meaningful way (e.g., \citealt{Bouwens:2017}). The HFF data have also provided a great opportunity to study the structure of the dark matter halos hosting these clusters (e.g., \citealt{Caminha:2017a, Natarajan:2017, Bergamini:2019,Meneghetti+2020}). 

The Reionization Lensing Cluster Survey (RELICS; PI: D. Coe; \citealt{Coe:2019}) consisted of 190 orbits of HST time to observe the cores of 41 galaxy clusters in 7 broadband filters, ranging from approximately 0.4 to 1.7 $\mu$m with the WFC3 and the ACS. This sample, spanning a wide redshift range ($z$~=~0.18-0.97), was carefully chosen to be representative of the most massive clusters ($M_{500}\gtrsim 8 \times 10^{14}$ M$_{\odot}$; thus also many of the most efficient cluster lenses) discovered so far, on the basis of their Sunyaev-Zel'dovich (SZ) mass estimates released by the Planck collaboration, and other known strong lenses. These include ACT-CLJ0102$-$49151 (El Gordo) (\citealt{Menanteau12,Zitrin:2013}), PLCK G287.0$+$32.9 with arcs 80\,\arcsec$\,$ from the core (\citealt{Gruen:2014}) and CL\,0152$-$1357, a massive merging cluster at $z=0.83$ \citep{Acebron:2019}. The HST imaging of the HFF combined with Spitzer imaging (390 hours; PIs: M. Bradac \& T. Soifer) yielded strongly lensed high-$z$ candidates among the best and brightest known. 

The latest contribution to the lensing data comes from The Beyond Ultra-deep Frontier Fields and Legacy Observations (BUFFALO) Program, with 101 orbits and 101 parallels as a part of Hubble Space Telescope's Treasury program, taking data from 2018-2020 \citep{sec2:steinhardt20}. BUFFALO expanded the existing coverage of the HFF fields in additonal filters: WFC3/IR F105W, F125W, and F160W and ACS/WFC F606W and F814W around each of the six HFF clusters and flanking fields. These fields are already covered by additional deep multi-wavelength datasets, including Spitzer and Chandra. Further important additions to the data are provided by the spectroscopic surveys, such as the Pilot-WINGS \citep{lagattuta22}.

\subsection{X-ray and SZ surveys for cluster lenses}
\label{sec2:lensing_X-ray}

X-ray selected clusters have been invaluable in identifying potential strong lensing clusters. \cite{horesh10} concluded that clusters from the the MAssive Cluster Survey \citep[MACS;][]{ebeling01} are significantly more---by a factor of $\sim\!6$---efficient as lenses than optically selected clusters. Many of the MACS clusters are now household names among those doing cluster lens modeling. X-ray clusters are also used in another capacity. Several efforts have been underway at X-ray wavelengths to obtain complementary information on the cluster hot gas component (e.g., \citealt{Ogrean:2015}). Owing to the adopted selection criteria, in all the clusters of the CLASH sample, measurements of the X-ray emission of the hot intracluster medium, taken with the Chandra telescope (and in some cases also with the XMM-Newton telescope) were available. These observations are in most cases sufficiently deep to estimate the hot gas and total mass profiles, under the assumption of hydrostatic equilibrium (\citealt{Donahue:2014}). In some clusters, a combined strong lensing, photometric and X-ray analysis has allowed a precise decomposition of the cluster total mass profiles into their stellar, hot-gas and dark-matter components (\citealt{Annunziatella:2017, Bonamigo:2017, Bonamigo:2018, Granata:2021}). Systematic searches for clusters at high redshifts, $z\sim 1$, has become possible using their Sunyev-Zeldovich (SZ) emission. The important surveys were done by the Planck CMB satellite and the South Pole Telescope (SPT). The masses of these clusters can be calibrated through weak lensing \citep{sereno17,raghu19,schrabback21}.

\subsection{Ground-based imaging surveys for cluster-scale strong lenses}

Wide optical and spectroscopic surveys such as the Sloan Digital Sky Survey \citep[SDSS; ][]{blanton2017} have been a fruitful hunting ground for several strong lensing surveys; from galaxy-lensed quasars, through galaxy-galaxy lenses, to the most massive cluster lenses. Teams have used different methods to identify candidate lenses in these public data. Some of the most well studied cluster lenses (e.g., the ``poster child'' cluster lens Abell 1689) are hardly unidentifiable as such in shallow SDSS imaging. The numerous multiple images in Abell 1689 are too faint to be robustly identified. Most shallow, ground-based surveys for cluster lenses aim to discover the lensing effect itself, i.e., an occurrence of highly distorted and magnified galaxy in the form of one of more giant arcs. The Sloan Giant Arcs Survey \citep[SGAS; PI: M. Gladders;][]{Bayliss2011,sharon2020} yielded hundreds of cluster lenses by mining the SDSS DR7 imaging data. Starting with the SDSS public catalog, SGAS first identified clusters and groups using the red sequence technique \citep{GladdersYee2000}. $g$, $r$, $i$, and $z$ cutouts around these fields were then examined by eye to identify lines of sight with bright arcs, and given scores. Lens and arc candidates were confirmed through imaging and spectroscopy with larger telescopes, reaching close to 100\% followup completeness of high-scoring candidates. The selection process was designed such that purity and completeness could be quantified, making the results useful for statistical studies of lensing occurrence (a.k.a arcs statistics, see section \ref{sec2:arc_statistics} and cluster studies.  Due to its selection function, the SGAS sample provided a bountiful supply of highly magnified background sources at ``cosmic noon'' -- those galaxies that reside in the epoch when the Universe formed most of its stars. The high magnification makes feasible detailed spectroscopic studies of the physical properties of these galaxies  \citep[see][and references therein]{sharon2020}, also leading to offshoot surveys such as M\textsc{eg}a\textsc{S}a\textsc{ura}  \citep{rigby18a}. Similar methods for discovery and confirmation of strong lensing clusters were employed in searching through the RCS-2 survey \citep{Gilbank2011} for lenses, and most recently in the COOL-LAMPS \citep{Khullar2021} survey. 

\subsection{Cosmological and large area surveys} 
\label{sec2:cosmological_large}

There are multiple on-going and planned surveys that will garner large samples of cluster lenses. While these are expected to be shallower than the HST cluster lens samples discussed above, the majority of them stand to detect and exploit weak lensing. One of recently concluded ground-based surveys is the Dark Energy Survey (DES). 

DES is a large survey of distant galaxies including clusters that aims to uncover the nature of dark energy that drives cosmic acceleration using multiple techniques, that include its impact on the abundance of galaxy clusters; weak gravitational lensing signals; type Ia supernovae and the detection of large-scale correlations between galaxies. The survey uses a powerful wide-field imaging camera called the Dark Energy Camera, or DECam, installed on the 4-meter Blanco telescope in Chile. During its five-year campaign (2013-2018), the DECam imaged approximately 5000 square degrees of the sky using five broadband filters, taking advantage of the excellent viewing conditions available on Cerro Tololo. The final DES dataset will consist of precise photometric and morphological information for over 200 million galaxies out to redshifts of 2.0. The survey is also periodically revisiting smaller patches of sky to find and study over 2500 Type Ia supernovae. Forecasts for expected cluster weak lensing results from the DES and other ground based surveys with the Rubin Observatory as well as from the Nancy Grace Roman Space Telescope are promising for constraining cosmological parameters \cite{Wu+2021}.

The recently successfully launched and deployed Euclid mission consists of a 1.2 m space telescope that is dedicated to study the imprint of dark energy and gravity via two powerful, complementary cosmological probes: weak gravitational lensing and galaxy clustering (via baryonic acoustic oscillations and redshift space distortion). These two complementary probes will capture signatures of the expansion rate of the Universe and the assembly of cosmic structures \citep{Amendola:2013}. The key instruments aboard the telescope are a high quality panoramic visible imager (VIS), a near infrared 3-filter (Y, J and H) photometer (NISP-P) and a slitless spectrograph (NISP-S). These instruments will permit accurate shape measurements and therefore quantify weak gravitational lensing effects of dark matter as well as three-dimensional mapping of structures with spectroscopic redshifts of galaxies and clusters of galaxies \citep{Sartoris:2016}. Euclid will observe 15,000 ${\rm deg}^2$ and is expected to find of the order of 60,000 galaxy clusters in the redshift range $z= 0.2-2$. It is forecast that $\sim$ 50,000 of them will have background galaxy densities $>15$ galaxies per sq. arcmin allowing accurate determination of cluster masses from weak lensing studies out to the virial radius and beyond \citep{Adam:2019}. The imaging and spectroscopic capabilities of Euclid will enable internal mass calibration from weak lensing and the study of the dynamics of cluster galaxies, in combination with ground based cluster surveys like DES \citep{Tutusaus:2020}. Euclid is also expected to detect several thousand lensed arcs, as predicted from the analysis with the Skylens simulation package, which has catalyzed the development of new automated arc-finder algorithms. Three ``Euclid Deep Fields'' covering around 40 ${\rm deg}^2$ in total will be also observed extending the scientific scope of the mission to access the very high-redshift universe.

\section{Science Results from modeling observed cluster lenses} \label{sec2:results_modelling}

\subsection{Insights on larger-scales}\label{sec:clscale}

One of the first clusters to have the mass profile derived from observed lensing was Abell 1689  \citep{Broadhurst05b}. Its profile is consistent with that expected in the LCDM universe, but with a somewhat larger concentration \citep{bro05a,coe10}, which can be explained by the lensing selection bias because higher concentration makes a cluster a stronger lens. A number of studies over the next decade looked at the distribution of properties of large sets of clusters \citep{richard2010,zitrin2011}. More recently, \cite{fox22} analysed a large sample (74) of RELICS clusters using their strong lensing data. Not all of these were found to be in equilibrium and relaxed, and consequently they display a large diversity in their density profiles, especially in the radial range 50-200 kpc.

In order to test the standard LCDM cosmological model, one needs to select clusters that are in equilibrium, because simulations \citep{Navarro:1997,nav04}, and theory \citep{hjo10,wil10,wag20b} make specific predictions for the spherically averaged density profiles of relaxed dark matter halos. To determine whether the cluster radial mass distribution is in good agreement with the predictions of , one needs a large radial range of the mass reconstructions. The strong lensing region of clusters, wherein the surface mass density exceeds the critical value, is relatively small, spanning at most one decade in radius. Therefore, it needs to be supplemented with weak lensing data. \cite{ume16} performed a joint analysis of 20 CLASH clusters using their strong, weak and magnification data, over two decades in radius, from $\sim 30-3000$ kpc, corresponding to approximately the virial radius of these clusters. They found that the stacked density profile in these clusters is well described by the fitting formulae from the cold dark matter simulations, i.e., NFW \citep{Navarro:1997} and Einasto \citep{nav04} profiles, as well theoretical DARKexp \citep{hjo10} profiles for relaxed, collisionless, self-gravitating systems. We re-iterate that massive galaxy clusters are uniquely suited for this important test of dark matter properties because their mass is dominated by dark matter over a wide range of spatial scales. In addition to radially averaged density profile, one can also use the median ellipticity of a set of relaxed clusters as a test of standard cosmology and the nature of dark matter. Using the same set of 20 CLASH clusters, \cite{umetsu2018} found that the median projected axis ratio of clusters' dark matter halos is in agreement with the predictions from recent numerical simulations of the standard collisionless cold dark matter model.

On much smaller scales than discussed above, the density profile in the very central regions of clusters, $\lesssim 50$ kpc, or a few percent of the virial radius, are often hard to estimate because the de-magnified central images (maxima in the Fermat potential) are usually not detected. A detection of two central images has been claimed in the cluster Abell 3827 by \citep{Massey+2018}. Yet, the question of the central profile---whether clusters have flat density cores or cusps---is very important for constraining of the nature of dark matter (also see Sect.~\ref{sec2:constraints_warm}). 

\cite{bib2:limousin22} modeled two unimodal clusters (i.e., with a single dominant center) AS 1063, and MACS J1206, and one bimodal cluster MACS J0416. While the last cluster is an ongoing merger, the dynamical status of the unimodal clusters is not clear; they show some evidence of disturbance, and so are probably not relaxed. A new metric utilizing the computation of the power spectrum of fluctuations in the mass and X-ray gas maps reveals that unimodal clusters that appear smooth may still be out of equilibrium \cite{Cerini+2023}. The authors conclude that cored profiles interior to $\sim 50$ kpc are favored over cusped ones in these 3 clusters. However, for the purposes of comparison with cosmological model predictions, and stress-testing the underlying structure formation model itself more clusters, and preferably relaxed ones, need to be analysed to make a conclusive statement about the mass distribution in the very central regions of clusters.

\subsection{Mapping dark matter substructure \& galaxy-galaxy lensing in clusters} 
\label{sec2:dark_matter}

The standard paradigm that describes structure formation in the Universe is the concordance cold dark matter model with a non-zero cosmological constant (LCDM). In simulations of the LCDM model, where dark matter is collisionless, the internal density distribution in dark-matter halos, as noted previously, converges to a roughly "universal" and cuspy density profile, the NFW profile (\citealt{Navarro:1996, Navarro:1997}) over several decades in mass. Moreover, the degree of central concentration of a halo in LCDM depends on its formation epoch, and hence on its total mass (\citealt{Wechsler:2002, Zhao:2003}). In this scenario, objects that virialize early are dense and compact if they get accreted and or bound into a larger halo during hierarchical assembly. These smaller scale collapsed structures are usually referred to as sub-halos that are held within the gravitational potential well of the more massive host halo. Spiraling in toward the center owing to dynamical friction, while they are truncated or disrupted by tidal forces, leads to changes in sub-halo masses, angular momentum and energy \citep{Ghigna:1998, DeLucia:2004}.

On cluster scales, observational tests of several concrete LCDM predictions using lensing mass models have been performed. The abundance of sub-halos in clusters, also referred to the mass function of substructure (or the sub-halo mass function) is predicted to be a power law of the form  $dn/dm \propto m^{-1.8}$. This prediction has been robustly tested with increasingly higher resolution HST data of cluster lenses and generations of state-of-the-art cosmological simulations ranging from the Millenium simulation \citep{Natarajan+2004,Natarajan:2007} to the Illustris simulation suite \citep{Natarajan:2017}. Predictions of tidal stripping of subhalos in LCDM have also been tested \citep{PNtidal2002,Sand:2002, Sand:2004, Natarajan:2009} with tidal radii estimated from lensing mass models. Overall, properties of cluster lenses and their sub-halo mass functions were found to be in good agreement with LCDM predictions. Recently more sophisticated and detailed comparisons with high fidelity mass reconstructions for clusters have become possible, so much so that we are now in a position to stress-test the LCDM model (\citealt{Grillo:2015, Natarajan:2017, Meneghetti+2020}). In fact, \citet{Natarajan:2017} report tension between the observed radial distribution of subhalos in the massive HFF cluster lenses compared to simulated clusters in LCDM, an issue that is replicated on galaxy scales as well, wherein the radial distribution of observed satellites is discrepant with theoretical predictions \cite{Carlsten+2020}. 

\citet{Meneghetti+2020} reported that the observed number of Galaxy-Galaxy Strong Lensing (GGSL; see the right panels of Fig.~\ref{fig2:ggsl}) events, i.e., strong lensing events observed on the scale of individual cluster galaxies on mass scales of $\sim 10^{11}\,M_{\odot}$ in cluster lenses, exceeds the expectations from LCDM simulations by more than one order of magnitude. These significant differences in the spatial and mass distributions of the cluster sub-halos do not seem to be fully fixed by the inclusion in cosmological simulations of baryonic physics, in the form of cooling, star formation, and feedback by active galactic nuclei. \cite{rob21} and \cite{bah21} argue that the reported under-abundance of GGSL events in simulations is due to their low resolution. These authors use considerably higher mass resolution runs, and also include several baryonic physical processes. They report that this results in a higher number density of substructures in lensing clusters, leading to comparable simulated and observed occurrence of GGSL events. GGSL is computed using the area enclosed within smaller scale secondary caustics. However, the majority of the contribution to the GGSL cross section computed from the simulations reported in \cite{rob21,bah21} arises from in-falling group-scale substructures that are more massive than those reported in \cite{Meneghetti+2020}. In addition, in recent work, \cite{Ragagnin+2022} note that the stellar masses of cluster galaxies in the suite of simulations studied by \cite{bah21,rob21} vastly exceed that of observed cluster galaxies which in turn artificially enhances their computed GGSL cross section.

In continuing work pursuing the origin of this discrepancy \cite{Meneghetti+2022} quantify the impact of the numerical resolution and AGN feedback scheme adopted on the predicted GGSL probability to alleviate the gap with observations. Improving the mass resolution by a factor of ten and twenty-five, while using the same galaxy formation model that includes AGN feedback does not affect the GGSL probability. Adopting an AGN feedback scheme that is less efficient at suppressing gas cooling and star formation leads to an increase in the GGSL probability by a factor between three and six, however, while such simulations form overly massive subhalos whose contribution to the GGSL would be higher, their Einstein radii are too large to be consistent with the observations. The primary contributors to the observed GGSL cross-sections are compact subhalos with lower masses (in the mass range of $\sim\,10^{10-11.5}\,M_{\odot}$), which are not present in the appropriate abundance as inferred from observations even in the current highest resolution LCDM simulations. Meanwhile, \cite{Ragagnin+2022} find that regardless of the resolution and galaxy formation model adopted, simulations are unable to simultaneously reproduce the observed stellar masses and compactness (or maximum circular velocities) of cluster galaxies and therefore account for GGSL as measured from observed lenses.

Substructure also exists further away from the centers of clusters and their distribution can also be compared to LCDM simulations. An excess of substructures in Abell~2744 as reconstructed from lensing effects compared to the galaxy clusters in the Millenium XXL $N$-body simulation, as discussed in \citet{sch17}, suggested a mis-match. However, this disagremment can be brought into concordance by more carefully matching the definitions of substructure in simulations and observations. Applying the same definition, reconciles their abundances in Abell~2744 \citep{sch18}. 

An alternative way to look at substructure within clusters is to analyse their projected mass power spectra. This recently proposed approach eliminates the need to identify individual substructures, and relies instead on characterizing the power spectrum of fluctuations as a function of scale  to compare the lensing reconstructed mass maps with that of simulated clusters \citep{moh16,Cerini+2023}.

Strong lensing when combined with kinematics has also made a significant impact on our understanding of the central galaxies in clusters, BCGs, and the dark matter profile slopes in these very central regions of clusters \cite{natarajan+1996}. \cite{Sand:2004,sand2008,newman2013b} found that once the BCG stellar mass of the BCG is subtracted, the dark matter slope is significantly shallower than NFW within $\sim 30$kpc, comparable to the effective radius of BCGs. \cite{newman2015} placed these findings in a larger context of central profile slopes as a function of halo mass. For groups and individual galaxies the central slope get progressively steeper. They conclude that dissipation-less processes are more important in the assembly of BCGs in larger halos, like clusters, leading to shallower slopes. Baryons, through gas cooling, are more important in lower mass halos, like groups and individual galaxies, and lead to steeper central density profiles.

\subsection{Cluster lenses as nature's telescopes}
\label{sec:SPmagnif}

One of the key benefits of gravitational lensing by clusters is the magnification they induce on the background galaxy population. Bringing into view faint background sources that would otherwise not be accessible even with the most powerful space-based telescopes, cluster lenses act as cosmic telescopes, that permit detailed studies of the magnified galaxies behind them. Lensing magnification stretches the area behind clusters by the magnification factor, thereby increasing the flux we receive from distant galaxies as lensing conserves surface brightness. Acting as nature's telescopes, clusters are expected to serendipitously reveal the earliest forming galaxies as magnification enhances their reach to probe the very early Universe.  

At intermediate redshifts, the high lensing magnification by galaxy clusters allows us to peer \textit{into} galaxies. A particularly valuable utility is in studying star forming galaxies at ``Cosmic Noon'', $z \sim 1 -3$, a critical epoch for the assembly of the massive galaxy population in the Universe.  Galaxies that would otherwise be too faint for detailed spectroscopy to be practical, even with the largest telescopes, are magnified by factors of tens to hundreds, making their detailed study feasible \citep{rigby18a}. This is thanks to the overall magnification, which significantly reduces the required observing time for mid- and high-resolution spectroscopy. The high linear magnification, which is manifested as (usually tangential) distortion, resolves the internal structure of galaxies, allowing us to measure the sizes and physical properties of even individual star forming clumps within them \citep[e.g.,][]{johnson2017}. Detailed lens models are then deployed to place the measurements in context of their intrinsic (source plane) values, reverse the magnification effect, and reconstruct the galaxy morphology \citep[e.g.,][]{sharon2022sunburst}.

\begin{figure}
    \centering
    \includegraphics[width=0.495\textwidth]{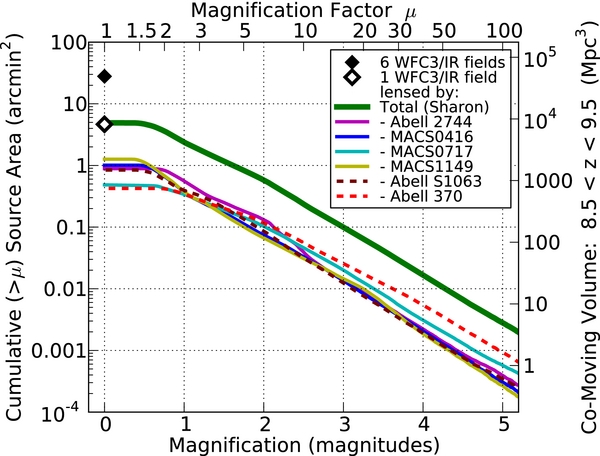}
    \caption{The cumulative area (left axis) and co-moving volume (right axis) of background sources at $8.5<z<9.5$ lensed by each of the HFF clusters to magnification $\mu$ and higher. We show, for reference, the corresponding area and co-moving volume of one and six WFC3/IR field of view regions (diamonds). (Plot taken from \cite{Johnson2014}.)}
    \label{fig:SPmagnif}
\end{figure}

A key output that lens mass modeling provides is the calculation of the area in the source plane that has been magnified by a given factor $\mu$. The area in the source plane that has (unsigned) magnification greater than $\mu$ scales as $\mu^{-2}$. As an example, we present in Fig.~\ref{fig:SPmagnif}, taken from \cite{Johnson2014}, the magnification power of the six HFF clusters modeled by \lenstool, plotted as the cumulative area (left axis) and volume (right axis) magnified above a certain value as a function of magnification. The source plane area in this illustrative case was computed by mapping background sources at $z = 9$, and summing the area that is magnified by more than a given magnification factor. To estimate the volume, the area is multiplied by the co-moving distance between $z = 8.5$ and $z = 9.5$. On average, the total $z = 9$ area that is observed through each cluster lens is about 20\% of the WFC3/IR FOV. Approximately 10\% of this area is magnified by more than a factor of six, which is equivalent to lowering the limiting magnitude by at least 2 magnitudes. 

Cluster lenses have brought into view intrinsically faint and extremely high redshift sources, including a candidate Lyman-break galaxy MACS0647-JD at $z=11$ in the CLASH survey \cite{Coe+2013} and most recently galaxy candidates detected by JWST at $z \sim 9 -13$ magnified by modest factors of 3--8 by the foreground cluster WHL0137 that was observed in 8 JWST filters spanning 0.8--5.0 microns and previously in 9 HST filters spanning 0.4--1.7 microns. Recent data from the JWST UNCOVER \& GLASS surveys have also revealed the presence of a high density of bright galaxies behind the cluster lens Abell 2744 \citep{Castellano+2022}. Other cluster surveys like RELICS have uncovered several highly magnified sources at $z \sim 6 - 7$ \citep{Acebron+2018}. In addition, the number counts of highly magnified sources obtained from cluster lenses permit constraining the  faint end of the galaxy luminosity function from moderate to high redshifts \citep[see for example results from the use of HFF cluster lenses: ][]{Alavi+2016,Bouwens+2017,Livermore:2017,Atek+2018,Yue+2018,Bouwens+2022}. The faint end slope encodes important information about the efficiency of star formation, a key unknown in our current understanding of galaxy formation. JWST with its access to the highest redshift, early forming galaxies is anticipated to transform these studies in the near future.

\subsection{Statistics of giant arcs in clusters}
\label{sec2:arc_statistics}

Bright, merging multiple images of extended background sources lensed by clusters often appear as curved, highly elongated arcs, called giant arcs. 
Originally, arc statistics were introduced in \citet{gro88} to investigate the abundance of giant arcs as a test of the hypothesis that these features were actually produced by gravitationally lensing. The abundance and the observable properties of giant arcs depend on the total mass and density distribution of clusters, their redshift evolution, the properties and evolution of sources, the nature of dark matter, as well as global cosmology \citep[e.g.,][]{bar98,wil99,Wambsganss+2004,Anson+2011,mah14}. In recent years the global cosmological parameters are estimated from other observations, mostly the CMB and baryonic acoustic oscillations (BOA), however giant arcs can be potentially used as a probe of redshift evolution of non-linear structure in the Universe, and the nature of dark matter.

The impact of the evolution of structure formation, in particular for different types of dark matter, on strong gravitational lensing observables is detailed in \citet{mah14} and \cite{ela14} in which warm and cold dark matter structure growth is compared with respect to their giant arc statistics. Both works conclude that warm dark matter (WDM) clusters show more extended and massive substructures than those of their cold dark matter (CDM) counterparts, thus enhancing the shear and convergence. In addition, these WDM substructures reach farther out into the cluster halo than the CDM ones. These effects lead to WDM clusters having enhanced giant arc production and not a reduced one as one would naively expect, if one considers only that small-scale structures are smoothed out in WDM cosmologies. Beyond the single lens plane, the impact of deflecting structures along the line of sight which are uncorrelated with the main lens are analysed in,  e.g., \citet{puc09, Bayliss+2014, li19}. They conclude that such structures need to be taken into account in order not to underestimate the abundance of giant arcs, because line-of-sight structures can boost the lensing cross-section for individual clusters by up to 50\%.  

Complementary to using giant arcs, highly magnified but undistorted images \citep[e.g.,][]{zit09b} can also be used to probe the structure of clusters \citep{wil98,mor15}.  Another approach to arc statistics is to search for observations of uncommon and irregular multiple image configurations deviating from the theoretically predicted ones (usually fold and cusp configurations) \citep{orb09,mee21}. Finding extremely rare image properties can refute the underlying assumptions, e.g., about the statistical distribution of dark matter in a given background cosmology. Though a potentially powerful diagnostic, statistics of giant arcs still needs further improvement in order to be turned into a reliable probe. The main difficulties---accounting for selection effects in sufficiently realistic lensing mass models \citep[e.g.,][]{li16, pla20}---will need to be tackled to make use the upcoming data: a 30-fold increase in the number of giant arcs from the full sky surveys like {\it Euclid} \citep{men13}.  Extensive reviews of the development of arc statistics are given in the introduction of \citet{cam13}, and in the overview paper by \citet{men13}. 

\subsection{Constraints on the nature of dark matter}
\label{sec2:constraints_warm}

Though in the standard cosmological model dark matter particles are cold and collisionless, and none of the existing observations rule that out, the possibility remains that dark matter, or some fraction thereof, is either warm (WDM) or self-interacting (SIDM). Cluster strong lensing can be used to test the properties of dark matter particles and constrain the level of self-interaction. 

{\it WDM:} predicts the existence of numerous sub-galactic scale dark matter halos, which would be largely invisible through in electromagnetic emission \citep{die07,spr08}. In WDM cosmologies, these low mass subhalos would be erased by the large streaming length of relativistic particles \citep{bos17,ros20}. Specifically,  and WDM predict very different abundances of subhalos with masses $\lesssim 10^8-10^9\msol$. Therefore, measuring subhalo abundances and properties down to this mass range will help indicate what cosmological model we live in. Isolated subhalos are very hard to detect gravitationally, however, in galaxies and clusters of galaxies their small individual lensing effects can be detected because they are superimposed and are amplified by the magnification effect of the underlying larger scale lens \cite{Cagan+2023}. In clusters, their presence will be especially easy to detect near the critical curves created by the cluster as whole, where magnifications are very high over $\sim$arcsec scales. These probes hold great promise \citep{kau19,dai20,gri21,diego22}, and as suggested in recent work, new lens modeling methodologies like the use of the curved arc bases might permit future detection \cite{Cagan+2023}.

{\it SIDM:} The recent interest in self-interacting dark matter was spurred by apparent discrepancies with the predictions of  on galaxy scales \citep{spe00} -- the so-called cusp-core problem. And even though most of these observations can now probably be accounted for with collisionless dark matter, the interest in dark matter self-interactions persists.  There are a few cluster-scale measurements that can be performed using strong lensing to place constraints on dark matter particle self-interaction cross-section per unit mass, $\sigma/m$. Some of the earliest proposed tests were cluster ellipticity \citep{mir02}: SIDM should result in rounder cluster cores and the sizes of tidal truncation radii for subhalos, that provide an estimate on the interaction cross-section \citep{Natarajan_2002}. However, it was later shown \citep{pet13} that ellipticity is likely subject to other dynamical effects, and cluster core sizes may likely provide a better diagnostic. The size of constant density cores in clusters can also place constraints on SIDM; a recent core estimate of $\lesssim 4\kpc$ in Abell 611 implies $\sigma/m\lesssim 0.1$g/cm$^2$ \citep{and19}. Using results of BAHAMAS-SIDM hydrodynamic simulations \cite{rob19} discuss lensing-based test of SIDM for future cluster surveys. For the current CLASH survey they conclude that cross-section $\sigma/m\gtrsim 1$g/cm$^2$ (for systems with velocity dispersions of $1000\,\kms$) can be ruled out. Baryons play an important role in determining the properties of clusters' central regions. Taking their distribution into account, \cite{bib2:despali2019} conclude that future wide-field surveys might be able to distinguish between CDM and SIDM models based on the distribution of the Einstein radii of cluster lenses. Because SIDM leads to flat density cores in clusters, the central bright galaxies are expected to wobble (i.e., wander away from the cluster center) and result in larger cluster-mass dependent offsets in SIDM cosmologies, compared to LCDM. Based on BAHAMAS-SIDM simulations, \cite{bib2:harvey2019} conclude that their sample of 10 observed clusters is consistent with CDM at $1.5\sigma$ level. One of the additional predictions of SIDM is that dark matter halos can undergo gravothermal collapse late in their evolution, similar to that of Globular Clusters, thereby increasing the central halo densities by a factor of several \citep{yan21}. In that case one could expect individual galaxies in galaxy clusters to act as exceptionally strong lenses, producing $\sim$ arcsecond scale multiple images. As noted previously many such GGSLs have recently been observed in galaxy clusters \citep{Meneghetti+2020}, and though their high abundance still remains an open question within collisionless , SIDM accompanied by gravothermal collapse may offer an alternative if warranted.

Clusters offer yet another powerful test of the nature of dark matter. Similar to gas ram pressure stripping, self-interactions between dark matter particles would tend to transfer momentum between two cluster-scale halos in merging clusters, or between cluster-scale halo and embedded galaxy-scale halo, resulting in offsets between baryonic and dark matter peaks. These effects have been investigated in simulations \citep{bib2:kahlhoefer2015},  and signatures of these interactions were searched for in galaxy clusters. After initial reports about finding offsets \citep{bib2:massey2015}, better data acquired later showed that the offsets were consistent with zero interaction cross-section \citep{bib2:massey2018}. \cite{bib2:robertson2017} use the Bullet Cluster and estimated that even an interaction cross-section as large as $\sigma/m=2\,$cm$^2$g$^{-1}$ is not ruled out by this cluster. However, they also caution that measuring offsets between baryonic and dark matter peaks must be done consistently in simulations and observations and this remains a challenge.

\subsection{Cosmography with cluster lenses} 
\label{sec2:cosmography_clusters}

If a source is strongly lensed into two images, the difference in time that light takes to reach the observer from the two different directions, i.e., the time delay ($\Delta t$) between the two images due to the difference in path length, is related to the gravitational potential of the lens ($\phi[{\theta}$]) and the so-called time-delay distance ($D_{\Delta t}$). The latter scales with the three angular-diameter distances, observer-deflector ($D_{\rm d}$), observer-source ($D_{\rm s}$), and deflector-source ($D_{\rm ds}$): $D_{\Delta t} \propto D_{\rm d}\times D_{\rm s}/D_{\rm ds}$. Since each of the distances is proportional to $c/H_0$, it implies that $D_{\Delta t} \propto 1/H_0$. The uncertainty in $D_{\Delta t}$ (and thus on $H_{0}$) is approximately the sum in quadrature of the uncertainty in time-delay ($\delta\Delta t$) and the uncertainty in differences in gravitational potential ($\delta \phi$). The time delays observed in the multiple images of time-varying sources strongly lensed by galaxy clusters have typical values of several months/years, so they can be measured with an approximate 1-3\% precision (e.g., \citealt{Fohlmeister:2013, Dahle:2015, Kelly:2016}). Most recently, \cite{mun22} report $<0.05\%$ precision on the 3 time delays of the quasar in cluster SDSS J1004+4114, which has been monitored for 14.5 years. Supernova, with their known lightcurves can provide precise estimates with considerably shorter monitoring. The main remaining uncertainty lies in the potential of the cluster lens. With sufficient photometric and spectroscopic observational data, it may be possible that $\delta \phi \lesssim$ 5\% can be reached in lens galaxy clusters \citep[e.g.,][]{Grillo:2015, Grillo:2016}. This corresponds to a remarkable precision of $\sim$6\% on the value of $H_0$ from a single lens cluster with measured time delays (e.g., \citealt{Grillo:2018, Grillo:2020}).

Time-delay distances are primarily sensitive to the value of $H_{0}$, and more mildly on those of other cosmological parameters. If a lens produces multiple images of two sources (1 and 2), located at different redshifts, the observed positions of the multiple images provide information about $\phi$ and the so-called family ratio $(D_{\rm ds1} \times D_{\rm s2})/(D_{\rm s1 }\times D_{\rm ds2})$. In galaxy clusters, usually showing several multiple images, different values of this ratio can be used at the same time to estimate the values of the cosmological matter ($\Omega_{\rm m}$) and dark-energy ($\Omega _{\Lambda}$) density parameters, and the dark-energy equation-of state ($w$) parameter \citep[e.g. ][]{Soucail:2004,GilmorePN2009}, defining the global geometry of the Universe. Using observed cluster lenses \cite{Jullo:2010,Caminha:2016,Magana+2018} have demonstrated that cluster strong lensing cosmography could be competitive with other cosmological probes once systematics are better quantified and controlled.

Multiple images of supernova Refsdal in HFF cluster MACS J1149 was recently used to constraint $H_0$ (\citealt{kelly+2023}). In addition to supernova, variable quasars can also be used as sources behind clusters. One of these is SDSS\,J1004$+$4112. The time delay of the fourth arriving image was predicted by different groups some years ago. After a nearly 15 year observing campaign that delay was finally measured to be $6.73\pm 0.003$ years \citep{mun22}, and agrees well with some of the predictions \citep{lie09,moh15}. Using the measured time delays in three clusters with background quasars and archival HST imaging, \cite{nap23} measured the value of $H_0$ which agrees with the one from Refsdal within $1\sigma$, but has somewhat larger modeling uncertainties, because the 3 clusters have fewer lensed images than HFF clusters. They estimate that of order of 50 cluster-lensed quasar systems would help reduce the uncertainties on $H_0$ to below the 1\% level, which is expected to be detected in the next decade \citep{robertson2020}. These results show that galaxy clusters are competitive in the measurement of the Hubble constant, once sufficient systems are discovered and analyzed.

Most of the baryons in galaxy clusters are in the form of hot, X-ray emitting gas. Massive galaxy clusters can be assumed to have formed from representative regions of the Universe, and hence the ratio of the mass in baryons they contain to that of dark matter (i.e., the gas mass fraction) should be a constant, not evolving with redshift. This assumption can be used to constrain global cosmological parameters, like $h^2\Omega_b$, and $\Omega_m$  \citep{applegate2016,allen2011,mantz2014}. Lensing plays a key role in this determination as it measures the cluster masses. One can also turn the argument around: assume cosmological parameters and determine if the baryon fraction in clusters evolves with redshift. \cite{holanda2017} estimate that there is no statistically significant evolution of the baryon mass fraction with $z$.

\subsection{Impact of Line-of-sight structures and multiple-lens planes}

\label{sec2:line-of-sight_structures}

Traditional single-plane lensing analyses assume that the mass of the deflector is concentrated in a single plane, at the redshift of a cluster. More advanced multi-plane lensing studies, taking into account the observed mass distributions at redshifts different from that of the cluster, are becoming possible, thanks to some recent software developments. In the former case, the light rays emitted from a distant source and received by an observer are deflected only once (i.e., when crossing the galaxy cluster); in the latter, multiple deflections have to be added properly \citep[see][]{Blandford1986,Schneider:2014}. The observed positions and fluxes of each galaxy contributing to the total deflection has to be re-scaled iteratively in the modeling process, according to the deflection and magnification values associated with the mass of the objects in the foreground. 

Studies by \citet{Bayliss+2014,Anson+2014,Chirivi:2018,wil18,McCully2017} found that line-of-sight (LOS) galaxies can have a significant impact on the reconstruction of some multiple image positions, but the inclusion of LOS mass distributions in the strong lensing models of galaxy clusters can reduce only modestly the rms offset between the observed and model-predicted positions of all multiple images. These studies also noted that the perturbing contribution of foreground galaxies is more important than that of the background ones and that accounting for LOS mass structures, as if they were at the cluster redshift, can partially lower this offset. Recently, \citet{Raney:2020} confirmed that LOS galaxies can account for a significant fraction of the typical image position residuals of current state-of-the-art strong lensing models of galaxy clusters and that the values of the image magnifications can be affected by a few percent. 

\subsection{Local properties of clusters around extended images} 
\label{sec2:local_properties}

Most of the work on cluster lensing presented in the literature involves reconstruction of the mass distribution over the whole lens plane region of the cluster. As described in earlier sections, this reconstruction can use simply parametrized models, free-form models, or hybrid ones. However, a very different type of analysis is also possible, one that does not involve modeling the whole contiguous central region of the cluster, and recovers cluster properties only {\it at} the locations of extended images.

The strong gravitational lensing formalism provides only \emph{local} information about the deflecting mass distribution in the lens plane at the positions of the multiple images. Assuming that the surface mass density and shear are constant over the area considered for evaluation around the multiple image locations, we can use the image morphology to retrieve ratios of convergences, $f^{(ij)}$ between multiple images $i$ and $j$ arising from the same individual source, $i,j=1,...,n_\mathrm{i}$ (with $n_\mathrm{i}$ being the total number of images from the same source), and the reduced shear components at the positions of all $n_\mathrm{i}$ images 
\be
f^{(ij)} \equiv \frac{1-\kappa(\tang^{(i)})}{1-\kappa(\tang^{(j)})} \;, \quad g^{(i)}_1 = \frac{\gamma_1(\tang^{(i)})}{1-\kappa(\tang^{(i)})} \;, \quad g^{(i)}_2 = \frac{\gamma_2(\tang^{(i)})}{1-\kappa(\tang^{(i)})} \;.
\label{eq2:fg}
\ee
These convergence ratios and reduced shear components can be determined by features in the brightness distributions of the multiple images, most often star forming regions.
At least three multiple images with three features that are not aligned in each image are required. \citet{wag19} details how to uniquely determine the $f^{(ij)}$ and $g^{(i)}$ from these feature vectors, \cite{lin22} and \cite{lin23} show the impact of non-unique feature matching when one image contains multiple candidate features that can be matched with one single feature showing in other images. \cite{lin22} also set up an algorithm to identify the positions of these features in the presence of noise.  Alternatively, the quadrupole moment of a featureless brightness profile around its centre of light can be employed\footnote{An example for the usage of the quadrupole ellipticity and direction is given in \citet{wag20} albeit for a galaxy-scale lens.}.

Inserting \eqref{eq2:fg} into the distortion matrix for each multiple image
\begin{equation}
\mathrm{A} = \begin{pmatrix}
  1 - \kappa - \gamma_1 &  \gamma_2 \\
  \gamma_2              &   1 - \kappa + \gamma_1
  \end{pmatrix} \;,
\end{equation} 
ratios of magnifications can also be determined based on these local lens properties. If flux measurements covering the same area of the multiple images are available, a comparison between the flux ratios and the magnification ratios yields a consistency check for the local lens properties in \eqref{eq2:fg}. Deviations may hint at dust extinction, additional micro-lensing, or lensing effects beyond convergence and shear. 

To investigate the impact of higher-order lensing effects in galaxy clusters, gradients of reduced shear and convergence maps in simulated clusters, like {\it Ares} and {\it Hera} \citep{men17}, are analysed, as further detailed in \cite{wag22}. On the resolution scale of the pixels used in these simulations, the results reveal that, in more than 90\% of all pixels covering the lensing area, the gradients in $\theta_1$- and $\theta_2$-directions are less than 10\% of the convergence and reduced shear values at these pixels. More than 67\% of all pixels in the {\it Ares} and {\it Hera} lensing area have gradients of less than 2\% of their respective distortion matrix entries.  Hence, at current observational precision, neglecting higher-order lensing effects on galaxy-cluster scales is valid to a good approximation.

\begin{figure}[h]
\centering
\includegraphics[width=0.48\textwidth]{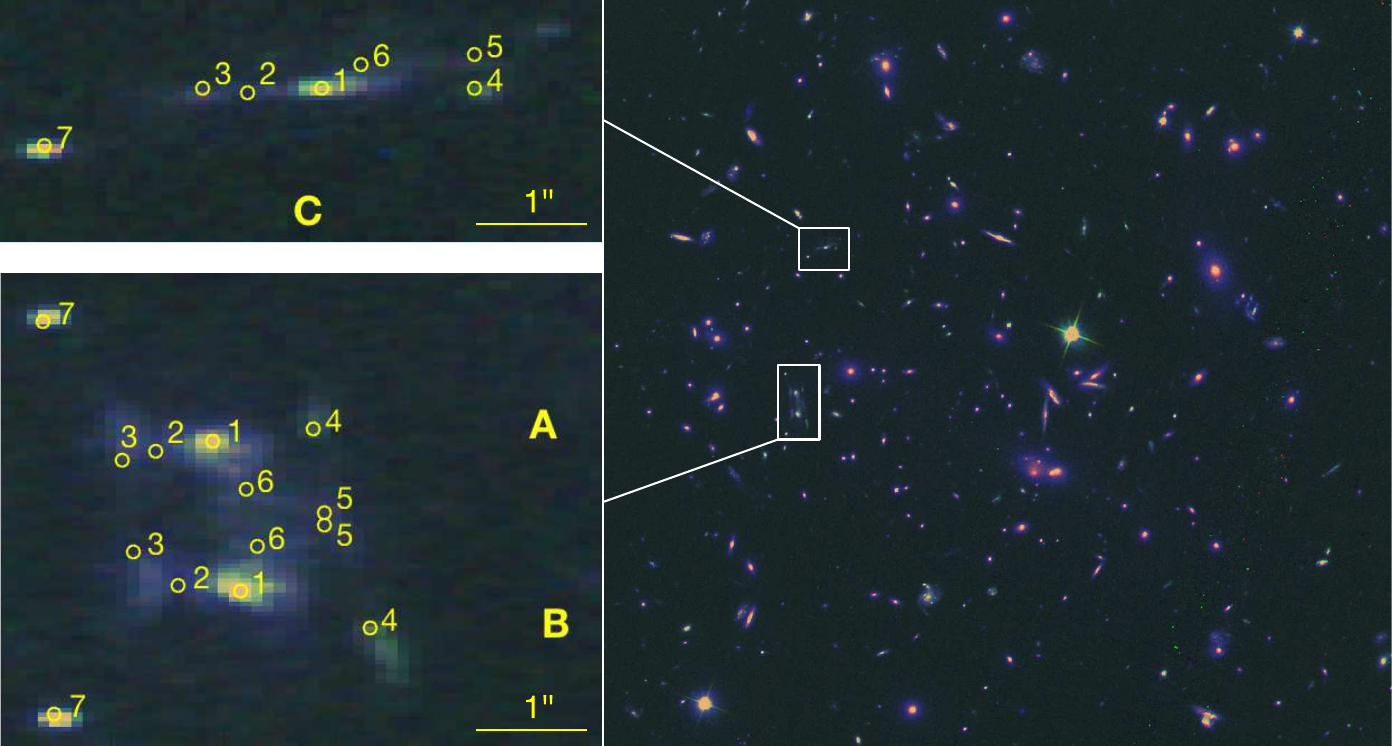}
\caption{Galaxy cluster RM J223013.1-080853.1 at $\zd=0.526 \pm 0.018$ with a cusp configuration of highly resolved images from a clumpy disk galaxy at $\zs=0.8200 \pm 0.0005$ (right). The features in the brightness profiles (numbered yellow circles in the details on the left) are used to reconstruct the local lens properties in \eqref{eq2:fg}.}
\label{fig2:hamiltons_object}
\end{figure}

In addition, assuming a smooth deflecting mass density in the vicinity of the multiple images that straddle a critical curve, we can use the observables in the multiple images to constrain the position of the critical curve between these images. To leading order, points on the critical curve between two images in a fold-configuration can be determined as the midpoint dividing the connection line of corresponding features in the two multiple images. Retrieving local lens properties is very useful in newly detected galaxy clusters in which only few spectroscopically confirmed multiple images have been identified. Using a synthetic cluster {\it Ares}, \citet{joh16}  estimate that one needs $>10$ spectroscopically confirmed image sets to precisely determine the mass enclosed within the Einstein radius of a cluster-lens. Thus, the lens modelling algorithms introduced in Sect.~\ref{sec2:modelling_methods} cannot be applied to galaxy clusters like RM J223013.1-080853.1 \citep{ryk16}, which was identified after a serendipitous discovery of a multiple-image system. So far, it only shows three spectroscopically confirmed multiple images for a common background galaxy at  $\zs =  0.8200$. Nevertheless, \citet{gri21} succeeded in constraining the smoothness scale of dark matter by applying the approach outlined above to the three highly resolved images. \figref{fig2:hamiltons_object} shows the galaxy cluster at $\zd=0.526$, determined by its $\sim$60 cluster member galaxies, and details of the confirmed multiple images. 

Using the \ptmatch-program\footnote{available at \url{https://github.com/ntessore/imagemap}} \citet{wag18b,wag18} determine the properties in \eqref{eq2:fg}. The fact that the algorithm obtains a solution supports the assumption of constant entries in the distortion matrix in the areas spanned by the brightness features shown in \figref{fig2:hamiltons_object} (left). The observed flux ratios in all five available \hst\ filters (F110W, F140W, F160W, F606W, F814W) also show a high degree of agreement with the magnification ratios of \ptmatch, which is a second consistency check corroborating the assumption. Each lens model has its underlying assumptions, for instance that light traces mass \citep[see][]{jul07,zit09}, or regularisation constraints on the smoothness of the mass density. Such assumptions are not employed in the observation-based approach outlined in Sect.~\ref{sec2:local_properties}. A comparison between the local lens properties (\eqref{eq2:fg}) extracted from the lens-model reconstructions at the multiple image positions and the same properties directly obtained from the multiple-image observables (according to the procedure described in Sect.~\ref{sec2:local_properties}) allows us to corroborate or refute lens model assumptions. 

As an example, \citet{wag18} use the strongly-lensing galaxy cluster CL0024+1654 at $\zd=0.39$ to investigate the differences between the lens-model-independent local lens properties, those obtained in a \lenstool\ and a \grale\ reconstruction. The local lens properties are extracted at the positions of five highly resolved images of a blue spiral background galaxy at $\zs=1.675$. In addition to these images, the \lenstool\ and \grale\ reconstructions employ the same set of 14 multiple image positions of 5 other background sources.  Images' extended nature is not taken into account, only their centres of light are used as image positions. 

\begin{figure}[ht]
\centering
\includegraphics[width=0.45\textwidth]{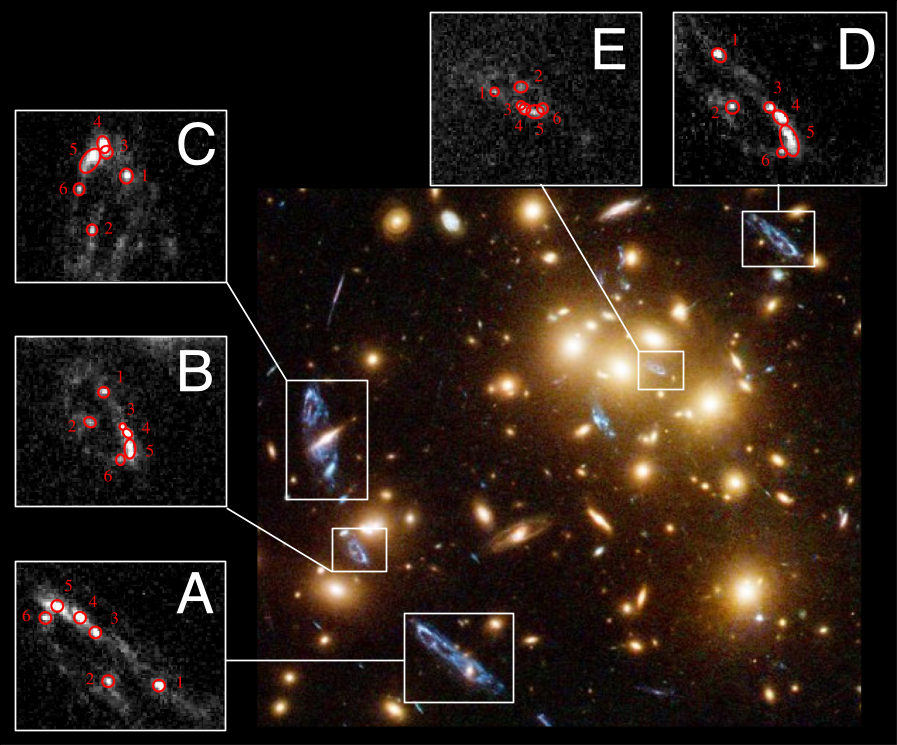}
\caption{The galaxy cluster strong gravitational lens CL0024+1654 at $\zd=0.39$ (background) and the five multiple images of a blue spiral background galaxy at $\zs=0.1675$ (details around CL0024) with their six brightness features, star-forming regions, that can be used to extract the local lens properties in \eqref{eq2:fg}. \citep[See ][for details.]{lin22}}
\label{fig2:cl0024_mis}
\end{figure}

The six brightness features in the five resolved multiple images shown in \figref{fig2:cl0024_mis} constrain the local lens properties of \eqref{eq2:fg} at the five positions in the cluster area for several filter bands, which shows overall consistency of these local lens properties \citep{lin22}. Extracting the same local lens properties determined by the \lenstool\ and \grale\ reconstructions, the authors find a high degree of agreement between the $f^{(ij)}$- and $g^{(j)}$-values of all three approaches. This implies that in the vicinity of the highly resolved multiple images, (i) the light-traces-mass assumption and the hypothesis of a constant mass-to-light-ratio for the brightest cluster member galaxies as used in \lenstool\ can both be corroborated for CL0024+1654; (ii) \grale\ arrives at similar conclusions via a completely different route.

\begin{figure*}[ht]
\centering
\includegraphics[width=0.37\textwidth]{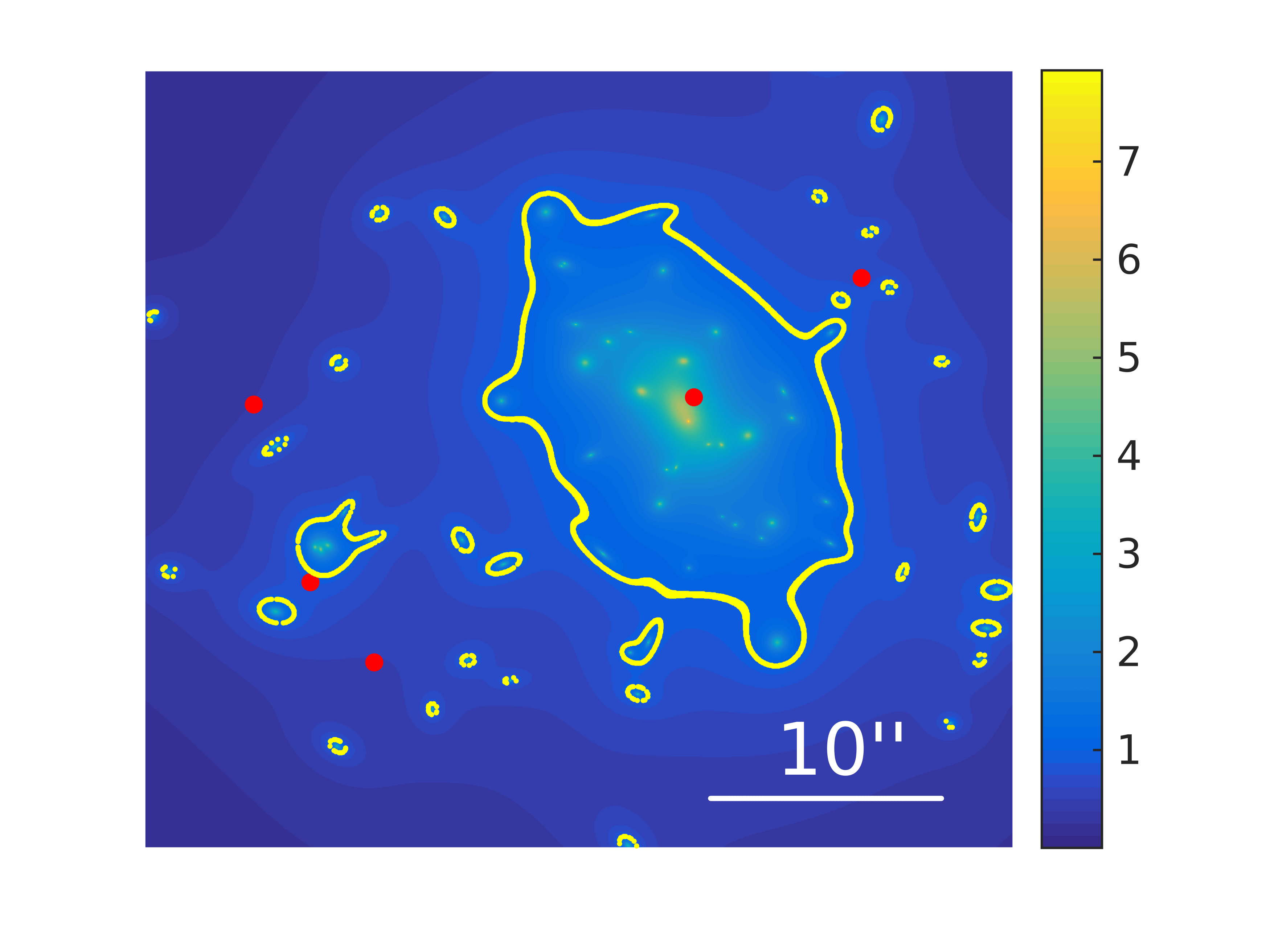} \hspace{5ex}
\includegraphics[width=0.37\textwidth]{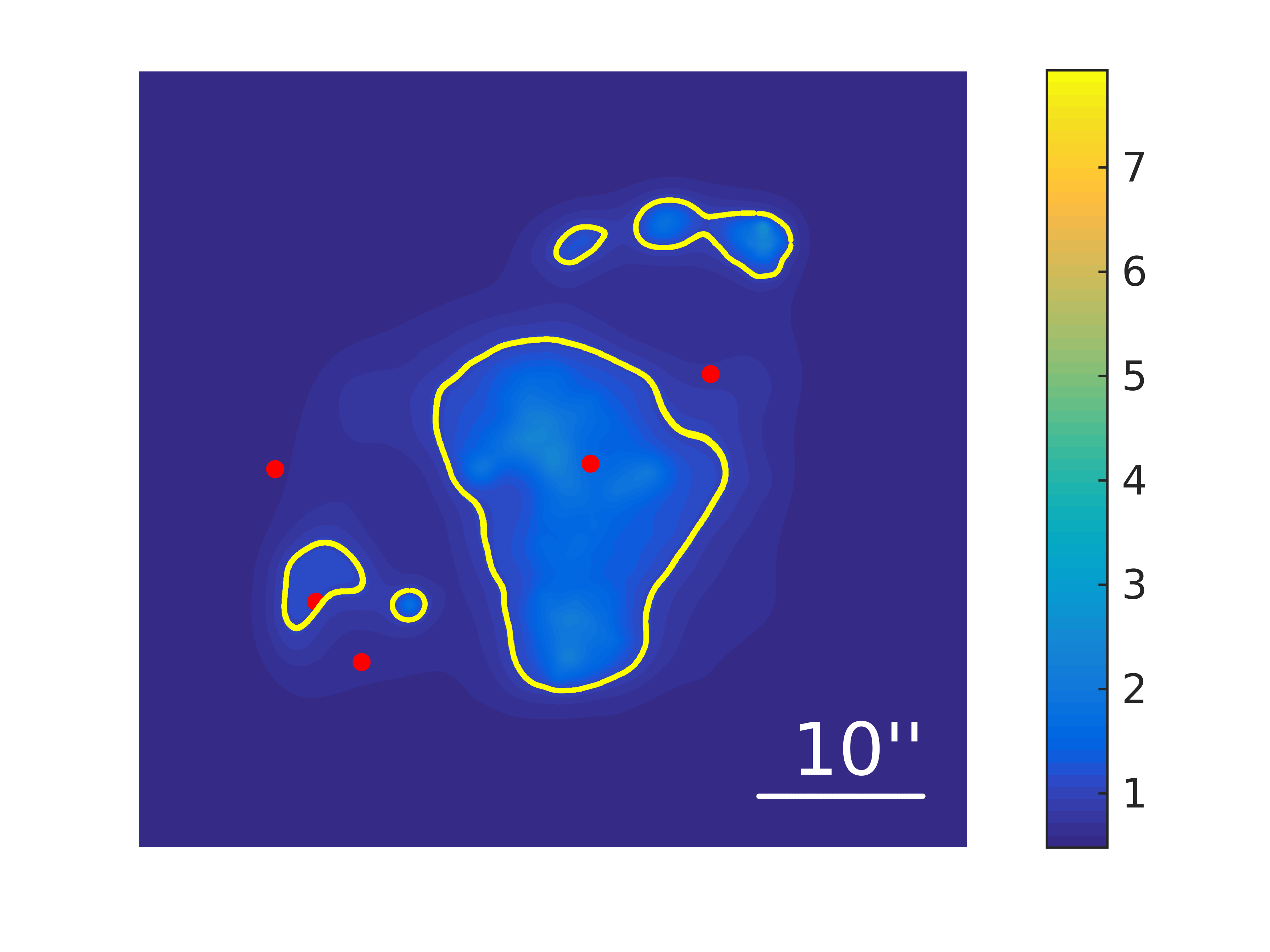} 
\caption{Convergence maps for CL0024+1654 obtained from \lenstool\ (left) and \grale\ (right) using the five images of the highly resolved spiral galaxy (red dots) and 14 additional images from five other source. 
The yellow curves mark the isocontour $\kappa~=~1$. 
\textit{(Image credits: \citet{wag18})}}
\label{fig2:cl0024_kappa}
\end{figure*}

Despite the fact that \lenstool\ and \grale\ generate the same \emph{local} lens properties at the five image positions, their \emph{global} convergence maps greatly differ, as shown in \figref{fig2:cl0024_kappa}.  This is to be expected because the lens models are dominated by the multiple image observables only at their positions. Farther away from the data points, lens-model specific additional assumptions determine the global morphology of the reconstruction, causing the differences in the resulting convergence maps. 

In the regime of several tens of multiple-image systems, the differences in the convergence maps between model-based lens reconstructions, as described above, occur because the sparse multiple-image constraints are not sufficient, and need to be complemented by additional assumptions to obtain a global lens morphology.  The observation-based, model-free method discussed in this Section can be used to compare and cross-check additional modeling assumptions. In the future, observations with JWST will increase the number of multiple images per cluster by an order of magnitude from about 100 to 1000 \citep{gho20}. Accordingly, the impact of model assumptions will decrease and the area covered by multiple images which is reconstructed with an accuracy above 90\% will increase from about 40-50\% to about 65\%. The use of model-free methods can also be extended to more images, further confirming the recovered mass distribution in clusters.

 \section{Lensing by Galaxy Groups}\label{sec2:groups}

Galaxy groups are similar to clusters, but have lower masses, $\sim 10^{13}-10^{15} M_\odot$, and so `fill' the mass gap between individual galaxies and clusters. Given that they are more numerous, galaxy groups can be important contributors to the study of background sources, the mass distribution in lenses, or cosmological parameters. For that reason, \cite{fox01} suggested that groups can be used to test for the NFW profile. They analytically compared the lensing efficiency of NFW vs. singular isothermal sphere (SIS) profiles, and concluded that NFW dark matter halos are $\sim 200$ less efficient than SIS, therefore strong lensing statistics by groups could provide a sensitive test of the inner density region of virialized structures. Their finding of low group cross-section to strong lensing is consistent with there being no instances of strong lensing by groups in the Hubble Deep Field. \cite{Newman_2015} addressed the same question of group's density profile. They used a sample of 10 strong lensing groups and found that the central regions are consistent with NFW, but are possibly somewhat steeper, which they ascribed to the contribution from the central elliptical galaxy. \cite{Verdugo_2014} analyzed how the Einstein radius of galaxy groups in the Strong Lensing Legacy Survey \citep{Limousin_2009,More_2012} relates to the group's properties, like velocity dispersion. They showed that all correlations between group properties have a considerable amount of scatter, but there is some evidence that the Einstein radius is decreasing with increasing redshift. \cite{Foex_2014} used weak lensing data for 80 strong lensing galaxy groups to estimate the groups' mass vs. concentration relation. The relation they found is steeper than that seen in simulations. They attribute this to the fact that strong lensing selected groups are preferentially aligned along our line of sight because that way they present a larger strong lensing cross-section, and are more easily detected. Detailed study of one group, SL2S\,J02140-0535 by \cite{Verdugo_2011}, whose measured concentration parameter is somewhat larger than predicted by simulations, is in agreement with this. Therefore galaxy groups selected by their strong lensing features may constitute a biased sample of groups in general.

\begin{figure}
    \centering
    \includegraphics[trim={0cm 0cm 0cm 0cm},clip,width=0.495\textwidth]{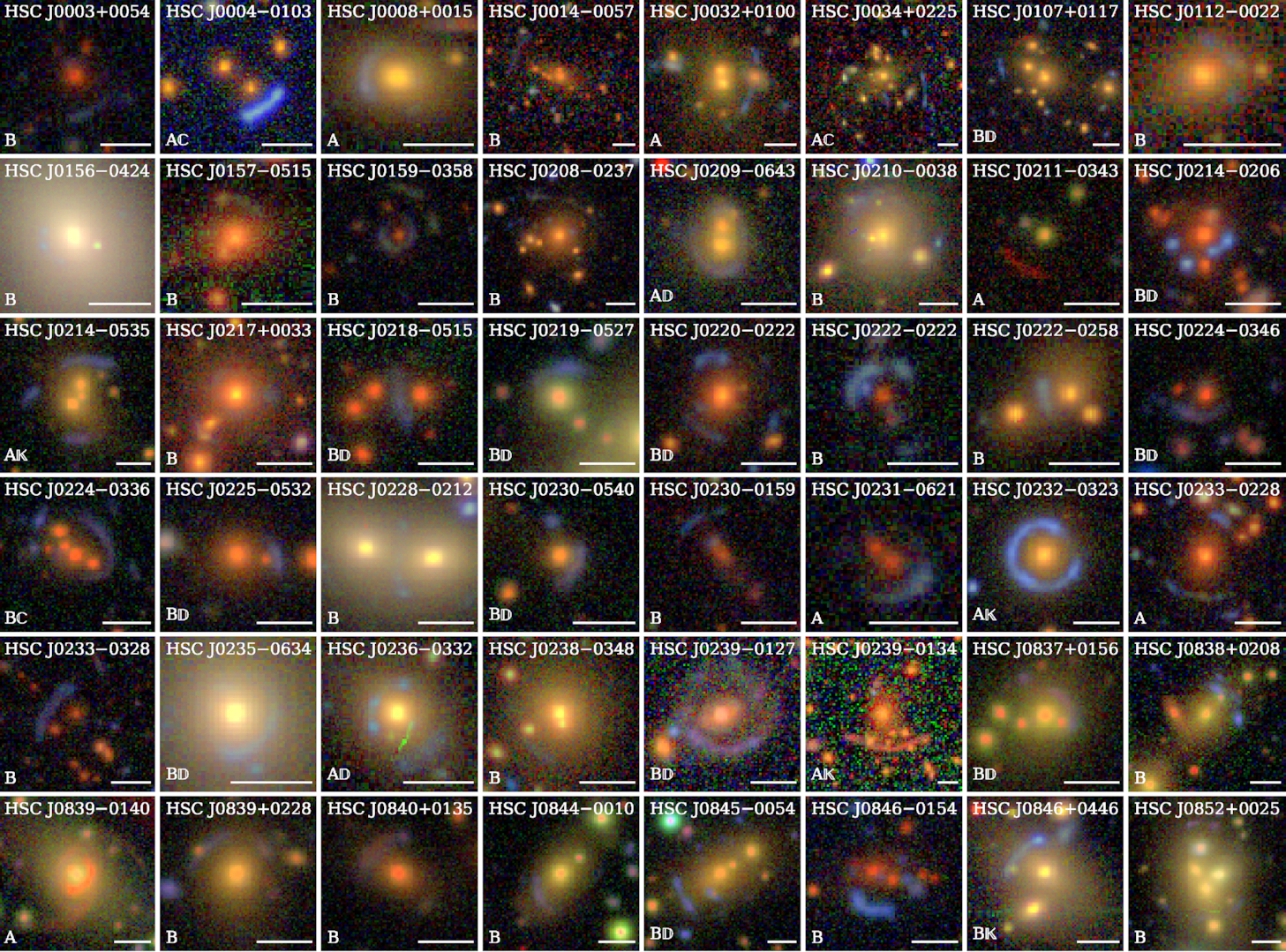}
    \caption{Examples of group-scale lensing candidates from the Hyper Supreme Cam on the {\it Subaru Telescope} \citep{jaelani20}. The letters A or B at the bottom of each panel indicate the confidence grade of this being a lens. The white horizontal bar at the bottom of each panel is 5 arcseconds long. North is up, East is to the left. }
    \label{fig:groups}
\end{figure}

In the recent years there have been a few dedicated searches for strong lensing in galaxy groups that aim to find and analyze or order 100 or more groups, instead of 10 or so used in earlier works. \cite{jaelani20} carried out a Survey of Gravitationally-lensed Objects in HSC Imaging using Hyper Suprime-Cam Subaru Strategic Program (HSC-SSP) Survey \citep{aihara18}, covering $\sim 1114$deg$^2$. They discovered 641 candidate lens systems, of which 47 are almost certainly bonafide lenses; see Fig.~\ref{fig:groups} for some examples. The future of group-scale lensing will definitely make use of such large surveys where the observational selection effects are well quantified, and resulting catalogues of groups can be used for statistical studies. But the other aspect of groups, detailed study of individual groups that have abundant data on both lensing and kinematics also promises to yield interesting results. For example, \cite{sec2:wang22} studies the mass distribution in CASSOWARY 31, using strong lensing data from HST imaging and kinematics from MUSE integral-field spectroscopy. They identified 5 sets of multiple images, on par with some clusters, and obtained detailed mass maps of the system, concluding that it is peculiar fossil group, strongly dominated by dark matter in its central regions. Systems like this can finally realize the prediction from 20 years ago \citep{fox01} that groups are excellent systems to probe the structure of dark matter halos.

\section{Prospects for cluster lensing from upcoming instruments and surveys} 
\label{sec2:prospects_upcoming}

JWST observations are expected to revolutionize the study of cluster lensing. The early results are very promising and exciting. \cite{Caminha2022,Golubchik2022,Mahler2022,Sharon2022} all present mass models of the first strong gravitational lens observed by the JWST, SMACS J0723.3-7327. In Fig.~\ref{fig:smacs} we show the lens model presented in \cite{Mahler2022}. The first deep-field observations of the JWST have yielded a surprisingly large number of very high redshift galaxy candidates, $z=10-16$ \citep{Furtak2022}. The JWST Prime Extragalactic Areas for Reionization and Lensing Science (PEARLS) project will look for high redshift galaxies behind 7 lensing clusters \citep{Windhorst2022}. Meanwhile \cite{Laporte2022} report finding a lensed protocluster candidate at $z=7.66$ behind a lensing galaxy cluster SMACS0723-7327. JWST data will allow testing our current cosmological paradigm via the detailed studies of substructure that it will provide for direct comparison with LCDM simulations and predictions.

The upcoming Euclid space mission with the expectation of detecting upto 60,000 galaxy clusters from $z\sim 0.2-2$, and of the order of several thousand strongly lensed arcs stands to transform our understanding of the assembly history of clusters. With accurate weak lensing mass estimates out to the virial radius, Euclid will enable tracing the mass build-up and growth of structure in the Universe. Euclid is expected to provide a stringent test of the cold dark matter paradigm as data from the mission will permit probing the radial density profile and the ellipticity of clusters, both of which are predicted by LCDM. The anticipated data deluge has catalyzed the development of many new automatic lensing detection algorithms. In combination with DES, Euclid data will provide tighter cosmological constraints from clusters. On the modeling side, improved existing lens inversion methods, as well as new reconstruction methodologies under development currently will rise to meet the challenge of JWST data and data from other planned surveys are expected to improve our understanding of the clusters and the high redshift Universe. 

\begin{acknowledgements}

We thank the International Space Science Institute in Bern (ISSI) for their hospitality for organizing the stimulating workshop on ``Strong Gravitational Lensing''. We thank Masamune Oguri and Anupreeta More for helpful comments and discussions.
\end{acknowledgements}

\section*{Conflict of Interest Statement}
The authors declare no competing interests.

\bibliographystyle{spbasic-FS} 
\bibliography{references}                

\begin{thebibliography}{304}
\expandafter\ifx\csname url\endcsname\relax
 \def\url#1{\burl{#1}}\fi
\expandafter\ifx\csname urlprefix\endcsname\relax\def\urlprefix{URL }\fi
\providecommand{\bibinfo}[2]{#2}
\providecommand{\eprint}[2][]{\url{#2}}
\providecommand{\doi}[1]{\urlstyle{rm}\url{https://doi.org/#1}}

\bibitem[{{Abdelsalam} et~al.(1998){Abdelsalam}, {Saha}, and
  {Williams}}]{abd98}
{Abdelsalam} HM, {Saha} P, {Williams} LLR (1998) {Nonparametric Reconstruction
  of Abell 2218 from Combined Weak and Strong Lensing}. \aj 116(4):1541--1552.
  \doi{10.1086/300546}.
  {\href{https://arxiv.org/abs/astro-ph/9806244}{{arXiv:astro-ph/9806244}}}
  {[astro-ph]}

\bibitem[{{Acebron} et~al.(2018){Acebron}, {Cibirka}, {Zitrin}, {Coe},
  {Agulli}, {Sharon}, {Brada{\v{c}}}, {Frye}, {Livermore}, {Mahler}, {Salmon},
  {Umetsu}, {Bradley}, {Andrade-Santos}, {Avila}, {Carrasco}, {Cerny},
  {Czakon}, {Dawson}, {Hoag}, {Huang}, {Johnson}, {Jones}, {Kikuchihara},
  {Lam}, {Lovisari}, {Mainali}, {Oesch}, {Ogaz}, {Ouchi}, {Past},
  {Paterno-Mahler}, {Peterson}, {Ryan}, {Sendra-Server}, {Stark}, {Strait},
  {Toft}, {Trenti}, and {Vulcani}}]{Acebron+2018}
{Acebron} A, {Cibirka} N, {Zitrin} A, {Coe} D, {Agulli} I, {Sharon} K,
  {Brada{\v{c}}} M, {Frye} B, {Livermore} RC, {Mahler} G, {Salmon} B, {Umetsu}
  K, {Bradley} L, {Andrade-Santos} F, {Avila} R, {Carrasco} D, {Cerny} C,
  {Czakon} NG, {Dawson} WA, {Hoag} AT, {Huang} KH, {Johnson} TL, {Jones} C,
  {Kikuchihara} S, {Lam} D, {Lovisari} L, {Mainali} R, {Oesch} PA, {Ogaz} S,
  {Ouchi} M, {Past} M, {Paterno-Mahler} R, {Peterson} A, {Ryan} RE,
  {Sendra-Server} I, {Stark} DP, {Strait} V, {Toft} S, {Trenti} M, {Vulcani} B
  (2018) {RELICS: Strong-lensing Analysis of the Massive Clusters MACS
  J0308.9+2645 and PLCK G171.9-40.7}. \apj 858(1):42.
  \doi{10.3847/1538-4357/aabe29}.
  {\href{https://arxiv.org/abs/1803.00560}{{arXiv:1803.00560}}} {[astro-ph.CO]}

\bibitem[{{Acebron} et~al.(2019){Acebron}, {Alon}, {Zitrin}, {Mahler}, {Coe},
  {Sharon}, {Cibirka}, {Brada{\v{c}}}, {Trenti}, {Umetsu}, {Andrade-Santos},
  {Avila}, {Bradley}, {Carrasco}, {Cerny}, {Czakon}, {Dawson}, {Frye}, {Hoag},
  {Huang}, {Johnson}, {Jones}, {Kikuchihara}, {Lam}, {Livermore}, {Lovisari},
  {Mainali}, {Oesch}, {Ogaz}, {Ouchi}, {Past}, {Paterno-Mahler}, {Peterson},
  {Ryan}, {Salmon}, {Sendra-Server}, {Stark}, {Strait}, {Toft}, and
  {Vulcani}}]{Acebron:2019}
{Acebron} A, {Alon} M, {Zitrin} A, {Mahler} G, {Coe} D, {Sharon} K, {Cibirka}
  N, {Brada{\v{c}}} M, {Trenti} M, {Umetsu} K, {Andrade-Santos} F, {Avila} RJ,
  {Bradley} L, {Carrasco} D, {Cerny} C, {Czakon} NG, {Dawson} WA, {Frye} B,
  {Hoag} AT, {Huang} KH, {Johnson} TL, {Jones} C, {Kikuchihara} S, {Lam} D,
  {Livermore} RC, {Lovisari} L, {Mainali} R, {Oesch} PA, {Ogaz} S, {Ouchi} M,
  {Past} M, {Paterno-Mahler} R, {Peterson} A, {Ryan} RE, {Salmon} B,
  {Sendra-Server} I, {Stark} DP, {Strait} V, {Toft} S, {Vulcani} B (2019)
  {RELICS: High-resolution Constraints on the Inner Mass Distribution of the z
  = 0.83 Merging Cluster RXJ0152.7-1357 from Strong Lensing}. \apj 874(2):132.
  \doi{10.3847/1538-4357/ab0adf}.
  {\href{https://arxiv.org/abs/1810.08122}{{arXiv:1810.08122}}} {[astro-ph.CO]}

\bibitem[{{Aihara} et~al.(2018){Aihara}, {Arimoto}, {Armstrong}
  et~al.}]{aihara18}
{Aihara} H, {Arimoto} N, {Armstrong} R, et~al. (2018) {The Hyper Suprime-Cam
  SSP Survey: Overview and survey design}. \pasj 70:S4.
  \doi{10.1093/pasj/psx066}.
  {\href{https://arxiv.org/abs/1704.05858}{{arXiv:1704.05858}}} {[astro-ph.IM]}

\bibitem[{{Alavi} et~al.(2016){Alavi}, {Siana}, {Richard}, {Rafelski},
  {Jauzac}, {Limousin}, {Freeman}, {Scarlata}, {Robertson}, {Stark}, {Teplitz},
  and {Desai}}]{Alavi+2016}
{Alavi} A, {Siana} B, {Richard} J, {Rafelski} M, {Jauzac} M, {Limousin} M,
  {Freeman} WR, {Scarlata} C, {Robertson} B, {Stark} DP, {Teplitz} HI, {Desai}
  V (2016) {The Evolution of the Faint End of the UV Luminosity Function during
  the Peak Epoch of Star Formation ($1 < z < 3)$}. \apj 832(1):56.
  \doi{10.3847/0004-637X/832/1/56}.
  {\href{https://arxiv.org/abs/1606.00469}{{arXiv:1606.00469}}} {[astro-ph.GA]}

\bibitem[{{Allen} et~al.(2011){Allen}, {Evrard}, and {Mantz}}]{allen2011}
{Allen} SW, {Evrard} AE, {Mantz} AB (2011) {Cosmological Parameters from
  Observations of Galaxy Clusters}. \araa 49(1):409--470.
  \doi{10.1146/annurev-astro-081710-102514}.
  {\href{https://arxiv.org/abs/1103.4829}{{arXiv:1103.4829}}} {[astro-ph.CO]}

\bibitem[{{Amendola} et~al.(2013){Amendola}, {Appleby}, {Bacon}, {Baker},
  {Baldi}, {Bartolo}, {Blanchard}, {Bonvin}, {Borgani}, {Branchini}, {Burrage},
  {Camera}, {Carbone}, {Casarini}, {Cropper}, {de Rham}, {Di Porto}, {Ealet},
  {Ferreira}, {Finelli}, {Garc{\'\i}a-Bellido}, {Giannantonio}, {Guzzo},
  {Heavens}, {Heisenberg}, {Heymans}, {Hoekstra}, {Hollenstein}, {Holmes},
  {Horst}, {Jahnke}, {Kitching}, {Koivisto}, {Kunz}, {La Vacca}, {March},
  {Majerotto}, {Markovic}, {Marsh}, {Marulli}, {Massey}, {Mellier}, {Mota},
  {Nunes}, {Percival}, {Pettorino}, {Porciani}, {Quercellini}, {Read},
  {Rinaldi}, {Sapone}, {Scaramella}, {Skordis}, {Simpson}, {Taylor}, {Thomas},
  {Trotta}, {Verde}, {Vernizzi}, {Vollmer}, {Wang}, {Weller}, and
  {Zlosnik}}]{Amendola:2013}
{Amendola} L, {Appleby} S, {Bacon} D, {Baker} T, {Baldi} M, {Bartolo} N,
  {Blanchard} A, {Bonvin} C, {Borgani} S, {Branchini} E, {Burrage} C, {Camera}
  S, {Carbone} C, {Casarini} L, {Cropper} M, {de Rham} C, {Di Porto} C, {Ealet}
  A, {Ferreira} PG, {Finelli} F, {Garc{\'\i}a-Bellido} J, {Giannantonio} T,
  {Guzzo} L, {Heavens} A, {Heisenberg} L, {Heymans} C, {Hoekstra} H,
  {Hollenstein} L, {Holmes} R, {Horst} O, {Jahnke} K, {Kitching} TD, {Koivisto}
  T, {Kunz} M, {La Vacca} G, {March} M, {Majerotto} E, {Markovic} K, {Marsh} D,
  {Marulli} F, {Massey} R, {Mellier} Y, {Mota} DF, {Nunes} NJ, {Percival} W,
  {Pettorino} V, {Porciani} C, {Quercellini} C, {Read} J, {Rinaldi} M, {Sapone}
  D, {Scaramella} R, {Skordis} C, {Simpson} F, {Taylor} A, {Thomas} S, {Trotta}
  R, {Verde} L, {Vernizzi} F, {Vollmer} A, {Wang} Y, {Weller} J, {Zlosnik} T
  (2013) {Cosmology and Fundamental Physics with the Euclid Satellite}. Living
  Rev Relativ 16(1):6. \doi{10.12942/lrr-2013-6}.
  {\href{https://arxiv.org/abs/1206.1225}{{arXiv:1206.1225}}} {[astro-ph.CO]}

\bibitem[{{Andrade} et~al.(2019){Andrade}, {Minor}, {Nierenberg}, and
  {Kaplinghat}}]{and19}
{Andrade} KE, {Minor} Q, {Nierenberg} A, {Kaplinghat} M (2019) {Detecting dark
  matter cores in galaxy clusters with strong lensing}. \mnras
  487(2):1905--1926. \doi{10.1093/mnras/stz1360}.
  {\href{https://arxiv.org/abs/1901.00507}{{arXiv:1901.00507}}} {[astro-ph.GA]}

\bibitem[{{Angus} et~al.(2006){Angus}, {Famaey}, and {Zhao}}]{angus06}
{Angus} GW, {Famaey} B, {Zhao} HS (2006) {Can MOND take a bullet? Analytical
  comparisons of three versions of MOND beyond spherical symmetry}. \mnras
  371(1):138--146. \doi{10.1111/j.1365-2966.2006.10668.x}.
  {\href{https://arxiv.org/abs/astro-ph/0606216}{{arXiv:astro-ph/0606216}}}
  {[astro-ph]}

\bibitem[{{Annunziatella} et~al.(2014){Annunziatella}, {Biviano}, {Mercurio},
  {Nonino}, {Rosati}, {Balestra}, {Presotto}, {Girardi}, {Gobat}, {Grillo},
  {Kelson}, {Medezinski}, {Postman}, {Scodeggio}, {Brescia}, {Demarco},
  {Fritz}, {Koekemoer}, {Lemze}, {Lombardi}, {Sartoris}, {Umetsu}, {Vanzella},
  {Bradley}, {Coe}, {Donahue}, {Infante}, {Kuchner}, {Maier}, {Reg{\H{o}}s},
  {Verdugo}, and {Ziegler}}]{Annunziatella:2014}
{Annunziatella} M, {Biviano} A, {Mercurio} A, {Nonino} M, {Rosati} P,
  {Balestra} I, {Presotto} V, {Girardi} M, {Gobat} R, {Grillo} C, {Kelson} D,
  {Medezinski} E, {Postman} M, {Scodeggio} M, {Brescia} M, {Demarco} R, {Fritz}
  A, {Koekemoer} A, {Lemze} D, {Lombardi} M, {Sartoris} B, {Umetsu} K,
  {Vanzella} E, {Bradley} L, {Coe} D, {Donahue} M, {Infante} L, {Kuchner} U,
  {Maier} C, {Reg{\H{o}}s} E, {Verdugo} M, {Ziegler} B (2014) {CLASH-VLT: The
  stellar mass function and stellar mass density profile of the z = 0.44
  cluster of galaxies MACS J1206.2-0847}. \aap 571:A80.
  \doi{10.1051/0004-6361/201424102}.
  {\href{https://arxiv.org/abs/1408.6356}{{arXiv:1408.6356}}} {[astro-ph.GA]}

\bibitem[{{Annunziatella} et~al.(2017){Annunziatella}, {Bonamigo}, {Grillo},
  {Mercurio}, {Rosati}, {Caminha}, {Biviano}, {Girardi}, {Gobat}, {Lombardi},
  and {Munari}}]{Annunziatella:2017}
{Annunziatella} M, {Bonamigo} M, {Grillo} C, {Mercurio} A, {Rosati} P,
  {Caminha} G, {Biviano} A, {Girardi} M, {Gobat} R, {Lombardi} M, {Munari} E
  (2017) {Mass Profile Decomposition of the Frontier Fields Cluster MACS
  J0416-2403: Insights on the Dark-matter Inner Profile}. \apj 851(2):81.
  \doi{10.3847/1538-4357/aa9845}.
  {\href{https://arxiv.org/abs/1711.02109}{{arXiv:1711.02109}}} {[astro-ph.CO]}

\bibitem[{{Applegate} et~al.(2016){Applegate}, {Mantz}, {Allen}, {von der
  Linden}, {Morris}, {Hilbert}, {Kelly}, {Burke}, {Ebeling}, {Rapetti}, and
  {Schmidt}}]{applegate2016}
{Applegate} DE, {Mantz} A, {Allen} SW, {von der Linden} A, {Morris} RG,
  {Hilbert} S, {Kelly} PL, {Burke} DL, {Ebeling} H, {Rapetti} DA, {Schmidt} RW
  (2016) {Cosmology and astrophysics from relaxed galaxy clusters - IV.
  Robustly calibrating hydrostatic masses with weak lensing}. \mnras
  457(2):1522--1534. \doi{10.1093/mnras/stw005}.
  {\href{https://arxiv.org/abs/1509.02162}{{arXiv:1509.02162}}} {[astro-ph.CO]}

\bibitem[{{Atek} et~al.(2018){Atek}, {Richard}, {Kneib}, and
  {Schaerer}}]{Atek+2018}
{Atek} H, {Richard} J, {Kneib} JP, {Schaerer} D (2018) {The extreme faint end
  of the UV luminosity function at z {\ensuremath{\sim}} 6 through
  gravitational telescopes: a comprehensive assessment of strong lensing
  uncertainties}. \mnras 479(4):5184--5195. \doi{10.1093/mnras/sty1820}.
  {\href{https://arxiv.org/abs/1803.09747}{{arXiv:1803.09747}}} {[astro-ph.GA]}

\bibitem[{{Bacon} et~al.(2006){Bacon}, {Goldberg}, {Rowe}, and
  {Taylor}}]{bacon2006}
{Bacon} DJ, {Goldberg} DM, {Rowe} BTP, {Taylor} AN (2006) {Weak gravitational
  flexion}. \mnras 365(2):414--428. \doi{10.1111/j.1365-2966.2005.09624.x}.
  {\href{https://arxiv.org/abs/astro-ph/0504478}{{arXiv:astro-ph/0504478}}}
  {[astro-ph]}

\bibitem[{{Bah{\'e}}(2021)}]{bah21}
{Bah{\'e}} YM (2021) {Strongly lensed cluster substructures are not in tension
  with {\ensuremath{\Lambda}}CDM}. \mnras \doi{10.1093/mnras/stab1392}.
  {\href{https://arxiv.org/abs/2101.12112}{{arXiv:2101.12112}}} {[astro-ph.GA]}

\bibitem[{{Balestra} et~al.(2013){Balestra}, {Vanzella}, {Rosati}, {Monna},
  {Grillo}, {Nonino}, {Mercurio}, {Biviano}, {Bradley}, {Coe}, {Fritz},
  {Postman}, {Seitz}, {Scodeggio}, {Tozzi}, {Zheng}, {Ziegler}, {Zitrin},
  {Annunziatella}, {Bartelmann}, {Benitez}, {Broadhurst}, {Bouwens}, {Czoske},
  {Donahue}, {Ford}, {Girardi}, {Infante}, {Jouvel}, {Kelson}, {Koekemoer},
  {Kuchner}, {Lemze}, {Lombardi}, {Maier}, {Medezinski}, {Melchior},
  {Meneghetti}, {Merten}, {Molino}, {Moustakas}, {Presotto}, {Smit}, and
  {Umetsu}}]{Balestra:2013}
{Balestra} I, {Vanzella} E, {Rosati} P, {Monna} A, {Grillo} C, {Nonino} M,
  {Mercurio} A, {Biviano} A, {Bradley} L, {Coe} D, {Fritz} A, {Postman} M,
  {Seitz} S, {Scodeggio} M, {Tozzi} P, {Zheng} W, {Ziegler} B, {Zitrin} A,
  {Annunziatella} M, {Bartelmann} M, {Benitez} N, {Broadhurst} T, {Bouwens} R,
  {Czoske} O, {Donahue} M, {Ford} H, {Girardi} M, {Infante} L, {Jouvel} S,
  {Kelson} D, {Koekemoer} A, {Kuchner} U, {Lemze} D, {Lombardi} M, {Maier} C,
  {Medezinski} E, {Melchior} P, {Meneghetti} M, {Merten} J, {Molino} A,
  {Moustakas} L, {Presotto} V, {Smit} R, {Umetsu} K (2013) {CLASH-VLT:
  spectroscopic confirmation of a z = 6.11 quintuply lensed galaxy in the
  Frontier Fields cluster RXC J2248.7-4431}. \aap 559:L9.
  \doi{10.1051/0004-6361/201322620}.
  {\href{https://arxiv.org/abs/1309.1593}{{arXiv:1309.1593}}} {[astro-ph.CO]}

\bibitem[{{Bartelmann} et~al.(1998){Bartelmann}, {Huss}, {Colberg}, {Jenkins},
  and {Pearce}}]{bar98}
{Bartelmann} M, {Huss} A, {Colberg} JM, {Jenkins} A, {Pearce} FR (1998) {Arc
  statistics with realistic cluster potentials. IV. Clusters in different
  cosmologies}. \aap 330:1--9.
  {\href{https://arxiv.org/abs/astro-ph/9707167}{{arXiv:astro-ph/9707167}}}
  {[astro-ph]}

\bibitem[{{Bayliss} et~al.(2011){Bayliss}, {Hennawi}, {Gladders}, {Koester},
  {Sharon}, {Dahle}, and {Oguri}}]{Bayliss2011}
{Bayliss} MB, {Hennawi} JF, {Gladders} MD, {Koester} BP, {Sharon} K, {Dahle} H,
  {Oguri} M (2011) {Gemini/GMOS Spectroscopy of 26 Strong-lensing-selected
  Galaxy Cluster Cores}. \apjs 193(1):8. \doi{10.1088/0067-0049/193/1/8}.
  {\href{https://arxiv.org/abs/1010.2714}{{arXiv:1010.2714}}} {[astro-ph.CO]}

\bibitem[{{Bayliss} et~al.(2014){Bayliss}, {Johnson}, {Gladders}, {Sharon}, and
  {Oguri}}]{Bayliss+2014}
{Bayliss} MB, {Johnson} T, {Gladders} MD, {Sharon} K, {Oguri} M (2014)
  {Line-of-sight Structure toward Strong Lensing Galaxy Clusters}. \apj
  783(1):41. \doi{10.1088/0004-637X/783/1/41}.
  {\href{https://arxiv.org/abs/1312.3637}{{arXiv:1312.3637}}} {[astro-ph.CO]}

\bibitem[{{Beauchesne} et~al.(2021){Beauchesne}, {Cl{\'e}ment}, {Richard}, and
  {Kneib}}]{bib2:beauchesne21}
{Beauchesne} B, {Cl{\'e}ment} B, {Richard} J, {Kneib} JP (2021) {Improving
  parametric mass modelling of lensing clusters through a perturbative
  approach}. \mnras \doi{10.1093/mnras/stab1684}.
  {\href{https://arxiv.org/abs/2106.05029}{{arXiv:2106.05029}}} {[astro-ph.CO]}

\bibitem[{{Bergamini} et~al.(2019){Bergamini}, {Rosati}, {Mercurio}, {Grillo},
  {Caminha}, {Meneghetti}, {Agnello}, {Biviano}, {Calura}, {Giocoli},
  {Lombardi}, {Rodighiero}, and {Vanzella}}]{Bergamini:2019}
{Bergamini} P, {Rosati} P, {Mercurio} A, {Grillo} C, {Caminha} GB, {Meneghetti}
  M, {Agnello} A, {Biviano} A, {Calura} F, {Giocoli} C, {Lombardi} M,
  {Rodighiero} G, {Vanzella} E (2019) {Enhanced cluster lensing models with
  measured galaxy kinematics}. \aap 631:A130.
  \doi{10.1051/0004-6361/201935974}.
  {\href{https://arxiv.org/abs/1905.13236}{{arXiv:1905.13236}}} {[astro-ph.GA]}

\bibitem[{{Bergamini} et~al.(2021){Bergamini}, {Rosati}, {Vanzella}, {Caminha},
  {Grillo}, {Mercurio}, {Meneghetti}, {Angora}, {Calura}, {Nonino}, and
  {Tozzi}}]{Bergamini:2021}
{Bergamini} P, {Rosati} P, {Vanzella} E, {Caminha} GB, {Grillo} C, {Mercurio}
  A, {Meneghetti} M, {Angora} G, {Calura} F, {Nonino} M, {Tozzi} P (2021) {A
  new high-precision strong lensing model of the galaxy cluster MACS
  J0416.1-2403. Robust characterization of the cluster mass distribution from
  VLT/MUSE deep observations}. \aap 645:A140.
  \doi{10.1051/0004-6361/202039564}.
  {\href{https://arxiv.org/abs/2010.00027}{{arXiv:2010.00027}}} {[astro-ph.GA]}

\bibitem[{{Biviano} et~al.(2013){Biviano}, {Rosati}, {Balestra}, {Mercurio},
  {Girardi}, {Nonino}, {Grillo}, {Scodeggio}, {Lemze}, {Kelson}, {Umetsu},
  {Postman}, {Zitrin}, {Czoske}, {Ettori}, {Fritz}, {Lombardi}, {Maier},
  {Medezinski}, {Mei}, {Presotto}, {Strazzullo}, {Tozzi}, {Ziegler},
  {Annunziatella}, {Bartelmann}, {Benitez}, {Bradley}, {Brescia}, {Broadhurst},
  {Coe}, {Demarco}, {Donahue}, {Ford}, {Gobat}, {Graves}, {Koekemoer},
  {Kuchner}, {Melchior}, {Meneghetti}, {Merten}, {Moustakas}, {Munari},
  {Reg{\H{o}}s}, {Sartoris}, {Seitz}, and {Zheng}}]{Biviano:2013}
{Biviano} A, {Rosati} P, {Balestra} I, {Mercurio} A, {Girardi} M, {Nonino} M,
  {Grillo} C, {Scodeggio} M, {Lemze} D, {Kelson} D, {Umetsu} K, {Postman} M,
  {Zitrin} A, {Czoske} O, {Ettori} S, {Fritz} A, {Lombardi} M, {Maier} C,
  {Medezinski} E, {Mei} S, {Presotto} V, {Strazzullo} V, {Tozzi} P, {Ziegler}
  B, {Annunziatella} M, {Bartelmann} M, {Benitez} N, {Bradley} L, {Brescia} M,
  {Broadhurst} T, {Coe} D, {Demarco} R, {Donahue} M, {Ford} H, {Gobat} R,
  {Graves} G, {Koekemoer} A, {Kuchner} U, {Melchior} P, {Meneghetti} M,
  {Merten} J, {Moustakas} L, {Munari} E, {Reg{\H{o}}s} E, {Sartoris} B, {Seitz}
  S, {Zheng} W (2013) {CLASH-VLT: The mass, velocity-anisotropy, and
  pseudo-phase-space density profiles of the z = 0.44 galaxy cluster MACS
  J1206.2-0847}. \aap 558:A1. \doi{10.1051/0004-6361/201321955}.
  {\href{https://arxiv.org/abs/1307.5867}{{arXiv:1307.5867}}} {[astro-ph.CO]}

\bibitem[{{Blandford} and {Narayan}(1986)}]{Blandford1986}
{Blandford} R, {Narayan} R (1986) {Fermat's Principle, Caustics, and the
  Classification of Gravitational Lens Images}. \apj 310:568.
  \doi{10.1086/164709}

\bibitem[{{Blandford} et~al.(1989){Blandford}, {Kochanek}, {Kovner}, and
  {Narayan}}]{Blandford1989}
{Blandford} RD, {Kochanek} CS, {Kovner} I, {Narayan} R (1989) {Gravitational
  Lens Optics}. Science 245(4920):824--830. \doi{10.1126/science.245.4920.824}

\bibitem[{{Blanton} et~al.(2017){Blanton}, {Bershady}, {Abolfathi}
  et~al.}]{blanton2017}
{Blanton} MR, {Bershady} MA, {Abolfathi} B, et~al. (2017) {Sloan Digital Sky
  Survey IV: Mapping the Milky Way, Nearby Galaxies, and the Distant Universe}.
  \aj 154(1):28. \doi{10.3847/1538-3881/aa7567}.
  {\href{https://arxiv.org/abs/1703.00052}{{arXiv:1703.00052}}} {[astro-ph.GA]}

\bibitem[{{Bonamigo} et~al.(2017){Bonamigo}, {Grillo}, {Ettori}, {Caminha},
  {Rosati}, {Mercurio}, {Annunziatella}, {Balestra}, and
  {Lombardi}}]{Bonamigo:2017}
{Bonamigo} M, {Grillo} C, {Ettori} S, {Caminha} GB, {Rosati} P, {Mercurio} A,
  {Annunziatella} M, {Balestra} I, {Lombardi} M (2017) {Joining X-Ray to
  Lensing: An Accurate Combined Analysis of MACS J0416.1-2403}. \apj
  842(2):132. \doi{10.3847/1538-4357/aa75cc}.
  {\href{https://arxiv.org/abs/1705.10322}{{arXiv:1705.10322}}} {[astro-ph.GA]}

\bibitem[{{Bonamigo} et~al.(2018){Bonamigo}, {Grillo}, {Ettori}, {Caminha},
  {Rosati}, {Mercurio}, {Munari}, {Annunziatella}, {Balestra}, and
  {Lombardi}}]{Bonamigo:2018}
{Bonamigo} M, {Grillo} C, {Ettori} S, {Caminha} GB, {Rosati} P, {Mercurio} A,
  {Munari} E, {Annunziatella} M, {Balestra} I, {Lombardi} M (2018) {Dissection
  of the Collisional and Collisionless Mass Components in a Mini Sample of
  CLASH and HFF Massive Galaxy Clusters at z {\ensuremath{\approx}} 0.4}. \apj
  864(1):98. \doi{10.3847/1538-4357/aad4a7}.
  {\href{https://arxiv.org/abs/1807.10286}{{arXiv:1807.10286}}} {[astro-ph.GA]}

\bibitem[{{Bose} et~al.(2017){Bose}, {Hellwing}, {Frenk}, {Jenkins}, {Lovell},
  {Helly}, {Li}, {Gonzalez-Perez}, and {Gao}}]{bos17}
{Bose} S, {Hellwing} WA, {Frenk} CS, {Jenkins} A, {Lovell} MR, {Helly} JC, {Li}
  B, {Gonzalez-Perez} V, {Gao} L (2017) {Substructure and galaxy formation in
  the Copernicus Complexio warm dark matter simulations}. \mnras
  464(4):4520--4533. \doi{10.1093/mnras/stw2686}.
  {\href{https://arxiv.org/abs/1604.07409}{{arXiv:1604.07409}}} {[astro-ph.CO]}

\bibitem[{{Bouwens} et~al.(2017{\natexlab{a}}){Bouwens}, {Illingworth},
  {Oesch}, {Atek}, {Lam}, and {Stefanon}}]{Bouwens:2017}
{Bouwens} RJ, {Illingworth} GD, {Oesch} PA, {Atek} H, {Lam} D, {Stefanon} M
  (2017{\natexlab{a}}) {Extremely Small Sizes for Faint z {\ensuremath{\sim}}
  2-8 Galaxies in the Hubble Frontier Fields: A Key Input for Establishing
  Their Volume Density and UV Emissivity}. \apj 843(1):41.
  \doi{10.3847/1538-4357/aa74e4}.
  {\href{https://arxiv.org/abs/1608.00966}{{arXiv:1608.00966}}} {[astro-ph.GA]}

\bibitem[{{Bouwens} et~al.(2017{\natexlab{b}}){Bouwens}, {Oesch},
  {Illingworth}, {Ellis}, and {Stefanon}}]{Bouwens+2017}
{Bouwens} RJ, {Oesch} PA, {Illingworth} GD, {Ellis} RS, {Stefanon} M
  (2017{\natexlab{b}}) {The z {\ensuremath{\sim}} 6 Luminosity Function Fainter
  than -15 mag from the Hubble Frontier Fields: The Impact of Magnification
  Uncertainties}. \apj 843(2):129. \doi{10.3847/1538-4357/aa70a4}.
  {\href{https://arxiv.org/abs/1610.00283}{{arXiv:1610.00283}}} {[astro-ph.GA]}

\bibitem[{{Bouwens} et~al.(2022){Bouwens}, {Illingworth}, {Ellis}, {Oesch}, and
  {Stefanon}}]{Bouwens+2022}
{Bouwens} RJ, {Illingworth} G, {Ellis} RS, {Oesch} P, {Stefanon} M (2022) {z
  2-9 Galaxies Magnified by the Hubble Frontier Field Clusters. II. Luminosity
  Functions and Constraints on a Faint-end Turnover}. \apj 940(1):55.
  \doi{10.3847/1538-4357/ac86d1}.
  {\href{https://arxiv.org/abs/2205.11526}{{arXiv:2205.11526}}} {[astro-ph.GA]}

\bibitem[{{Brada{\v{c}}} et~al.(2006){Brada{\v{c}}}, {Clowe}, {Gonzalez},
  {Marshall}, {Forman}, {Jones}, {Markevitch}, {Randall}, {Schrabback}, and
  {Zaritsky}}]{Bradac:06}
{Brada{\v{c}}} M, {Clowe} D, {Gonzalez} AH, {Marshall} P, {Forman} W, {Jones}
  C, {Markevitch} M, {Randall} S, {Schrabback} T, {Zaritsky} D (2006) {Strong
  and Weak Lensing United. III. Measuring the Mass Distribution of the Merging
  Galaxy Cluster 1ES 0657-558}. \apj 652:937--947. \doi{10.1086/508601}.
  {\href{https://arxiv.org/abs/astro-ph/0608408}{{arXiv:astro-ph/0608408}}}
  {[astro-ph]}

\bibitem[{{Brada{\v{c}}} et~al.(2009){Brada{\v{c}}}, {Treu}, {Applegate},
  {Gonzalez}, {Clowe}, {Forman}, {Jones}, {Marshall}, {Schneider}, and
  {Zaritsky}}]{Bradac:09}
{Brada{\v{c}}} M, {Treu} T, {Applegate} D, {Gonzalez} AH, {Clowe} D, {Forman}
  W, {Jones} C, {Marshall} P, {Schneider} P, {Zaritsky} D (2009) {Focusing
  Cosmic Telescopes: Exploring Redshift $z \sim$ 5--6 Galaxies with the Bullet
  Cluster 1E0657 - 56}. \apj 706(2):1201--1212.
  \doi{10.1088/0004-637X/706/2/1201}.
  {\href{https://arxiv.org/abs/0910.2708}{{arXiv:0910.2708}}} {[astro-ph.CO]}

\bibitem[{{Broadhurst} et~al.(2005{\natexlab{a}}){Broadhurst}, {Ben{\'\i}tez},
  {Coe}, {Sharon}, {Zekser}, {White}, {Ford}, {Bouwens}, {Blakeslee},
  {Clampin}, {Cross}, {Franx}, {Frye}, {Hartig}, {Illingworth}, {Infante},
  {Menanteau}, {Meurer}, {Postman}, {Ardila}, {Bartko}, {Brown}, {Burrows},
  {Cheng}, {Feldman}, {Golimowski}, {Goto}, {Gronwall}, {Herranz}, {Holden},
  {Homeier}, {Krist}, {Lesser}, {Martel}, {Miley}, {Rosati}, {Sirianni},
  {Sparks}, {Steindling}, {Tran}, {Tsvetanov}, and {Zheng}}]{Broadhurst05b}
{Broadhurst} T, {Ben{\'\i}tez} N, {Coe} D, {Sharon} K, {Zekser} K, {White} R,
  {Ford} H, {Bouwens} R, {Blakeslee} J, {Clampin} M, {Cross} N, {Franx} M,
  {Frye} B, {Hartig} G, {Illingworth} G, {Infante} L, {Menanteau} F, {Meurer}
  G, {Postman} M, {Ardila} DR, {Bartko} F, {Brown} RA, {Burrows} CJ, {Cheng}
  ES, {Feldman} PD, {Golimowski} DA, {Goto} T, {Gronwall} C, {Herranz} D,
  {Holden} B, {Homeier} N, {Krist} JE, {Lesser} MP, {Martel} AR, {Miley} GK,
  {Rosati} P, {Sirianni} M, {Sparks} WB, {Steindling} S, {Tran} HD, {Tsvetanov}
  ZI, {Zheng} W (2005{\natexlab{a}}) {Strong-Lensing Analysis of A1689 from
  Deep Advanced Camera Images}. \apj 621(1):53--88. \doi{10.1086/426494}.
  {\href{https://arxiv.org/abs/astro-ph/0409132}{{arXiv:astro-ph/0409132}}}
  {[astro-ph]}

\bibitem[{{Broadhurst} et~al.(2005{\natexlab{b}}){Broadhurst}, {Takada},
  {Umetsu}, {Kong}, {Arimoto}, {Chiba}, and {Futamase}}]{bro05a}
{Broadhurst} T, {Takada} M, {Umetsu} K, {Kong} X, {Arimoto} N, {Chiba} M,
  {Futamase} T (2005{\natexlab{b}}) {The Surprisingly Steep Mass Profile of
  A1689, from a Lensing Analysis of Subaru Images}. \apjl 619(2):L143--L146.
  \doi{10.1086/428122}.
  {\href{https://arxiv.org/abs/astro-ph/0412192}{{arXiv:astro-ph/0412192}}}
  {[astro-ph]}

\bibitem[{{Cain} et~al.(2016){Cain}, {Brada{\v{c}}}, and {Levinson}}]{cain16}
{Cain} B, {Brada{\v{c}}} M, {Levinson} R (2016) {Reconstruction of small-scale
  galaxy cluster substructure with lensing flexion}. \mnras 463(4):4287--4300.
  \doi{10.1093/mnras/stw2270}.
  {\href{https://arxiv.org/abs/1503.08218}{{arXiv:1503.08218}}} {[astro-ph.CO]}

\bibitem[{{Caminha} et~al.(2013){Caminha}, {Estrada}, and {Makler}}]{cam13}
{Caminha} GB, {Estrada} J, {Makler} M (2013) {Magnification Bias in
  Gravitational Arc Statistics}. arXiv e-prints
  {\href{https://arxiv.org/abs/1308.6569}{{arXiv:1308.6569}}} {[astro-ph.CO]}

\bibitem[{{Caminha} et~al.(2016){Caminha}, {Grillo}, {Rosati}, {Balestra},
  {Karman}, {Lombardi}, {Mercurio}, {Nonino}, {Tozzi}, {Zitrin}, {Biviano},
  {Girardi}, {Koekemoer}, {Melchior}, {Meneghetti}, {Munari}, {Suyu}, {Umetsu},
  {Annunziatella}, {Borgani}, {Broadhurst}, {Caputi}, {Coe}, {Delgado-Correal},
  {Ettori}, {Fritz}, {Frye}, {Gobat}, {Maier}, {Monna}, {Postman}, {Sartoris},
  {Seitz}, {Vanzella}, and {Ziegler}}]{Caminha:2016}
{Caminha} GB, {Grillo} C, {Rosati} P, {Balestra} I, {Karman} W, {Lombardi} M,
  {Mercurio} A, {Nonino} M, {Tozzi} P, {Zitrin} A, {Biviano} A, {Girardi} M,
  {Koekemoer} AM, {Melchior} P, {Meneghetti} M, {Munari} E, {Suyu} SH, {Umetsu}
  K, {Annunziatella} M, {Borgani} S, {Broadhurst} T, {Caputi} KI, {Coe} D,
  {Delgado-Correal} C, {Ettori} S, {Fritz} A, {Frye} B, {Gobat} R, {Maier} C,
  {Monna} A, {Postman} M, {Sartoris} B, {Seitz} S, {Vanzella} E, {Ziegler} B
  (2016) {CLASH-VLT: A highly precise strong lensing model of the galaxy
  cluster RXC J2248.7-4431 (Abell S1063) and prospects for cosmography}. \aap
  587:A80. \doi{10.1051/0004-6361/201527670}.
  {\href{https://arxiv.org/abs/1512.04555}{{arXiv:1512.04555}}} {[astro-ph.CO]}

\bibitem[{{Caminha} et~al.(2017){Caminha}, {Grillo}, {Rosati}, {Balestra},
  {Mercurio}, {Vanzella}, {Biviano}, {Caputi}, {Delgado-Correal}, {Karman},
  {Lombardi}, {Meneghetti}, {Sartoris}, and {Tozzi}}]{Caminha:2017a}
{Caminha} GB, {Grillo} C, {Rosati} P, {Balestra} I, {Mercurio} A, {Vanzella} E,
  {Biviano} A, {Caputi} KI, {Delgado-Correal} C, {Karman} W, {Lombardi} M,
  {Meneghetti} M, {Sartoris} B, {Tozzi} P (2017) {A refined mass distribution
  of the cluster MACS J0416.1-2403 from a new large set of spectroscopic
  multiply lensed sources}. \aap 600:A90. \doi{10.1051/0004-6361/201629297}.
  {\href{https://arxiv.org/abs/1607.03462}{{arXiv:1607.03462}}} {[astro-ph.GA]}

\bibitem[{{Caminha} et~al.(2019){Caminha}, {Rosati}, {Grillo}, {Rosani},
  {Caputi}, {Meneghetti}, {Mercurio}, {Balestra}, {Bergamini}, {Biviano},
  {Nonino}, {Umetsu}, {Vanzella}, {Annunziatella}, {Broadhurst},
  {Delgado-Correal}, {Demarco}, {Koekemoer}, {Lombardi}, {Maier}, {Verdugo},
  and {Zitrin}}]{Caminha:2019}
{Caminha} GB, {Rosati} P, {Grillo} C, {Rosani} G, {Caputi} KI, {Meneghetti} M,
  {Mercurio} A, {Balestra} I, {Bergamini} P, {Biviano} A, {Nonino} M, {Umetsu}
  K, {Vanzella} E, {Annunziatella} M, {Broadhurst} T, {Delgado-Correal} C,
  {Demarco} R, {Koekemoer} AM, {Lombardi} M, {Maier} C, {Verdugo} M, {Zitrin} A
  (2019) {Strong lensing models of eight CLASH clusters from extensive
  spectroscopy: Accurate total mass reconstructions in the cores}. \aap
  632:A36. \doi{10.1051/0004-6361/201935454}.
  {\href{https://arxiv.org/abs/1903.05103}{{arXiv:1903.05103}}} {[astro-ph.GA]}

\bibitem[{{Caminha} et~al.(2022{\natexlab{a}}){Caminha}, {Suyu}, {Grillo}, and
  {Rosati}}]{Caminha+2022}
{Caminha} GB, {Suyu} SH, {Grillo} C, {Rosati} P (2022{\natexlab{a}}) {Galaxy
  cluster strong lensing cosmography. Cosmological constraints from a sample of
  regular galaxy clusters}. \aap 657:A83. \doi{10.1051/0004-6361/202141994}.
  {\href{https://arxiv.org/abs/2110.06232}{{arXiv:2110.06232}}} {[astro-ph.CO]}

\bibitem[{{Caminha} et~al.(2022{\natexlab{b}}){Caminha}, {Suyu}, {Mercurio},
  {Brammer}, {Bergamini}, {Acebron}, and {Vanzella}}]{Caminha2022}
{Caminha} GB, {Suyu} SH, {Mercurio} A, {Brammer} G, {Bergamini} P, {Acebron} A,
  {Vanzella} E (2022{\natexlab{b}}) {First JWST observations of a gravitational
  lens. Mass model from new multiple images with near-infrared observations of
  SMACS J0723.3{\ensuremath{-}}7327}. \aap 666:L9.
  \doi{10.1051/0004-6361/202244517}.
  {\href{https://arxiv.org/abs/2207.07567}{{arXiv:2207.07567}}} {[astro-ph.GA]}

\bibitem[{{Carlsten} et~al.(2020){Carlsten}, {Greene}, {Peter}, {Greco}, and
  {Beaton}}]{Carlsten+2020}
{Carlsten} SG, {Greene} JE, {Peter} AHG, {Greco} JP, {Beaton} RL (2020) {Radial
  Distributions of Dwarf Satellite Systems in the Local Volume}. \apj
  902(2):124. \doi{10.3847/1538-4357/abb60b}.
  {\href{https://arxiv.org/abs/2006.02444}{{arXiv:2006.02444}}} {[astro-ph.GA]}

\bibitem[{{Castellano} et~al.(2023){Castellano}, {Fontana}, {Treu}, {Merlin},
  {Santini}, {Bergamini}, {Grillo}, {Rosati}, {Acebron}, {Leethochawalit},
  {Paris}, {Bonchi}, {Belfiori}, {Calabr{\`o}}, {Correnti}, {Nonino},
  {Polenta}, {Trenti}, {Boyett}, {Brammer}, {Broadhurst}, {Caminha}, {Chen},
  {Filippenko}, {Fortuni}, {Glazebrook}, {Mascia}, {Mason}, {Menci},
  {Meneghetti}, {Mercurio}, {Metha}, {Morishita}, {Nanayakkara}, {Pentericci},
  {Roberts-Borsani}, {Roy}, {Vanzella}, {Vulcani}, {Yang}, and
  {Wang}}]{Castellano+2022}
{Castellano} M, {Fontana} A, {Treu} T, {Merlin} E, {Santini} P, {Bergamini} P,
  {Grillo} C, {Rosati} P, {Acebron} A, {Leethochawalit} N, {Paris} D, {Bonchi}
  A, {Belfiori} D, {Calabr{\`o}} A, {Correnti} M, {Nonino} M, {Polenta} G,
  {Trenti} M, {Boyett} K, {Brammer} G, {Broadhurst} T, {Caminha} GB, {Chen} W,
  {Filippenko} AV, {Fortuni} F, {Glazebrook} K, {Mascia} S, {Mason} CA, {Menci}
  N, {Meneghetti} M, {Mercurio} A, {Metha} B, {Morishita} T, {Nanayakkara} T,
  {Pentericci} L, {Roberts-Borsani} G, {Roy} N, {Vanzella} E, {Vulcani} B,
  {Yang} L, {Wang} X (2023) {Early Results from GLASS-JWST. XIX. A High Density
  of Bright Galaxies at z {\ensuremath{\approx}} 10 in the A2744 Region}. \apjl
  948(2):L14. \doi{10.3847/2041-8213/accea5}.
  {\href{https://arxiv.org/abs/2212.06666}{{arXiv:2212.06666}}} {[astro-ph.GA]}

\bibitem[{{Cerini} et~al.(2023){Cerini}, {Cappelluti}, and
  {Natarajan}}]{Cerini+2023}
{Cerini} G, {Cappelluti} N, {Natarajan} P (2023) {New Metrics to Probe the
  Dynamical State of Galaxy Clusters}. \apj 945(2):152.
  \doi{10.3847/1538-4357/acbccb}.
  {\href{https://arxiv.org/abs/2209.06831}{{arXiv:2209.06831}}} {[astro-ph.CO]}

\bibitem[{{Cha} and {Jee}(2022)}]{cha22}
{Cha} S, {Jee} MJ (2022) {MARS: A New Maximum-entropy-regularized Strong
  Lensing Mass Reconstruction Method}. \apj 931(2):127.
  \doi{10.3847/1538-4357/ac69df}.
  {\href{https://arxiv.org/abs/2202.10489}{{arXiv:2202.10489}}} {[astro-ph.CO]}

\bibitem[{{Chiriv{\`\i}} et~al.(2018){Chiriv{\`\i}}, {Suyu}, {Grillo},
  {Halkola}, {Balestra}, {Caminha}, {Mercurio}, and {Rosati}}]{Chirivi:2018}
{Chiriv{\`\i}} G, {Suyu} SH, {Grillo} C, {Halkola} A, {Balestra} I, {Caminha}
  GB, {Mercurio} A, {Rosati} P (2018) {MACS J0416.1-2403: Impact of
  line-of-sight structures on strong gravitational lensing modelling of galaxy
  clusters}. \aap 614:A8. \doi{10.1051/0004-6361/201731433}.
  {\href{https://arxiv.org/abs/1706.07815}{{arXiv:1706.07815}}} {[astro-ph.CO]}

\bibitem[{{Clowe} et~al.(2006{\natexlab{a}}){Clowe}, {Brada{\v{c}}},
  {Gonzalez}, {Markevitch}, {Randall}, {Jones}, and {Zaritsky}}]{clowe06}
{Clowe} D, {Brada{\v{c}}} M, {Gonzalez} AH, {Markevitch} M, {Randall} SW,
  {Jones} C, {Zaritsky} D (2006{\natexlab{a}}) {A Direct Empirical Proof of the
  Existence of Dark Matter}. \apjl 648(2):L109--L113. \doi{10.1086/508162}.
  {\href{https://arxiv.org/abs/astro-ph/0608407}{{arXiv:astro-ph/0608407}}}
  {[astro-ph]}

\bibitem[{{Clowe} et~al.(2006{\natexlab{b}}){Clowe}, {Schneider},
  {Arag{\'o}n-Salamanca}, {Bremer}, {De Lucia}, {Halliday}, {Jablonka},
  {Milvang-Jensen}, {Pell{\'o}}, {Poggianti}, {Rudnick}, {Saglia}, {Simard},
  {White}, and {Zaritsky}}]{clowe06b}
{Clowe} D, {Schneider} P, {Arag{\'o}n-Salamanca} A, {Bremer} M, {De Lucia} G,
  {Halliday} C, {Jablonka} P, {Milvang-Jensen} B, {Pell{\'o}} R, {Poggianti} B,
  {Rudnick} G, {Saglia} R, {Simard} L, {White} S, {Zaritsky} D
  (2006{\natexlab{b}}) {Weak lensing mass reconstructions of the ESO Distant
  Cluster Survey}. \aap 451(2):395--408. \doi{10.1051/0004-6361:20041787}.
  {\href{https://arxiv.org/abs/astro-ph/0511746}{{arXiv:astro-ph/0511746}}}
  {[astro-ph]}

\bibitem[{{Coe} et~al.(2010){Coe}, {Ben{\'\i}tez}, {Broadhurst}, and
  {Moustakas}}]{coe10}
{Coe} D, {Ben{\'\i}tez} N, {Broadhurst} T, {Moustakas} LA (2010) {A
  High-resolution Mass Map of Galaxy Cluster Substructure: LensPerfect Analysis
  of A1689}. \apj 723(2):1678--1702. \doi{10.1088/0004-637X/723/2/1678}.
  {\href{https://arxiv.org/abs/1005.0398}{{arXiv:1005.0398}}} {[astro-ph.CO]}

\bibitem[{{Coe} et~al.(2012){Coe}, {Umetsu}, {Zitrin}, {Donahue}, {Medezinski},
  {Postman}, {Carrasco}, {Anguita}, {Geller}, {Rines}, {Diaferio}, {Kurtz},
  {Bradley}, {Koekemoer}, {Zheng}, {Nonino}, {Molino}, {Mahdavi}, {Lemze},
  {Infante}, {Ogaz}, {Melchior}, {Host}, {Ford}, {Grillo}, {Rosati},
  {Jim{\'e}nez-Teja}, {Moustakas}, {Broadhurst}, {Ascaso}, {Lahav},
  {Bartelmann}, {Ben{\'\i}tez}, {Bouwens}, {Graur}, {Graves}, {Jha}, {Jouvel},
  {Kelson}, {Moustakas}, {Maoz}, {Meneghetti}, {Merten}, {Riess}, {Rodney}, and
  {Seitz}}]{Coe:2012}
{Coe} D, {Umetsu} K, {Zitrin} A, {Donahue} M, {Medezinski} E, {Postman} M,
  {Carrasco} M, {Anguita} T, {Geller} MJ, {Rines} KJ, {Diaferio} A, {Kurtz} MJ,
  {Bradley} L, {Koekemoer} A, {Zheng} W, {Nonino} M, {Molino} A, {Mahdavi} A,
  {Lemze} D, {Infante} L, {Ogaz} S, {Melchior} P, {Host} O, {Ford} H, {Grillo}
  C, {Rosati} P, {Jim{\'e}nez-Teja} Y, {Moustakas} J, {Broadhurst} T, {Ascaso}
  B, {Lahav} O, {Bartelmann} M, {Ben{\'\i}tez} N, {Bouwens} R, {Graur} O,
  {Graves} G, {Jha} S, {Jouvel} S, {Kelson} D, {Moustakas} L, {Maoz} D,
  {Meneghetti} M, {Merten} J, {Riess} A, {Rodney} S, {Seitz} S (2012) {CLASH:
  Precise New Constraints on the Mass Profile of the Galaxy Cluster A2261}.
  \apj 757(1):22. \doi{10.1088/0004-637X/757/1/22}.
  {\href{https://arxiv.org/abs/1201.1616}{{arXiv:1201.1616}}} {[astro-ph.CO]}

\bibitem[{{Coe} et~al.(2013){Coe}, {Zitrin}, {Carrasco}, {Shu}, {Zheng},
  {Postman}, {Bradley}, {Koekemoer}, {Bouwens}, {Broadhurst}, {Monna}, {Host},
  {Moustakas}, {Ford}, {Moustakas}, {van der Wel}, {Donahue}, {Rodney},
  {Ben{\'\i}tez}, {Jouvel}, {Seitz}, {Kelson}, and {Rosati}}]{Coe+2013}
{Coe} D, {Zitrin} A, {Carrasco} M, {Shu} X, {Zheng} W, {Postman} M, {Bradley}
  L, {Koekemoer} A, {Bouwens} R, {Broadhurst} T, {Monna} A, {Host} O,
  {Moustakas} LA, {Ford} H, {Moustakas} J, {van der Wel} A, {Donahue} M,
  {Rodney} SA, {Ben{\'\i}tez} N, {Jouvel} S, {Seitz} S, {Kelson} DD, {Rosati} P
  (2013) {CLASH: Three Strongly Lensed Images of a Candidate z
  {\ensuremath{\approx}} 11 Galaxy}. \apj 762(1):32.
  \doi{10.1088/0004-637X/762/1/32}.
  {\href{https://arxiv.org/abs/1211.3663}{{arXiv:1211.3663}}} {[astro-ph.CO]}

\bibitem[{{Coe} et~al.(2019){Coe}, {Salmon}, {Brada{\v{c}}} et~al.}]{Coe:2019}
{Coe} D, {Salmon} B, {Brada{\v{c}}} M, et~al. (2019) {RELICS: Reionization
  Lensing Cluster Survey}. \apj 884(1):85. \doi{10.3847/1538-4357/ab412b}.
  {\href{https://arxiv.org/abs/1903.02002}{{arXiv:1903.02002}}} {[astro-ph.GA]}

\bibitem[{{{\c{S}}eng{\"u}l} et~al.(2023){{\c{S}}eng{\"u}l}, {Birrer},
  {Natarajan}, and {Dvorkin}}]{Cagan+2023}
{{\c{S}}eng{\"u}l} A{\c{C}}, {Birrer} S, {Natarajan} P, {Dvorkin} C (2023)
  {Detecting low-mass perturbers in cluster lenses using curved arc bases}.
  \mnras 526(2):2525--2541. \doi{10.1093/mnras/stad2784}.
  {\href{https://arxiv.org/abs/2303.14786}{{arXiv:2303.14786}}} {[astro-ph.GA]}

\bibitem[{{Dahle} et~al.(2015){Dahle}, {Gladders}, {Sharon}, {Bayliss}, and
  {Rigby}}]{Dahle:2015}
{Dahle} H, {Gladders} MD, {Sharon} K, {Bayliss} MB, {Rigby} JR (2015) {Time
  Delay Measurements for the Cluster-lensed Sextuple Quasar SDSS J2222+2745}.
  \apj 813(1):67. \doi{10.1088/0004-637X/813/1/67}.
  {\href{https://arxiv.org/abs/1505.06187}{{arXiv:1505.06187}}} {[astro-ph.CO]}

\bibitem[{{Dai} et~al.(2020){Dai}, {Kaurov}, {Sharon}, {Florian},
  {Miralda-Escud{\'e}}, {Venumadhav}, {Frye}, {Rigby}, and {Bayliss}}]{dai20}
{Dai} L, {Kaurov} AA, {Sharon} K, {Florian} M, {Miralda-Escud{\'e}} J,
  {Venumadhav} T, {Frye} B, {Rigby} JR, {Bayliss} M (2020) {Asymmetric surface
  brightness structure of caustic crossing arc in SDSS J1226+2152: a case for
  dark matter substructure}. \mnras 495(3):3192--3208.
  \doi{10.1093/mnras/staa1355}.
  {\href{https://arxiv.org/abs/2001.00261}{{arXiv:2001.00261}}} {[astro-ph.GA]}

\bibitem[{{D'Aloisio} and {Natarajan}(2011)}]{Anson+2011}
{D'Aloisio} A, {Natarajan} P (2011) {The effects of primordial non-Gaussianity
  on giant-arc statistics}. \mnras 415(2):1913--1927.
  \doi{10.1111/j.1365-2966.2011.18837.x}.
  {\href{https://arxiv.org/abs/1102.5097}{{arXiv:1102.5097}}} {[astro-ph.CO]}

\bibitem[{{D'Aloisio} et~al.(2014){D'Aloisio}, {Natarajan}, and
  {Shapiro}}]{Anson+2014}
{D'Aloisio} A, {Natarajan} P, {Shapiro} PR (2014) {The effect of large-scale
  structure on the magnification of high-redshift sources by cluster lenses}.
  \mnras 445(4):3581--3591. \doi{10.1093/mnras/stu1931}.
  {\href{https://arxiv.org/abs/1311.1614}{{arXiv:1311.1614}}} {[astro-ph.CO]}

\bibitem[{{De Lucia} et~al.(2004){De Lucia}, {Kauffmann}, {Springel}, {White},
  {Lanzoni}, {Stoehr}, {Tormen}, and {Yoshida}}]{DeLucia:2004}
{De Lucia} G, {Kauffmann} G, {Springel} V, {White} SDM, {Lanzoni} B, {Stoehr}
  F, {Tormen} G, {Yoshida} N (2004) {Substructures in cold dark matter haloes}.
  \mnras 348(1):333--344. \doi{10.1111/j.1365-2966.2004.07372.x}.
  {\href{https://arxiv.org/abs/astro-ph/0306205}{{arXiv:astro-ph/0306205}}}
  {[astro-ph]}

\bibitem[{{Despali} et~al.(2019){Despali}, {Sparre}, {Vegetti}, {Vogelsberger},
  {Zavala}, and {Marinacci}}]{bib2:despali2019}
{Despali} G, {Sparre} M, {Vegetti} S, {Vogelsberger} M, {Zavala} J, {Marinacci}
  F (2019) {The interplay of self-interacting dark matter and baryons in
  shaping the halo evolution}. \mnras 484(4):4563--4573.
  \doi{10.1093/mnras/stz273}.
  {\href{https://arxiv.org/abs/1811.02569}{{arXiv:1811.02569}}} {[astro-ph.GA]}

\bibitem[{{Diego} et~al.(2005){Diego}, {Protopapas}, {Sandvik}, and
  {Tegmark}}]{Diego:05}
{Diego} JM, {Protopapas} P, {Sandvik} HB, {Tegmark} M (2005) {Non-parametric
  inversion of strong lensing systems}. \mnras 360(2):477--491.
  \doi{10.1111/j.1365-2966.2005.09021.x}.
  {\href{https://arxiv.org/abs/astro-ph/0408418}{{arXiv:astro-ph/0408418}}}
  {[astro-ph]}

\bibitem[{{Diego} et~al.(2007){Diego}, {Tegmark}, {Protopapas}, and {Sand
  vik}}]{Diego:07}
{Diego} JM, {Tegmark} M, {Protopapas} P, {Sand vik} HB (2007) {Combined
  reconstruction of weak and strong lensing data with WSLAP}. \mnras
  375(3):958--970. \doi{10.1111/j.1365-2966.2007.11380.x}.
  {\href{https://arxiv.org/abs/astro-ph/0509103}{{arXiv:astro-ph/0509103}}}
  {[astro-ph]}

\bibitem[{{Diego} et~al.(2016{\natexlab{a}}){Diego}, {Broadhurst}, {Chen},
  {Lim}, {Zitrin}, {Chan}, {Coe}, {Ford}, {Lam}, and {Zheng}}]{Diego+2016}
{Diego} JM, {Broadhurst} T, {Chen} C, {Lim} J, {Zitrin} A, {Chan} B, {Coe} D,
  {Ford} HC, {Lam} D, {Zheng} W (2016{\natexlab{a}}) {A free-form prediction
  for the reappearance of supernova Refsdal in the Hubble Frontier Fields
  cluster MACSJ1149.5+2223}. \mnras 456(1):356--365.
  \doi{10.1093/mnras/stv2638}.
  {\href{https://arxiv.org/abs/1504.05953}{{arXiv:1504.05953}}} {[astro-ph.CO]}

\bibitem[{{Diego} et~al.(2016{\natexlab{b}}){Diego}, {Broadhurst}, {Wong},
  {Silk}, {Lim}, {Zheng}, {Lam}, and {Ford}}]{Diego:16}
{Diego} JM, {Broadhurst} T, {Wong} J, {Silk} J, {Lim} J, {Zheng} W, {Lam} D,
  {Ford} H (2016{\natexlab{b}}) {A free-form mass model of the Hubble Frontier
  Fields cluster AS1063 (RXC J2248.7-4431) with over one hundred constraints}.
  \mnras 459(4):3447--3459. \doi{10.1093/mnras/stw865}.
  {\href{https://arxiv.org/abs/1512.07916}{{arXiv:1512.07916}}} {[astro-ph.CO]}

\bibitem[{{Diego} et~al.(2022){Diego}, {Pascale}, {Kavanagh}, {Kelly}, {Dai},
  {Frye}, and {Broadhurst}}]{diego22}
{Diego} JM, {Pascale} M, {Kavanagh} BJ, {Kelly} P, {Dai} L, {Frye} B,
  {Broadhurst} T (2022) {Godzilla, a monster lurks in the Sunburst galaxy}.
  \aap 665:A134. \doi{10.1051/0004-6361/202243605}.
  {\href{https://arxiv.org/abs/2203.08158}{{arXiv:2203.08158}}} {[astro-ph.GA]}

\bibitem[{{Diemand} et~al.(2007){Diemand}, {Kuhlen}, and {Madau}}]{die07}
{Diemand} J, {Kuhlen} M, {Madau} P (2007) {Formation and Evolution of Galaxy
  Dark Matter Halos and Their Substructure}. \apj 667(2):859--877.
  \doi{10.1086/520573}.
  {\href{https://arxiv.org/abs/astro-ph/0703337}{{arXiv:astro-ph/0703337}}}
  {[astro-ph]}

\bibitem[{{Donahue} et~al.(2014){Donahue}, {Voit}, {Mahdavi}, {Umetsu},
  {Ettori}, {Merten}, {Postman}, {Hoffer}, {Baldi}, {Coe}, {Czakon},
  {Bartelmann}, {Benitez}, {Bouwens}, {Bradley}, {Broadhurst}, {Ford},
  {Gastaldello}, {Grillo}, {Infante}, {Jouvel}, {Koekemoer}, {Kelson}, {Lahav},
  {Lemze}, {Medezinski}, {Melchior}, {Meneghetti}, {Molino}, {Moustakas},
  {Moustakas}, {Nonino}, {Rosati}, {Sayers}, {Seitz}, {Van der Wel}, {Zheng},
  and {Zitrin}}]{Donahue:2014}
{Donahue} M, {Voit} GM, {Mahdavi} A, {Umetsu} K, {Ettori} S, {Merten} J,
  {Postman} M, {Hoffer} A, {Baldi} A, {Coe} D, {Czakon} N, {Bartelmann} M,
  {Benitez} N, {Bouwens} R, {Bradley} L, {Broadhurst} T, {Ford} H,
  {Gastaldello} F, {Grillo} C, {Infante} L, {Jouvel} S, {Koekemoer} A, {Kelson}
  D, {Lahav} O, {Lemze} D, {Medezinski} E, {Melchior} P, {Meneghetti} M,
  {Molino} A, {Moustakas} J, {Moustakas} LA, {Nonino} M, {Rosati} P, {Sayers}
  J, {Seitz} S, {Van der Wel} A, {Zheng} W, {Zitrin} A (2014) {CLASH-X: A
  Comparison of Lensing and X-Ray Techniques for Measuring the Mass Profiles of
  Galaxy Clusters}. \apj 794(2):136. \doi{10.1088/0004-637X/794/2/136}.
  {\href{https://arxiv.org/abs/1405.7876}{{arXiv:1405.7876}}} {[astro-ph.CO]}

\bibitem[{{Ebeling} et~al.(2001){Ebeling}, {Edge}, and {Henry}}]{ebeling01}
{Ebeling} H, {Edge} AC, {Henry} JP (2001) {MACS: A Quest for the Most Massive
  Galaxy Clusters in the Universe}. \apj 553(2):668--676. \doi{10.1086/320958}.
  {\href{https://arxiv.org/abs/astro-ph/0009101}{{arXiv:astro-ph/0009101}}}
  {[astro-ph]}

\bibitem[{{Eichner} et~al.(2013){Eichner}, {Seitz}, {Suyu}, {Halkola},
  {Umetsu}, {Zitrin}, {Coe}, {Monna}, {Rosati}, {Grillo}, {Balestra},
  {Postman}, {Koekemoer}, {Zheng}, {H{\o}st}, {Lemze}, {Broadhurst},
  {Moustakas}, {Bradley}, {Molino}, {Nonino}, {Mercurio}, {Scodeggio},
  {Bartelmann}, {Benitez}, {Bouwens}, {Donahue}, {Infante}, {Jouvel}, {Kelson},
  {Lahav}, {Medezinski}, {Melchior}, {Merten}, and {Riess}}]{Eichner:2013}
{Eichner} T, {Seitz} S, {Suyu} SH, {Halkola} A, {Umetsu} K, {Zitrin} A, {Coe}
  D, {Monna} A, {Rosati} P, {Grillo} C, {Balestra} I, {Postman} M, {Koekemoer}
  A, {Zheng} W, {H{\o}st} O, {Lemze} D, {Broadhurst} T, {Moustakas} L,
  {Bradley} L, {Molino} A, {Nonino} M, {Mercurio} A, {Scodeggio} M,
  {Bartelmann} M, {Benitez} N, {Bouwens} R, {Donahue} M, {Infante} L, {Jouvel}
  S, {Kelson} D, {Lahav} O, {Medezinski} E, {Melchior} P, {Merten} J, {Riess} A
  (2013) {Galaxy Halo Truncation and Giant Arc Surface Brightness
  Reconstruction in the Cluster MACSJ1206.2-0847}. \apj 774(2):124.
  \doi{10.1088/0004-637X/774/2/124}.
  {\href{https://arxiv.org/abs/1306.5240}{{arXiv:1306.5240}}} {[astro-ph.CO]}

\bibitem[{{Elahi} et~al.(2014){Elahi}, {Mahdi}, {Power}, and {Lewis}}]{ela14}
{Elahi} PJ, {Mahdi} HS, {Power} C, {Lewis} GF (2014) {Warm dark haloes
  accretion histories and their gravitational signatures}. \mnras
  444(3):2333--2345. \doi{10.1093/mnras/stu1614}.
  {\href{https://arxiv.org/abs/1406.3413}{{arXiv:1406.3413}}} {[astro-ph.CO]}

\bibitem[{{Euclid Collaboration} et~al.(2019){Euclid Collaboration}, {Adam},
  {Vannier}, {Maurogordato}, {Biviano}, {Adami}, {Ascaso}, {Bellagamba},
  {Benoist}, {Cappi}, {D{\'\i}az-S{\'a}nchez}, {Durret}, {Farrens}, {Gonzalez},
  {Iovino}, {Licitra}, {Maturi}, {Mei}, {Merson}, {Munari}, {Pell{\'o}},
  {Ricci}, {Rocci}, {Roncarelli}, {Sarron}, {Amoura}, {Andreon}, {Apostolakos},
  {Arnaud}, {Bardelli}, {Bartlett}, {Baugh}, {Borgani}, {Brodwin}, {Castander},
  {Castignani}, {Cucciati}, {De Lucia}, {Dubath}, {Fosalba}, {Giocoli},
  {Hoekstra}, {Mamon}, {Melin}, {Moscardini}, {Paltani}, {Radovich},
  {Sartoris}, {Schultheis}, {Sereno}, {Weller}, {Burigana}, {Carvalho},
  {Corcione}, {Kurki-Suonio}, {Lilje}, {Sirri}, {Toledo-Moreo}, and
  {Zamorani}}]{Adam:2019}
{Euclid Collaboration}, {Adam} R, {Vannier} M, {Maurogordato} S, {Biviano} A,
  {Adami} C, {Ascaso} B, {Bellagamba} F, {Benoist} C, {Cappi} A,
  {D{\'\i}az-S{\'a}nchez} A, {Durret} F, {Farrens} S, {Gonzalez} AH, {Iovino}
  A, {Licitra} R, {Maturi} M, {Mei} S, {Merson} A, {Munari} E, {Pell{\'o}} R,
  {Ricci} M, {Rocci} PF, {Roncarelli} M, {Sarron} F, {Amoura} Y, {Andreon} S,
  {Apostolakos} N, {Arnaud} M, {Bardelli} S, {Bartlett} J, {Baugh} CM,
  {Borgani} S, {Brodwin} M, {Castander} F, {Castignani} G, {Cucciati} O, {De
  Lucia} G, {Dubath} P, {Fosalba} P, {Giocoli} C, {Hoekstra} H, {Mamon} GA,
  {Melin} JB, {Moscardini} L, {Paltani} S, {Radovich} M, {Sartoris} B,
  {Schultheis} M, {Sereno} M, {Weller} J, {Burigana} C, {Carvalho} CS,
  {Corcione} L, {Kurki-Suonio} H, {Lilje} PB, {Sirri} G, {Toledo-Moreo} R,
  {Zamorani} G (2019) {Euclid preparation. III. Galaxy cluster detection in the
  wide photometric survey, performance and algorithm selection}. \aap 627:A23.
  \doi{10.1051/0004-6361/201935088}.
  {\href{https://arxiv.org/abs/1906.04707}{{arXiv:1906.04707}}} {[astro-ph.CO]}

\bibitem[{{Finney} et~al.(2018){Finney}, {Brada{\v{c}}}, {Huang}, {Hoag},
  {Morishita}, {Schrabback}, {Treu}, {Borello Schmidt}, {Lemaux}, {Wang}, and
  {Mason}}]{finney18}
{Finney} EQ, {Brada{\v{c}}} M, {Huang} KH, {Hoag} A, {Morishita} T,
  {Schrabback} T, {Treu} T, {Borello Schmidt} K, {Lemaux} BC, {Wang} X, {Mason}
  C (2018) {Mass Modeling of Frontier Fields Cluster MACS J1149.5+2223 Using
  Strong and Weak Lensing}. \apj 859:58. \doi{10.3847/1538-4357/aabf97}

\bibitem[{{Fo{\"e}x} et~al.(2014){Fo{\"e}x}, {Motta}, {Jullo}, {Limousin}, and
  {Verdugo}}]{Foex_2014}
{Fo{\"e}x} G, {Motta} V, {Jullo} E, {Limousin} M, {Verdugo} T (2014) {SARCS
  strong-lensing galaxy groups. II. Mass-concentration relation and
  strong-lensing bias}. \aap 572:A19. \doi{10.1051/0004-6361/201424706}.
  {\href{https://arxiv.org/abs/1409.5905}{{arXiv:1409.5905}}} {[astro-ph.CO]}

\bibitem[{{Fohlmeister} et~al.(2013){Fohlmeister}, {Kochanek}, {Falco},
  {Wambsganss}, {Oguri}, and {Dai}}]{Fohlmeister:2013}
{Fohlmeister} J, {Kochanek} CS, {Falco} EE, {Wambsganss} J, {Oguri} M, {Dai} X
  (2013) {A Two-year Time Delay for the Lensed Quasar SDSS J1029+2623}. \apj
  764(2):186. \doi{10.1088/0004-637X/764/2/186}.
  {\href{https://arxiv.org/abs/1207.5776}{{arXiv:1207.5776}}} {[astro-ph.CO]}

\bibitem[{{Fox} et~al.(2022){Fox}, {Mahler}, {Sharon}, and {Remolina
  Gonz{\'a}lez}}]{fox22}
{Fox} C, {Mahler} G, {Sharon} K, {Remolina Gonz{\'a}lez} JD (2022) {The
  Strongest Cluster Lenses: An Analysis of the Relation between Strong
  Gravitational Lensing Strength and the Physical Properties of Galaxy
  Clusters}. \apj 928(1):87. \doi{10.3847/1538-4357/ac5024}.
  {\href{https://arxiv.org/abs/2104.05585}{{arXiv:2104.05585}}} {[astro-ph.CO]}

\bibitem[{{Fox} and {Pen}(2001)}]{fox01}
{Fox} DC, {Pen} UL (2001) {Gravitational Lensing by Galaxy Groups in the Hubble
  Deep Field}. \apj 546(1):35--46. \doi{10.1086/318242}.
  {\href{https://arxiv.org/abs/astro-ph/0009068}{{arXiv:astro-ph/0009068}}}
  {[astro-ph]}

\bibitem[{{Furtak} et~al.(2023){Furtak}, {Shuntov}, {Atek}, {Zitrin},
  {Richard}, {Lehnert}, and {Chevallard}}]{Furtak2022}
{Furtak} LJ, {Shuntov} M, {Atek} H, {Zitrin} A, {Richard} J, {Lehnert} MD,
  {Chevallard} J (2023) {Constraining the physical properties of the first
  lensed $z \sim$ 9--16 galaxy candidates with JWST}. \mnras 519(2):3064--3075.
  \doi{10.1093/mnras/stac3717}.
  {\href{https://arxiv.org/abs/2208.05473}{{arXiv:2208.05473}}} {[astro-ph.GA]}

\bibitem[{{Ghigna} et~al.(1998){Ghigna}, {Moore}, {Governato}, {Lake}, {Quinn},
  and {Stadel}}]{Ghigna:1998}
{Ghigna} S, {Moore} B, {Governato} F, {Lake} G, {Quinn} T, {Stadel} J (1998)
  {Dark matter haloes within clusters}. \mnras 300(1):146--162.
  \doi{10.1046/j.1365-8711.1998.01918.x}.
  {\href{https://arxiv.org/abs/astro-ph/9801192}{{arXiv:astro-ph/9801192}}}
  {[astro-ph]}

\bibitem[{{Ghosh} et~al.(2020){Ghosh}, {Williams}, and {Liesenborgs}}]{gho20}
{Ghosh} A, {Williams} LLR, {Liesenborgs} J (2020) {Free-form grale lens
  inversion of galaxy clusters with up to 1000 multiple images}. \mnras
  494(3):3998--4014. \doi{10.1093/mnras/staa962}.
  {\href{https://arxiv.org/abs/2004.01724}{{arXiv:2004.01724}}} {[astro-ph.CO]}

\bibitem[{{Ghosh} et~al.(2021){Ghosh}, {Williams}, {Liesenborgs}, {Acebron},
  {Jauzac}, {Koekemoer}, {Mahler}, {Niemiec}, {Steinhardt}, {Faisst},
  {Lagattuta}, and {Natarajan}}]{gho21}
{Ghosh} A, {Williams} LLR, {Liesenborgs} J, {Acebron} A, {Jauzac} M,
  {Koekemoer} AM, {Mahler} G, {Niemiec} A, {Steinhardt} C, {Faisst} AL,
  {Lagattuta} D, {Natarajan} P (2021) {Further support for a trio of
  mass-to-light deviations in Abell 370: free-form grale lens inversion using
  BUFFALO strong lensing data}. \mnras 506(4):6144--6158.
  \doi{10.1093/mnras/stab1196}.
  {\href{https://arxiv.org/abs/2104.11781}{{arXiv:2104.11781}}} {[astro-ph.CO]}

\bibitem[{{Ghosh} et~al.(2023){Ghosh}, {Adams}, {Williams}, {Liesenborgs},
  {Alavi}, and {Scarlata}}]{gho22}
{Ghosh} A, {Adams} D, {Williams} LLR, {Liesenborgs} J, {Alavi} A, {Scarlata} C
  (2023) {An excursion into the core of the cluster lens Abell 1689}. \mnras
  525(2):2519--2534. \doi{10.1093/mnras/stad2418}.
  {\href{https://arxiv.org/abs/2206.08584}{{arXiv:2206.08584}}} {[astro-ph.CO]}

\bibitem[{{Gilbank} et~al.(2011){Gilbank}, {Gladders}, {Yee}, and
  {Hsieh}}]{Gilbank2011}
{Gilbank} DG, {Gladders} MD, {Yee} HKC, {Hsieh} BC (2011) {The Red-sequence
  Cluster Survey-2 (RCS-2): Survey Details and Photometric Catalog
  Construction}. \aj 141(3):94. \doi{10.1088/0004-6256/141/3/94}.
  {\href{https://arxiv.org/abs/1012.3470}{{arXiv:1012.3470}}} {[astro-ph.CO]}

\bibitem[{{Gilmore} and {Natarajan}(2009)}]{GilmorePN2009}
{Gilmore} J, {Natarajan} P (2009) {Cosmography with cluster strong lensing}.
  \mnras 396(1):354--364. \doi{10.1111/j.1365-2966.2009.14612.x}.
  {\href{https://arxiv.org/abs/astro-ph/0605245}{{arXiv:astro-ph/0605245}}}
  {[astro-ph]}

\bibitem[{{Gladders} and {Yee}(2000)}]{GladdersYee2000}
{Gladders} MD, {Yee} HKC (2000) {A New Method For Galaxy Cluster Detection. I.
  The Algorithm}. \aj 120(4):2148--2162. \doi{10.1086/301557}.
  {\href{https://arxiv.org/abs/astro-ph/0004092}{{arXiv:astro-ph/0004092}}}
  {[astro-ph]}

\bibitem[{{Goldberg} and {Natarajan}(2002)}]{goldberg+2002}
{Goldberg} DM, {Natarajan} P (2002) {The Galaxy Octopole Moment as a Probe of
  Weak-Lensing Shear Fields}. \apj 564(1):65--72. \doi{10.1086/324202}.
  {\href{https://arxiv.org/abs/astro-ph/0107187}{{arXiv:astro-ph/0107187}}}
  {[astro-ph]}

\bibitem[{{Golubchik} et~al.(2022){Golubchik}, {Furtak}, {Meena}, and
  {Zitrin}}]{Golubchik2022}
{Golubchik} M, {Furtak} LJ, {Meena} AK, {Zitrin} A (2022) {HST Strong-lensing
  Model for the First JWST Galaxy Cluster SMACS J0723.3-7327}. \apj 938(1):14.
  \doi{10.3847/1538-4357/ac8ff1}.
  {\href{https://arxiv.org/abs/2207.05007}{{arXiv:2207.05007}}} {[astro-ph.CO]}

\bibitem[{{Gonzalez} et~al.(2005){Gonzalez}, {Zabludoff}, and
  {Zaritsky}}]{Gonzalez+2005}
{Gonzalez} AH, {Zabludoff} AI, {Zaritsky} D (2005) {Intracluster Light in
  Nearby Galaxy Clusters: Relationship to the Halos of Brightest Cluster
  Galaxies}. \apj 618(1):195--213. \doi{10.1086/425896}.
  {\href{https://arxiv.org/abs/astro-ph/0406244}{{arXiv:astro-ph/0406244}}}
  {[astro-ph]}

\bibitem[{{Gonz{\'a}lez-L{\'o}pez} et~al.(2017){Gonz{\'a}lez-L{\'o}pez},
  {Bauer}, {Romero-Ca{\~n}izales}, {Kneissl}, {Villard}, {Carvajal}, {Kim},
  {Laporte}, {Anguita}, {Aravena}, {Bouwens}, {Bradley}, {Carrasco}, {Demarco},
  {Ford}, {Ibar}, {Infante}, {Messias}, {Mu{\~n}oz Arancibia}, {Nagar},
  {Padilla}, {Treister}, {Troncoso}, and {Zitrin}}]{gl2017}
{Gonz{\'a}lez-L{\'o}pez} J, {Bauer} FE, {Romero-Ca{\~n}izales} C, {Kneissl} R,
  {Villard} E, {Carvajal} R, {Kim} S, {Laporte} N, {Anguita} T, {Aravena} M,
  {Bouwens} RJ, {Bradley} L, {Carrasco} M, {Demarco} R, {Ford} H, {Ibar} E,
  {Infante} L, {Messias} H, {Mu{\~n}oz Arancibia} AM, {Nagar} N, {Padilla} N,
  {Treister} E, {Troncoso} P, {Zitrin} A (2017) {The ALMA Frontier Fields
  Survey. I. 1.1 mm continuum detections in Abell 2744, MACS J0416.1-2403 and
  MACS J1149.5+2223}. \aap 597:A41. \doi{10.1051/0004-6361/201628806}.
  {\href{https://arxiv.org/abs/1607.03808}{{arXiv:1607.03808}}} {[astro-ph.GA]}

\bibitem[{{Granata} et~al.(2022){Granata}, {Mercurio}, {Grillo}, {Tortorelli},
  {Bergamini}, {Meneghetti}, {Rosati}, {Caminha}, and {Nonino}}]{Granata:2021}
{Granata} G, {Mercurio} A, {Grillo} C, {Tortorelli} L, {Bergamini} P,
  {Meneghetti} M, {Rosati} P, {Caminha} GB, {Nonino} M (2022) {Improved strong
  lensing modelling of galaxy clusters using the Fundamental Plane: Detailed
  mapping of the baryonic and dark matter mass distribution of Abell S1063}.
  \aap 659:A24. \doi{10.1051/0004-6361/202141817}.
  {\href{https://arxiv.org/abs/2107.09079}{{arXiv:2107.09079}}} {[astro-ph.GA]}

\bibitem[{{Griffiths} et~al.(2021){Griffiths}, {Rudisel}, {Wagner}, {Hamilton},
  {Huang}, and {Villforth}}]{gri21}
{Griffiths} RE, {Rudisel} M, {Wagner} J, {Hamilton} T, {Huang} PC, {Villforth}
  C (2021) {Hamilton's Object - a clumpy galaxy straddling the gravitational
  caustic of a galaxy cluster: constraints on dark matter clumping}. \mnras
  506(2):1595--1608. \doi{10.1093/mnras/stab1375}.
  {\href{https://arxiv.org/abs/2105.04562}{{arXiv:2105.04562}}} {[astro-ph.CO]}

\bibitem[{{Grillo} et~al.(2014){Grillo}, {Gobat}, {Presotto}, {Balestra},
  {Mercurio}, {Rosati}, {Nonino}, {Vanzella}, {Christensen}, {Graves},
  {Biviano}, {Lemze}, {Bartelmann}, {Benitez}, {Bouwens}, {Bradley},
  {Broadhurst}, {Coe}, {Donahue}, {Ford}, {Infante}, {Jouvel}, {Kelson},
  {Koekemoer}, {Lahav}, {Medezinski}, {Melchior}, {Meneghetti}, {Merten},
  {Molino}, {Monna}, {Moustakas}, {Moustakas}, {Postman}, {Seitz}, {Umetsu},
  {Zheng}, and {Zitrin}}]{Grillo:2014}
{Grillo} C, {Gobat} R, {Presotto} V, {Balestra} I, {Mercurio} A, {Rosati} P,
  {Nonino} M, {Vanzella} E, {Christensen} L, {Graves} G, {Biviano} A, {Lemze}
  D, {Bartelmann} M, {Benitez} N, {Bouwens} R, {Bradley} L, {Broadhurst} T,
  {Coe} D, {Donahue} M, {Ford} H, {Infante} L, {Jouvel} S, {Kelson} D,
  {Koekemoer} A, {Lahav} O, {Medezinski} E, {Melchior} P, {Meneghetti} M,
  {Merten} J, {Molino} A, {Monna} A, {Moustakas} J, {Moustakas} LA, {Postman}
  M, {Seitz} S, {Umetsu} K, {Zheng} W, {Zitrin} A (2014) {CLASH: Extending
  Galaxy Strong Lensing to Small Physical Scales with Distant Sources Highly
  Magnified by Galaxy Cluster Members}. \apj 786(1):11.
  \doi{10.1088/0004-637X/786/1/11}.
  {\href{https://arxiv.org/abs/1403.0573}{{arXiv:1403.0573}}} {[astro-ph.CO]}

\bibitem[{{Grillo} et~al.(2015){Grillo}, {Suyu}, {Rosati}, {Mercurio},
  {Balestra}, {Munari}, {Nonino}, {Caminha}, {Lombardi}, {De Lucia}, {Borgani},
  {Gobat}, {Biviano}, {Girardi}, {Umetsu}, {Coe}, {Koekemoer}, {Postman},
  {Zitrin}, {Halkola}, {Broadhurst}, {Sartoris}, {Presotto}, {Annunziatella},
  {Maier}, {Fritz}, {Vanzella}, and {Frye}}]{Grillo:2015}
{Grillo} C, {Suyu} SH, {Rosati} P, {Mercurio} A, {Balestra} I, {Munari} E,
  {Nonino} M, {Caminha} GB, {Lombardi} M, {De Lucia} G, {Borgani} S, {Gobat} R,
  {Biviano} A, {Girardi} M, {Umetsu} K, {Coe} D, {Koekemoer} AM, {Postman} M,
  {Zitrin} A, {Halkola} A, {Broadhurst} T, {Sartoris} B, {Presotto} V,
  {Annunziatella} M, {Maier} C, {Fritz} A, {Vanzella} E, {Frye} B (2015)
  {CLASH-VLT: Insights on the Mass Substructures in the Frontier Fields Cluster
  MACS J0416.1-2403 through Accurate Strong Lens Modeling}. \apj 800(1):38.
  \doi{10.1088/0004-637X/800/1/38}.
  {\href{https://arxiv.org/abs/1407.7866}{{arXiv:1407.7866}}} {[astro-ph.CO]}

\bibitem[{{Grillo} et~al.(2016){Grillo}, {Karman}, {Suyu}, {Rosati},
  {Balestra}, {Mercurio}, {Lombardi}, {Treu}, {Caminha}, {Halkola}, {Rodney},
  {Gavazzi}, and {Caputi}}]{Grillo:2016}
{Grillo} C, {Karman} W, {Suyu} SH, {Rosati} P, {Balestra} I, {Mercurio} A,
  {Lombardi} M, {Treu} T, {Caminha} GB, {Halkola} A, {Rodney} SA, {Gavazzi} R,
  {Caputi} KI (2016) {The Story of Supernova
  {\textquotedblleft}Refsdal{\textquotedblright} Told by Muse}. \apj 822(2):78.
  \doi{10.3847/0004-637X/822/2/78}.
  {\href{https://arxiv.org/abs/1511.04093}{{arXiv:1511.04093}}} {[astro-ph.GA]}

\bibitem[{{Grillo} et~al.(2018){Grillo}, {Rosati}, {Suyu}, {Balestra},
  {Caminha}, {Halkola}, {Kelly}, {Lombardi}, {Mercurio}, {Rodney}, and
  {Treu}}]{Grillo:2018}
{Grillo} C, {Rosati} P, {Suyu} SH, {Balestra} I, {Caminha} GB, {Halkola} A,
  {Kelly} PL, {Lombardi} M, {Mercurio} A, {Rodney} SA, {Treu} T (2018)
  {Measuring the Value of the Hubble Constant {\textquotedblleft}{\`a} la
  Refsdal{\textquotedblright}}. \apj 860(2):94. \doi{10.3847/1538-4357/aac2c9}.
  {\href{https://arxiv.org/abs/1802.01584}{{arXiv:1802.01584}}} {[astro-ph.CO]}

\bibitem[{{Grillo} et~al.(2020){Grillo}, {Rosati}, {Suyu}, {Caminha},
  {Mercurio}, and {Halkola}}]{Grillo:2020}
{Grillo} C, {Rosati} P, {Suyu} SH, {Caminha} GB, {Mercurio} A, {Halkola} A
  (2020) {On the Accuracy of Time-delay Cosmography in the Frontier Fields
  Cluster MACS J1149.5+2223 with Supernova Refsdal}. \apj 898(1):87.
  \doi{10.3847/1538-4357/ab9a4c}.
  {\href{https://arxiv.org/abs/2001.02232}{{arXiv:2001.02232}}} {[astro-ph.CO]}

\bibitem[{{Grossman} and {Narayan}(1988)}]{gro88}
{Grossman} SA, {Narayan} R (1988) {Arcs from Gravitational Lensing}. \apjl
  324:L37. \doi{10.1086/185086}

\bibitem[{{Gruen} et~al.(2014){Gruen}, {Seitz}, {Brimioulle}, {Kosyra},
  {Koppenhoefer}, {Lee}, {Bender}, {Riffeser}, {Eichner}, {Weidinger}, and
  {Bierschenk}}]{Gruen:2014}
{Gruen} D, {Seitz} S, {Brimioulle} F, {Kosyra} R, {Koppenhoefer} J, {Lee} CH,
  {Bender} R, {Riffeser} A, {Eichner} T, {Weidinger} T, {Bierschenk} M (2014)
  {Weak lensing analysis of SZ-selected clusters of galaxies from the SPT and
  Planck surveys}. \mnras 442(2):1507--1544. \doi{10.1093/mnras/stu949}.
  {\href{https://arxiv.org/abs/1310.6744}{{arXiv:1310.6744}}} {[astro-ph.CO]}

\bibitem[{{Harvey} et~al.(2019){Harvey}, {Robertson}, {Massey}, and
  {McCarthy}}]{bib2:harvey2019}
{Harvey} D, {Robertson} A, {Massey} R, {McCarthy} IG (2019) {Observable tests
  of self-interacting dark matter in galaxy clusters: BCG wobbles in a constant
  density core}. \mnras 488(2):1572--1579. \doi{10.1093/mnras/stz1816}.
  {\href{https://arxiv.org/abs/1812.06981}{{arXiv:1812.06981}}} {[astro-ph.CO]}

\bibitem[{{Hjorth} and {Williams}(2010)}]{hjo10}
{Hjorth} J, {Williams} LLR (2010) {Statistical Mechanics of Collisionless
  Orbits. I. Origin of Central Cusps in Dark-matter Halos}. \apj
  722(1):851--855. \doi{10.1088/0004-637X/722/1/851}.
  {\href{https://arxiv.org/abs/1010.0265}{{arXiv:1010.0265}}} {[astro-ph.CO]}

\bibitem[{{Hoag} et~al.(2016){Hoag}, {Huang}, {Treu}, {Brada{\v c}}, {Schmidt},
  {Wang}, {Brammer}, {Broussard}, {Amorin}, {Castellano}, {Fontana}, {Merlin},
  {Schrabback}, {Trenti}, and {Vulcani}}]{hoag16}
{Hoag} A, {Huang} KH, {Treu} T, {Brada{\v c}} M, {Schmidt} KB, {Wang} X,
  {Brammer} GB, {Broussard} A, {Amorin} R, {Castellano} M, {Fontana} A,
  {Merlin} E, {Schrabback} T, {Trenti} M, {Vulcani} B (2016) {The Grism
  Lens-Amplified Survey from Space (GLASS). VI. Comparing the Mass and Light in
  MACS J0416.1-2403 Using Frontier Field Imaging and GLASS Spectroscopy}. \apj
  831:182. \doi{10.3847/0004-637X/831/2/182}.
  {\href{https://arxiv.org/abs/1603.00505}{{arXiv:1603.00505}}}

\bibitem[{{Holanda} et~al.(2017){Holanda}, {Busti}, {Gonzalez},
  {Andrade-Santos}, and {Alcaniz}}]{holanda2017}
{Holanda} RFL, {Busti} VC, {Gonzalez} JE, {Andrade-Santos} F, {Alcaniz} JS
  (2017) {Cosmological constraints on the gas depletion factor in galaxy
  clusters}. \jcap 2017(12):016. \doi{10.1088/1475-7516/2017/12/016}.
  {\href{https://arxiv.org/abs/1706.07321}{{arXiv:1706.07321}}} {[astro-ph.CO]}

\bibitem[{{Horesh} et~al.(2010){Horesh}, {Maoz}, {Ebeling}, {Seidel}, and
  {Bartelmann}}]{horesh10}
{Horesh} A, {Maoz} D, {Ebeling} H, {Seidel} G, {Bartelmann} M (2010) {The
  lensing efficiencies of MACS X-ray-selected versus RCS optically selected
  galaxy clusters}. \mnras 406(2):1318--1336.
  \doi{10.1111/j.1365-2966.2010.16763.x}.
  {\href{https://arxiv.org/abs/1004.2048}{{arXiv:1004.2048}}} {[astro-ph.CO]}

\bibitem[{{Ishigaki} et~al.(2015){Ishigaki}, {Kawamata}, {Ouchi}, {Oguri},
  {Shimasaku}, and {Ono}}]{Ishigaki:15}
{Ishigaki} M, {Kawamata} R, {Ouchi} M, {Oguri} M, {Shimasaku} K, {Ono} Y (2015)
  {Hubble Frontier Fields First Complete Cluster Data: Faint Galaxies at $z
  \sim $5--10 for UV Luminosity Functions and Cosmic Reionization}. \apj
  799(1):12. \doi{10.1088/0004-637X/799/1/12}.
  {\href{https://arxiv.org/abs/1408.6903}{{arXiv:1408.6903}}} {[astro-ph.GA]}

\bibitem[{{Jaelani} et~al.(2020){Jaelani}, {More}, {Oguri}, {Sonnenfeld},
  {Suyu}, {Rusu}, {Wong}, {Chan}, {Kayo}, {Lee}, {Chao}, {Coupon}, {Inoue}, and
  {Futamase}}]{jaelani20}
{Jaelani} AT, {More} A, {Oguri} M, {Sonnenfeld} A, {Suyu} SH, {Rusu} CE, {Wong}
  KC, {Chan} JHH, {Kayo} I, {Lee} CH, {Chao} DCY, {Coupon} J, {Inoue} KT,
  {Futamase} T (2020) {Survey of Gravitationally lensed Objects in HSC Imaging
  (SuGOHI) - V. Group-to-cluster scale lens search from the HSC-SSP Survey}.
  \mnras 495(1):1291--1310. \doi{10.1093/mnras/staa1062}.
  {\href{https://arxiv.org/abs/2002.01611}{{arXiv:2002.01611}}} {[astro-ph.GA]}

\bibitem[{{Jauzac} et~al.(2016){Jauzac}, {Eckert}, {Schwinn}, {Harvey},
  {Baugh}, {Robertson}, {Bose}, {Massey}, {Owers}, {Ebeling}, {Shan}, {Jullo},
  {Kneib}, {Richard}, {Atek}, {Cl{\'e}ment}, {Egami}, {Israel}, {Knowles},
  {Limousin}, {Natarajan}, {Rexroth}, {Taylor}, and {Tchernin}}]{Jauzac+2016}
{Jauzac} M, {Eckert} D, {Schwinn} J, {Harvey} D, {Baugh} CM, {Robertson} A,
  {Bose} S, {Massey} R, {Owers} M, {Ebeling} H, {Shan} HY, {Jullo} E, {Kneib}
  JP, {Richard} J, {Atek} H, {Cl{\'e}ment} B, {Egami} E, {Israel} H, {Knowles}
  K, {Limousin} M, {Natarajan} P, {Rexroth} M, {Taylor} P, {Tchernin} C (2016)
  {The extraordinary amount of substructure in the Hubble Frontier Fields
  cluster Abell 2744}. \mnras 463(4):3876--3893. \doi{10.1093/mnras/stw2251}.
  {\href{https://arxiv.org/abs/1606.04527}{{arXiv:1606.04527}}} {[astro-ph.CO]}

\bibitem[{{Johnson} and {Sharon}(2016)}]{joh16}
{Johnson} TL, {Sharon} K (2016) {The Systematics of Strong Lens Modeling
  Quantified: The Effects of Constraint Selection and Redshift Information on
  Magnification, Mass, and Multiple Image Predictability}. \apj 832(1):82.
  \doi{10.3847/0004-637X/832/1/82}.
  {\href{https://arxiv.org/abs/1608.08713}{{arXiv:1608.08713}}} {[astro-ph.CO]}

\bibitem[{{Johnson} et~al.(2014){Johnson}, {Sharon}, {Bayliss}, {Gladders},
  {Coe}, and {Ebeling}}]{Johnson2014}
{Johnson} TL, {Sharon} K, {Bayliss} MB, {Gladders} MD, {Coe} D, {Ebeling} H
  (2014) {Lens Models and Magnification Maps of the Six Hubble Frontier Fields
  Clusters}. \apj 797(1):48. \doi{10.1088/0004-637X/797/1/48}.
  {\href{https://arxiv.org/abs/1405.0222}{{arXiv:1405.0222}}} {[astro-ph.CO]}

\bibitem[{{Johnson} et~al.(2017){Johnson}, {Rigby}, {Sharon}, {Gladders},
  {Florian}, {Bayliss}, {Wuyts}, {Whitaker}, {Livermore}, and
  {Murray}}]{johnson2017}
{Johnson} TL, {Rigby} JR, {Sharon} K, {Gladders} MD, {Florian} M, {Bayliss} MB,
  {Wuyts} E, {Whitaker} KE, {Livermore} R, {Murray} KT (2017) {Star Formation
  at z = 2.481 in the Lensed Galaxy SDSS J1110+6459: Star Formation Down to 30
  pc Scales}. \apjl 843(2):L21. \doi{10.3847/2041-8213/aa7516}.
  {\href{https://arxiv.org/abs/1707.00706}{{arXiv:1707.00706}}} {[astro-ph.GA]}

\bibitem[{{Jullo} and {Kneib}(2009)}]{Jullo:09}
{Jullo} E, {Kneib} JP (2009) {Multiscale cluster lens mass mapping - I. Strong
  lensing modelling}. \mnras 395(3):1319--1332.
  \doi{10.1111/j.1365-2966.2009.14654.x}.
  {\href{https://arxiv.org/abs/0901.3792}{{arXiv:0901.3792}}} {[astro-ph.CO]}

\bibitem[{{Jullo} et~al.(2007{\natexlab{a}}){Jullo}, {Kneib}, {Limousin},
  {El{\'\i}asd{\'o}ttir}, {Marshall}, and {Verdugo}}]{Jullo:07}
{Jullo} E, {Kneib} JP, {Limousin} M, {El{\'\i}asd{\'o}ttir} {\'A}, {Marshall}
  PJ, {Verdugo} T (2007{\natexlab{a}}) {A Bayesian approach to strong lensing
  modelling of galaxy clusters}. New J Phys 9(12):447.
  \doi{10.1088/1367-2630/9/12/447}.
  {\href{https://arxiv.org/abs/0706.0048}{{arXiv:0706.0048}}} {[astro-ph]}

\bibitem[{{Jullo} et~al.(2007{\natexlab{b}}){Jullo}, {Kneib}, {Limousin},
  {El{\'\i}asd{\'o}ttir}, {Marshall}, and {Verdugo}}]{jul07}
{Jullo} E, {Kneib} JP, {Limousin} M, {El{\'\i}asd{\'o}ttir} {\'A}, {Marshall}
  PJ, {Verdugo} T (2007{\natexlab{b}}) {A Bayesian approach to strong lensing
  modelling of galaxy clusters}. New J Phys 9(12):447.
  \doi{10.1088/1367-2630/9/12/447}.
  {\href{https://arxiv.org/abs/0706.0048}{{arXiv:0706.0048}}} {[astro-ph]}

\bibitem[{{Jullo} et~al.(2010){Jullo}, {Natarajan}, {Kneib}, {D'Aloisio},
  {Limousin}, {Richard}, and {Schimd}}]{Jullo:2010}
{Jullo} E, {Natarajan} P, {Kneib} JP, {D'Aloisio} A, {Limousin} M, {Richard} J,
  {Schimd} C (2010) {Cosmological constraints from strong gravitational lensing
  in clusters of galaxies.} Science 329:924--927.
  \doi{10.1126/science.1185759}.
  {\href{https://arxiv.org/abs/1008.4802}{{arXiv:1008.4802}}} {[astro-ph.CO]}

\bibitem[{Jullo et~al.(2015)Jullo, Acebron, Limousin, Giocoli, Despali, and
  Bonamigo}]{jullo+2015}
Jullo E, Acebron A, Limousin M, Giocoli C, Despali G, Bonamigo M (2015) Strong
  lensing cosmography in the frontier fields. Proceedings of the International
  Astronomical Union 11(A29B):801--803. \doi{10.1017/S1743921316006888}

\bibitem[{{Kahlhoefer} et~al.(2015){Kahlhoefer}, {Schmidt-Hoberg}, {Kummer},
  and {Sarkar}}]{bib2:kahlhoefer2015}
{Kahlhoefer} F, {Schmidt-Hoberg} K, {Kummer} J, {Sarkar} S (2015) {On the
  interpretation of dark matter self-interactions in Abell 3827}. \mnras
  452(1):L54--L58. \doi{10.1093/mnrasl/slv088}.
  {\href{https://arxiv.org/abs/1504.06576}{{arXiv:1504.06576}}} {[astro-ph.CO]}

\bibitem[{{Kaiser}(1992)}]{Kaiser1992}
{Kaiser} N (1992) {Weak Gravitational Lensing of Distant Galaxies}. \apj
  388:272. \doi{10.1086/171151}

\bibitem[{{Kaiser} and {Squires}(1993)}]{Kaiser1993}
{Kaiser} N, {Squires} G (1993) {Mapping the Dark Matter with Weak Gravitational
  Lensing}. \apj 404:441. \doi{10.1086/172297}

\bibitem[{{Karman} et~al.(2015){Karman}, {Caputi}, {Grillo}, {Balestra},
  {Rosati}, {Vanzella}, {Coe}, {Christensen}, {Koekemoer}, {Kr{\"u}hler},
  {Lombardi}, {Mercurio}, {Nonino}, and {van der Wel}}]{Karman:2015}
{Karman} W, {Caputi} KI, {Grillo} C, {Balestra} I, {Rosati} P, {Vanzella} E,
  {Coe} D, {Christensen} L, {Koekemoer} AM, {Kr{\"u}hler} T, {Lombardi} M,
  {Mercurio} A, {Nonino} M, {van der Wel} A (2015) {MUSE integral-field
  spectroscopy towards the Frontier Fields cluster Abell S1063. I. Data
  products and redshift identifications}. \aap 574:A11.
  \doi{10.1051/0004-6361/201424962}.
  {\href{https://arxiv.org/abs/1409.3507}{{arXiv:1409.3507}}} {[astro-ph.GA]}

\bibitem[{{Kaurov} et~al.(2019){Kaurov}, {Dai}, {Venumadhav},
  {Miralda-Escud{\'e}}, and {Frye}}]{kau19}
{Kaurov} AA, {Dai} L, {Venumadhav} T, {Miralda-Escud{\'e}} J, {Frye} B (2019)
  {Highly Magnified Stars in Lensing Clusters: New Evidence in a Galaxy Lensed
  by MACS J0416.1-2403}. \apj 880(1):58. \doi{10.3847/1538-4357/ab2888}.
  {\href{https://arxiv.org/abs/1902.10090}{{arXiv:1902.10090}}} {[astro-ph.GA]}

\bibitem[{{Kawamata} et~al.(2016){Kawamata}, {Oguri}, {Ishigaki}, {Shimasaku},
  and {Ouchi}}]{Kawamata:16}
{Kawamata} R, {Oguri} M, {Ishigaki} M, {Shimasaku} K, {Ouchi} M (2016) {Precise
  Strong Lensing Mass Modeling of Four Hubble Frontier Field Clusters and a
  Sample of Magnified High-redshift Galaxies}. \apj 819(2):114.
  \doi{10.3847/0004-637X/819/2/114}.
  {\href{https://arxiv.org/abs/1510.06400}{{arXiv:1510.06400}}} {[astro-ph.GA]}

\bibitem[{{Keeton}(2010)}]{Keeton:10}
{Keeton} CR (2010) {On modeling galaxy-scale strong lens systems}. General
  Relativity and Gravitation 42(9):2151--2176. \doi{10.1007/s10714-010-1041-1}

\bibitem[{{Kelly} et~al.(2015){Kelly}, {Rodney}, {Treu}, {Foley}, {Brammer},
  {Schmidt}, {Zitrin}, {Sonnenfeld}, {Strolger}, {Graur}, {Filippenko}, {Jha},
  {Riess}, {Bradac}, {Weiner}, {Scolnic}, {Malkan}, {von der Linden}, {Trenti},
  {Hjorth}, {Gavazzi}, {Fontana}, {Merten}, {McCully}, {Jones}, {Postman},
  {Dressler}, {Patel}, {Cenko}, {Graham}, and {Tucker}}]{Kelly+2015}
{Kelly} PL, {Rodney} SA, {Treu} T, {Foley} RJ, {Brammer} G, {Schmidt} KB,
  {Zitrin} A, {Sonnenfeld} A, {Strolger} LG, {Graur} O, {Filippenko} AV, {Jha}
  SW, {Riess} AG, {Bradac} M, {Weiner} BJ, {Scolnic} D, {Malkan} MA, {von der
  Linden} A, {Trenti} M, {Hjorth} J, {Gavazzi} R, {Fontana} A, {Merten} JC,
  {McCully} C, {Jones} T, {Postman} M, {Dressler} A, {Patel} B, {Cenko} SB,
  {Graham} ML, {Tucker} BE (2015) {Multiple images of a highly magnified
  supernova formed by an early-type cluster galaxy lens}. Science
  347(6226):1123--1126. \doi{10.1126/science.aaa3350}.
  {\href{https://arxiv.org/abs/1411.6009}{{arXiv:1411.6009}}} {[astro-ph.CO]}

\bibitem[{{Kelly} et~al.(2016){Kelly}, {Rodney}, {Treu}, {Strolger}, {Foley},
  {Jha}, {Selsing}, {Brammer}, {Brada{\v{c}}}, {Cenko}, {Graur}, {Filippenko},
  {Hjorth}, {McCully}, {Molino}, {Nonino}, {Riess}, {Schmidt}, {Tucker}, {von
  der Linden}, {Weiner}, and {Zitrin}}]{Kelly:2016}
{Kelly} PL, {Rodney} SA, {Treu} T, {Strolger} LG, {Foley} RJ, {Jha} SW,
  {Selsing} J, {Brammer} G, {Brada{\v{c}}} M, {Cenko} SB, {Graur} O,
  {Filippenko} AV, {Hjorth} J, {McCully} C, {Molino} A, {Nonino} M, {Riess} AG,
  {Schmidt} KB, {Tucker} B, {von der Linden} A, {Weiner} BJ, {Zitrin} A (2016)
  {Deja Vu All Over Again: The Reappearance of Supernova Refsdal}. \apjl
  819(1):L8. \doi{10.3847/2041-8205/819/1/L8}.
  {\href{https://arxiv.org/abs/1512.04654}{{arXiv:1512.04654}}} {[astro-ph.CO]}

\bibitem[{{Kelly} et~al.(2023){Kelly}, {Rodney}, {Treu}, {Oguri}, {Chen},
  {Zitrin}, {Birrer}, {Bonvin}, {Dessart}, {Diego}, {Filippenko}, {Foley},
  {Gilman}, {Hjorth}, {Jauzac}, {Mandel}, {Millon}, {Pierel}, {Sharon},
  {Thorp}, {Williams}, {Broadhurst}, {Dressler}, {Graur}, {Jha}, {McCully},
  {Postman}, {Schmidt}, {Tucker}, and {von der Linden}}]{kelly+2023}
{Kelly} PL, {Rodney} S, {Treu} T, {Oguri} M, {Chen} W, {Zitrin} A, {Birrer} S,
  {Bonvin} V, {Dessart} L, {Diego} JM, {Filippenko} AV, {Foley} RJ, {Gilman} D,
  {Hjorth} J, {Jauzac} M, {Mandel} K, {Millon} M, {Pierel} J, {Sharon} K,
  {Thorp} S, {Williams} L, {Broadhurst} T, {Dressler} A, {Graur} O, {Jha} S,
  {McCully} C, {Postman} M, {Schmidt} KB, {Tucker} BE, {von der Linden} A
  (2023) {Constraints on the Hubble constant from supernova Refsdal's
  reappearance}. Science 380(6649):abh1322. \doi{10.1126/science.abh1322}.
  {\href{https://arxiv.org/abs/2305.06367}{{arXiv:2305.06367}}} {[astro-ph.CO]}

\bibitem[{{Khullar} et~al.(2021){Khullar}, {Gozman}, {Lin}, {Martinez},
  {Matthews Acu{\~n}a}, {Medina}, {Merz}, {Sanchez}, {Sisco}, {Kavin Stein},
  {Sukay}, {Tavangar}, {Bayliss}, {Bleem}, {Brownsberger}, {Dahle}, {Florian},
  {Gladders}, {Mahler}, {Rigby}, {Sharon}, and {Stark}}]{Khullar2021}
{Khullar} G, {Gozman} K, {Lin} JJ, {Martinez} MN, {Matthews Acu{\~n}a} OS,
  {Medina} E, {Merz} K, {Sanchez} JA, {Sisco} EE, {Kavin Stein} DJ, {Sukay} EO,
  {Tavangar} K, {Bayliss} MB, {Bleem} LE, {Brownsberger} S, {Dahle} H,
  {Florian} MK, {Gladders} MD, {Mahler} G, {Rigby} JR, {Sharon} K, {Stark} AA
  (2021) {COOL-LAMPS. I. An Extraordinarily Bright Lensed Galaxy at Redshift
  5.04}. \apj 906(2):107. \doi{10.3847/1538-4357/abcb86}.
  {\href{https://arxiv.org/abs/2011.06601}{{arXiv:2011.06601}}} {[astro-ph.GA]}

\bibitem[{{Kneib} and {Natarajan}(2011)}]{JPK-PN2011}
{Kneib} JP, {Natarajan} P (2011) {Cluster lenses}. \aapr 19:47.
  \doi{10.1007/s00159-011-0047-3}.
  {\href{https://arxiv.org/abs/1202.0185}{{arXiv:1202.0185}}} {[astro-ph.CO]}

\bibitem[{{Kneib} et~al.(1993){Kneib}, {Mellier}, {Fort}, and {Mathez}}]{kne93}
{Kneib} JP, {Mellier} Y, {Fort} B, {Mathez} G (1993) {The distribution of dark
  matter in distant cluster-lenses : modelling modelling A 370.} \aap 273:367

\bibitem[{{Kneib} et~al.(1996){Kneib}, {Ellis}, {Smail}, {Couch}, and
  {Sharples}}]{Kneib:96}
{Kneib} JP, {Ellis} RS, {Smail} I, {Couch} WJ, {Sharples} RM (1996) {Hubble
  Space Telescope Observations of the Lensing Cluster Abell 2218}. \apj
  471:643. \doi{10.1086/177995}.
  {\href{https://arxiv.org/abs/astro-ph/9511015}{{arXiv:astro-ph/9511015}}}
  {[astro-ph]}

\bibitem[{{Kochanek}(1990)}]{Kochanek1990}
{Kochanek} CS (1990) {Inverting cluster gravitational lenses.} \mnras 247:135

\bibitem[{{K{\"o}hlinger} and {Schmidt}(2014)}]{koh14}
{K{\"o}hlinger} F, {Schmidt} RW (2014) {Strong lensing in RX J1347.5-1145
  revisited}. \mnras 437(2):1858--1871. \doi{10.1093/mnras/stt2017}.
  {\href{https://arxiv.org/abs/1310.0021}{{arXiv:1310.0021}}} {[astro-ph.CO]}

\bibitem[{{Kovner}(1989)}]{kov89}
{Kovner} I (1989) {Diagnostics of Compact Clusters of Galaxies by Giant
  Luminous Arcs}. \apj 337:621. \doi{10.1086/167133}

\bibitem[{{Lagattuta} et~al.(2022){Lagattuta}, {Richard}, {Bauer}, {Cerny},
  {Claeyssens}, {Guaita}, {Jauzac}, {Jeanneau}, {Koekemoer}, {Mahler}, {Prieto
  Lyon}, {Acebron}, {Meneghetti}, {Niemiec}, {Zitrin}, {Bianconi}, {Connor},
  {Cen}, {Edge}, {Faisst}, {Limousin}, {Massey}, {Sereno}, {Sharon}, and
  {Weaver}}]{lagattuta22}
{Lagattuta} DJ, {Richard} J, {Bauer} FE, {Cerny} C, {Claeyssens} A, {Guaita} L,
  {Jauzac} M, {Jeanneau} A, {Koekemoer} AM, {Mahler} G, {Prieto Lyon} G,
  {Acebron} A, {Meneghetti} M, {Niemiec} A, {Zitrin} A, {Bianconi} M, {Connor}
  T, {Cen} R, {Edge} A, {Faisst} AL, {Limousin} M, {Massey} R, {Sereno} M,
  {Sharon} K, {Weaver} JR (2022) {Pilot-WINGS: An extended MUSE view of the
  structure of Abell 370}. \mnras 514(1):497--517. \doi{10.1093/mnras/stac418}.
  {\href{https://arxiv.org/abs/2202.04663}{{arXiv:2202.04663}}} {[astro-ph.GA]}

\bibitem[{{Lage} and {Farrar}(2015)}]{lage15}
{Lage} C, {Farrar} GR (2015) {The bullet cluster is not a cosmological
  anomaly}. \jcap 2015(2):038. \doi{10.1088/1475-7516/2015/02/038}.
  {\href{https://arxiv.org/abs/1406.6703}{{arXiv:1406.6703}}} {[astro-ph.GA]}

\bibitem[{{Lam}(2019)}]{lam19}
{Lam} D (2019) {A New Approach to Free-form Cluster Lens Modeling Inspired by
  the JPEG Image Compression Method}. \pasp 131(1005):114505.
  \doi{10.1088/1538-3873/ab35c0}.
  {\href{https://arxiv.org/abs/1906.00006}{{arXiv:1906.00006}}} {[astro-ph.CO]}

\bibitem[{{Lanusse} et~al.(2016){Lanusse}, {Starck}, {Leonard}, and
  {Pires}}]{lanusse2016}
{Lanusse} F, {Starck} JL, {Leonard} A, {Pires} S (2016) {High resolution weak
  lensing mass mapping combining shear and flexion}. \aap 591:A2.
  \doi{10.1051/0004-6361/201628278}.
  {\href{https://arxiv.org/abs/1603.01599}{{arXiv:1603.01599}}} {[astro-ph.CO]}

\bibitem[{{Laporte} et~al.(2022){Laporte}, {Zitrin}, {Dole}, {Roberts-Borsani},
  {Furtak}, and {Witten}}]{Laporte2022}
{Laporte} N, {Zitrin} A, {Dole} H, {Roberts-Borsani} G, {Furtak} LJ, {Witten} C
  (2022) {A lensed protocluster candidate at z = 7.66 identified in JWST
  observations of the galaxy cluster SMACS0723{\ensuremath{-}}7327}. \aap
  667:L3. \doi{10.1051/0004-6361/202244719}.
  {\href{https://arxiv.org/abs/2208.04930}{{arXiv:2208.04930}}} {[astro-ph.GA]}

\bibitem[{{Lasko} et~al.(2023){Lasko}, {Williams}, and {Ghosh}}]{lasko2023}
{Lasko} K, {Williams} LLR, {Ghosh} A (2023) {What multiple images say about the
  large-scale mass maps of galaxy clusters}. \mnras 525(4):5423--5436.
  \doi{10.1093/mnras/stad2622}.
  {\href{https://arxiv.org/abs/2309.05730}{{arXiv:2309.05730}}}

\bibitem[{{Leonard} et~al.(2007){Leonard}, {Goldberg}, {Haaga}, and
  {Massey}}]{leonard07}
{Leonard} A, {Goldberg} DM, {Haaga} JL, {Massey} R (2007) {Gravitational Shear,
  Flexion, and Strong Lensing in Abell 1689}. \apj 666(1):51--63.
  \doi{10.1086/520109}.
  {\href{https://arxiv.org/abs/astro-ph/0702242}{{arXiv:astro-ph/0702242}}}
  {[astro-ph]}

\bibitem[{{Li} et~al.(2016){Li}, {Gladders}, {Rangel}, {Florian}, {Bleem},
  {Heitmann}, {Habib}, and {Fasel}}]{li16}
{Li} N, {Gladders} MD, {Rangel} EM, {Florian} MK, {Bleem} LE, {Heitmann} K,
  {Habib} S, {Fasel} P (2016) {PICS: Simulations of Strong Gravitational
  Lensing in Galaxy Clusters}. \apj 828(1):54.
  \doi{10.3847/0004-637X/828/1/54}.
  {\href{https://arxiv.org/abs/1511.03673}{{arXiv:1511.03673}}} {[astro-ph.CO]}

\bibitem[{{Li} et~al.(2019){Li}, {Gladders}, {Heitmann}, {Rangel}, {Child},
  {Florian}, {Bleem}, {Habib}, and {Finkel}}]{li19}
{Li} N, {Gladders} MD, {Heitmann} K, {Rangel} EM, {Child} HL, {Florian} MK,
  {Bleem} LE, {Habib} S, {Finkel} HJ (2019) {The Importance of Secondary Halos
  for Strong Lensing in Massive Galaxy Clusters across Redshift}. \apj
  878(2):122. \doi{10.3847/1538-4357/ab1f74}.
  {\href{https://arxiv.org/abs/1810.13330}{{arXiv:1810.13330}}} {[astro-ph.CO]}

\bibitem[{{Liesenborgs} et~al.(2006){Liesenborgs}, {De Rijcke}, and
  {Dejonghe}}]{Liesenborgs:06}
{Liesenborgs} J, {De Rijcke} S, {Dejonghe} H (2006) {A genetic algorithm for
  the non-parametric inversion of strong lensing systems}. \mnras
  367(3):1209--1216. \doi{10.1111/j.1365-2966.2006.10040.x}.
  {\href{https://arxiv.org/abs/astro-ph/0601124}{{arXiv:astro-ph/0601124}}}
  {[astro-ph]}

\bibitem[{{Liesenborgs} et~al.(2009){Liesenborgs}, {de Rijcke}, {Dejonghe}, and
  {Bekaert}}]{lie09}
{Liesenborgs} J, {de Rijcke} S, {Dejonghe} H, {Bekaert} P (2009)
  {Non-parametric strong lens inversion of SDSS J1004+4112}. \mnras
  397(1):341--349. \doi{10.1111/j.1365-2966.2009.14912.x}.
  {\href{https://arxiv.org/abs/0904.2382}{{arXiv:0904.2382}}} {[astro-ph.CO]}

\bibitem[{{Liesenborgs} et~al.(2020){Liesenborgs}, {Williams}, {Wagner}, and
  {De Rijcke}}]{lie20}
{Liesenborgs} J, {Williams} LLR, {Wagner} J, {De Rijcke} S (2020) {Extended
  lens reconstructions with grale: exploiting time-domain, substructural, and
  weak lensing information}. \mnras 494(3):3253--3274.
  \doi{10.1093/mnras/staa842}.
  {\href{https://arxiv.org/abs/2003.10377}{{arXiv:2003.10377}}} {[astro-ph.CO]}

\bibitem[{{Limousin} et~al.(2009){Limousin}, {Cabanac}, {Gavazzi}, {Kneib},
  {Motta}, {Richard}, {Thanjavur}, {Foex}, {Pello}, {Crampton}, {Faure},
  {Fort}, {Jullo}, {Marshall}, {Mellier}, {More}, {Soucail}, {Suyu},
  {Swinbank}, {Sygnet}, {Tu}, {Valls-Gabaud}, {Verdugo}, and
  {Willis}}]{Limousin_2009}
{Limousin} M, {Cabanac} R, {Gavazzi} R, {Kneib} JP, {Motta} V, {Richard} J,
  {Thanjavur} K, {Foex} G, {Pello} R, {Crampton} D, {Faure} C, {Fort} B,
  {Jullo} E, {Marshall} P, {Mellier} Y, {More} A, {Soucail} G, {Suyu} S,
  {Swinbank} M, {Sygnet} JF, {Tu} H, {Valls-Gabaud} D, {Verdugo} T, {Willis} J
  (2009) {A new window of exploration in the mass spectrum: strong lensing by
  galaxy groups in the SL2S}. \aap 502(2):445--456.
  \doi{10.1051/0004-6361/200811473}.
  {\href{https://arxiv.org/abs/0812.1033}{{arXiv:0812.1033}}} {[astro-ph]}

\bibitem[{{Limousin} et~al.(2022){Limousin}, {Beauchesne}, and
  {Jullo}}]{bib2:limousin22}
{Limousin} M, {Beauchesne} B, {Jullo} E (2022) {Dark matter in galaxy clusters:
  Parametric strong-lensing approach}. \aap 664:A90.
  \doi{10.1051/0004-6361/202243278}.
  {\href{https://arxiv.org/abs/2202.02992}{{arXiv:2202.02992}}} {[astro-ph.CO]}

\bibitem[{{Lin} et~al.(2022){Lin}, {Wagner}, and {Griffiths}}]{lin22}
{Lin} J, {Wagner} J, {Griffiths} RE (2022) {Generalised model-independent
  characterisation of strong gravitational lenses VIII. automated multi-band
  feature detection to constrain local lens properties}. \mnras
  517(2):1821--1836. \doi{10.1093/mnras/stac2576}

\bibitem[{Lin et~al.(2023)Lin, Wagner, and Griffiths}]{lin23}
Lin J, Wagner J, Griffiths RE (2023) {Much ado about no offset --
  characterizing the anomalous multiple-image configuration and the
  model-driven displacement between light and mass in the multiplane strong
  lens Abell 3827}. \mnras 526(2):2776--2794. \doi{10.1093/mnras/stad2800}

\bibitem[{{Livermore} et~al.(2017){Livermore}, {Finkelstein}, and
  {Lotz}}]{Livermore:2017}
{Livermore} RC, {Finkelstein} SL, {Lotz} JM (2017) {Directly Observing the
  Galaxies Likely Responsible for Reionization}. \apj 835(2):113.
  \doi{10.3847/1538-4357/835/2/113}.
  {\href{https://arxiv.org/abs/1604.06799}{{arXiv:1604.06799}}} {[astro-ph.GA]}

\bibitem[{{Lotz} et~al.(2017){Lotz}, {Koekemoer}, {Coe}, {Grogin}, {Capak},
  {Mack}, {Anderson}, {Avila}, {Barker}, {Borncamp}, {Brammer}, {Durbin},
  {Gunning}, {Hilbert}, {Jenkner}, {Khandrika}, {Levay}, {Lucas}, {MacKenty},
  {Ogaz}, {Porterfield}, {Reid}, {Robberto}, {Royle}, {Smith},
  {Storrie-Lombardi}, {Sunnquist}, {Surace}, {Taylor}, {Williams}, {Bullock},
  {Dickinson}, {Finkelstein}, {Natarajan}, {Richard}, {Robertson}, {Tumlinson},
  {Zitrin}, {Flanagan}, {Sembach}, {Soifer}, and {Mountain}}]{Lotz:2017}
{Lotz} JM, {Koekemoer} A, {Coe} D, {Grogin} N, {Capak} P, {Mack} J, {Anderson}
  J, {Avila} R, {Barker} EA, {Borncamp} D, {Brammer} G, {Durbin} M, {Gunning}
  H, {Hilbert} B, {Jenkner} H, {Khandrika} H, {Levay} Z, {Lucas} RA, {MacKenty}
  J, {Ogaz} S, {Porterfield} B, {Reid} N, {Robberto} M, {Royle} P, {Smith} LJ,
  {Storrie-Lombardi} LJ, {Sunnquist} B, {Surace} J, {Taylor} DC, {Williams} R,
  {Bullock} J, {Dickinson} M, {Finkelstein} S, {Natarajan} P, {Richard} J,
  {Robertson} B, {Tumlinson} J, {Zitrin} A, {Flanagan} K, {Sembach} K, {Soifer}
  BT, {Mountain} M (2017) {The Frontier Fields: Survey Design and Initial
  Results}. \apj 837(1):97. \doi{10.3847/1538-4357/837/1/97}.
  {\href{https://arxiv.org/abs/1605.06567}{{arXiv:1605.06567}}} {[astro-ph.GA]}

\bibitem[{{Lynds} and {Petrosian}(1989)}]{Lynds1989}
{Lynds} R, {Petrosian} V (1989) {Luminous Arcs in Clusters of Galaxies}. \apj
  336:1. \doi{10.1086/166989}

\bibitem[{Magana et~al.(2018)Magana, Acebron, Motta, Verdugo, Jullo, and
  Limousin}]{Magana+2018}
Magana J, Acebron A, Motta V, Verdugo T, Jullo E, Limousin M (2018) {Strong
  Lensing Modeling in Galaxy Clusters as a Promising Method to Test
  Cosmography. I. Parametric Dark Energy Models}. \apj 865(2):122.
  \doi{10.3847/1538-4357/aada7d}

\bibitem[{{Mahdi} et~al.(2014){Mahdi}, {van Beek}, {Elahi}, {Lewis}, {Power},
  and {Killedar}}]{mah14}
{Mahdi} HS, {van Beek} M, {Elahi} PJ, {Lewis} GF, {Power} C, {Killedar} M
  (2014) {Gravitational lensing in WDM cosmologies: the cross-section for giant
  arcs}. \mnras 441(3):1954--1963. \doi{10.1093/mnras/stu705}.
  {\href{https://arxiv.org/abs/1404.1644}{{arXiv:1404.1644}}} {[astro-ph.CO]}

\bibitem[{{Mahler} et~al.(2023){Mahler}, {Jauzac}, {Richard}, {Beauchesne},
  {Ebeling}, {Lagattuta}, {Natarajan}, {Sharon}, {Atek}, {Claeyssens},
  {Cl{\'e}ment}, {Eckert}, {Edge}, {Kneib}, and {Niemiec}}]{Mahler2022}
{Mahler} G, {Jauzac} M, {Richard} J, {Beauchesne} B, {Ebeling} H, {Lagattuta}
  D, {Natarajan} P, {Sharon} K, {Atek} H, {Claeyssens} A, {Cl{\'e}ment} B,
  {Eckert} D, {Edge} A, {Kneib} JP, {Niemiec} A (2023) {Precision Modeling of
  JWST's First Cluster Lens SMACS J0723.3-7327}. \apj 945(1):49.
  \doi{10.3847/1538-4357/acaea9}.
  {\href{https://arxiv.org/abs/2207.07101}{{arXiv:2207.07101}}} {[astro-ph.GA]}

\bibitem[{{Mantz} et~al.(2014){Mantz}, {Allen}, {Morris}, {Rapetti},
  {Applegate}, {Kelly}, {von der Linden}, and {Schmidt}}]{mantz2014}
{Mantz} AB, {Allen} SW, {Morris} RG, {Rapetti} DA, {Applegate} DE, {Kelly} PL,
  {von der Linden} A, {Schmidt} RW (2014) {Cosmology and astrophysics from
  relaxed galaxy clusters - II. Cosmological constraints}. \mnras
  440(3):2077--2098. \doi{10.1093/mnras/stu368}.
  {\href{https://arxiv.org/abs/1402.6212}{{arXiv:1402.6212}}} {[astro-ph.CO]}

\bibitem[{{Massey} et~al.(2015){Massey}, {Williams}, {Smit}, {Swinbank},
  {Kitching}, {Harvey}, {Jauzac}, {Israel}, {Clowe}, {Edge}, {Hilton}, {Jullo},
  {Leonard}, {Liesenborgs}, {Merten}, {Mohammed}, {Nagai}, {Richard},
  {Robertson}, {Saha}, {Santana}, {Stott}, and {Tittley}}]{bib2:massey2015}
{Massey} R, {Williams} L, {Smit} R, {Swinbank} M, {Kitching} TD, {Harvey} D,
  {Jauzac} M, {Israel} H, {Clowe} D, {Edge} A, {Hilton} M, {Jullo} E, {Leonard}
  A, {Liesenborgs} J, {Merten} J, {Mohammed} I, {Nagai} D, {Richard} J,
  {Robertson} A, {Saha} P, {Santana} R, {Stott} J, {Tittley} E (2015) {The
  behaviour of dark matter associated with four bright cluster galaxies in the
  10 kpc core of Abell 3827}. \mnras 449(4):3393--3406.
  \doi{10.1093/mnras/stv467}.
  {\href{https://arxiv.org/abs/1504.03388}{{arXiv:1504.03388}}} {[astro-ph.CO]}

\bibitem[{{Massey} et~al.(2018{\natexlab{a}}){Massey}, {Harvey}, {Liesenborgs},
  {Richard}, {Stach}, {Swinbank}, {Taylor}, {Williams}, {Clowe}, {Courbin},
  {Edge}, {Israel}, {Jauzac}, {Joseph}, {Jullo}, {Kitching}, {Leonard},
  {Merten}, {Nagai}, {Nightingale}, {Robertson}, {Romualdez}, {Saha}, {Smit},
  {Tam}, and {Tittley}}]{Massey+2018}
{Massey} R, {Harvey} D, {Liesenborgs} J, {Richard} J, {Stach} S, {Swinbank} M,
  {Taylor} P, {Williams} L, {Clowe} D, {Courbin} F, {Edge} A, {Israel} H,
  {Jauzac} M, {Joseph} R, {Jullo} E, {Kitching} TD, {Leonard} A, {Merten} J,
  {Nagai} D, {Nightingale} J, {Robertson} A, {Romualdez} LJ, {Saha} P, {Smit}
  R, {Tam} SI, {Tittley} E (2018{\natexlab{a}}) {Dark matter dynamics in Abell
  3827: new data consistent with standard cold dark matter}. \mnras
  477(1):669--677. \doi{10.1093/mnras/sty630}.
  {\href{https://arxiv.org/abs/1708.04245}{{arXiv:1708.04245}}} {[astro-ph.CO]}

\bibitem[{{Massey} et~al.(2018{\natexlab{b}}){Massey}, {Harvey}, {Liesenborgs},
  {Richard}, {Stach}, {Swinbank}, {Taylor}, {Williams}, {Clowe}, {Courbin},
  {Edge}, {Israel}, {Jauzac}, {Joseph}, {Jullo}, {Kitching}, {Leonard},
  {Merten}, {Nagai}, {Nightingale}, {Robertson}, {Romualdez}, {Saha}, {Smit},
  {Tam}, and {Tittley}}]{bib2:massey2018}
{Massey} R, {Harvey} D, {Liesenborgs} J, {Richard} J, {Stach} S, {Swinbank} M,
  {Taylor} P, {Williams} L, {Clowe} D, {Courbin} F, {Edge} A, {Israel} H,
  {Jauzac} M, {Joseph} R, {Jullo} E, {Kitching} TD, {Leonard} A, {Merten} J,
  {Nagai} D, {Nightingale} J, {Robertson} A, {Romualdez} LJ, {Saha} P, {Smit}
  R, {Tam} SI, {Tittley} E (2018{\natexlab{b}}) {Dark matter dynamics in Abell
  3827: new data consistent with standard cold dark matter}. \mnras
  477(1):669--677. \doi{10.1093/mnras/sty630}.
  {\href{https://arxiv.org/abs/1708.04245}{{arXiv:1708.04245}}} {[astro-ph.CO]}

\bibitem[{{McCully} et~al.(2014){McCully}, {Keeton}, {Wong}, and
  {Zabludoff}}]{McCully:14}
{McCully} C, {Keeton} CR, {Wong} KC, {Zabludoff} AI (2014) {A new hybrid
  framework to efficiently model lines of sight to gravitational lenses}.
  \mnras 443(4):3631--3642. \doi{10.1093/mnras/stu1316}.
  {\href{https://arxiv.org/abs/1401.0197}{{arXiv:1401.0197}}} {[astro-ph.CO]}

\bibitem[{{McCully} et~al.(2017){McCully}, {Keeton}, {Wong}, and
  {Zabludoff}}]{McCully2017}
{McCully} C, {Keeton} CR, {Wong} KC, {Zabludoff} AI (2017) {Quantifying
  Environmental and Line-of-sight Effects in Models of Strong Gravitational
  Lens Systems}. \apj 836(1):141. \doi{10.3847/1538-4357/836/1/141}.
  {\href{https://arxiv.org/abs/1601.05417}{{arXiv:1601.05417}}} {[astro-ph.CO]}

\bibitem[{{Medezinski} et~al.(2013){Medezinski}, {Umetsu}, {Nonino}, {Merten},
  {Zitrin}, {Broadhurst}, {Donahue}, {Sayers}, {Waizmann}, {Koekemoer}, {Coe},
  {Molino}, {Melchior}, {Mroczkowski}, {Czakon}, {Postman}, {Meneghetti},
  {Lemze}, {Ford}, {Grillo}, {Kelson}, {Bradley}, {Moustakas}, {Bartelmann},
  {Ben{\'\i}tez}, {Biviano}, {Bouwens}, {Golwala}, {Graves}, {Infante},
  {Jim{\'e}nez-Teja}, {Jouvel}, {Lahav}, {Moustakas}, {Ogaz}, {Rosati},
  {Seitz}, and {Zheng}}]{Medezinski:2013}
{Medezinski} E, {Umetsu} K, {Nonino} M, {Merten} J, {Zitrin} A, {Broadhurst} T,
  {Donahue} M, {Sayers} J, {Waizmann} JC, {Koekemoer} A, {Coe} D, {Molino} A,
  {Melchior} P, {Mroczkowski} T, {Czakon} N, {Postman} M, {Meneghetti} M,
  {Lemze} D, {Ford} H, {Grillo} C, {Kelson} D, {Bradley} L, {Moustakas} J,
  {Bartelmann} M, {Ben{\'\i}tez} N, {Biviano} A, {Bouwens} R, {Golwala} S,
  {Graves} G, {Infante} L, {Jim{\'e}nez-Teja} Y, {Jouvel} S, {Lahav} O,
  {Moustakas} L, {Ogaz} S, {Rosati} P, {Seitz} S, {Zheng} W (2013) {CLASH:
  Complete Lensing Analysis of the Largest Cosmic Lens MACS J0717.5+3745 and
  Surrounding Structures}. \apj 777(1):43. \doi{10.1088/0004-637X/777/1/43}.
  {\href{https://arxiv.org/abs/1304.1223}{{arXiv:1304.1223}}} {[astro-ph.CO]}

\bibitem[{{Meena} and {Bagla}(2021)}]{mee21}
{Meena} AK, {Bagla} JS (2021) {Exotic image formation in strong gravitational
  lensing by clusters of galaxies - I. Cross-section}. \mnras
  503(2):2097--2107. \doi{10.1093/mnras/stab577}.
  {\href{https://arxiv.org/abs/2009.13418}{{arXiv:2009.13418}}} {[astro-ph.CO]}

\bibitem[{{Menanteau} et~al.(2012){Menanteau}, {Hughes}, {Sif{\'o}n}, {Hilton},
  {Gonz{\'a}lez}, {Infante}, {Barrientos}, {Baker}, {Bond}, {Das}, {Devlin},
  {Dunkley}, {Hajian}, {Hincks}, {Kosowsky}, {Marsden}, {Marriage}, {Moodley},
  {Niemack}, {Nolta}, {Page}, {Reese}, {Sehgal}, {Sievers}, {Spergel},
  {Staggs}, and {Wollack}}]{Menanteau12}
{Menanteau} F, {Hughes} JP, {Sif{\'o}n} C, {Hilton} M, {Gonz{\'a}lez} J,
  {Infante} L, {Barrientos} LF, {Baker} AJ, {Bond} JR, {Das} S, {Devlin} MJ,
  {Dunkley} J, {Hajian} A, {Hincks} AD, {Kosowsky} A, {Marsden} D, {Marriage}
  TA, {Moodley} K, {Niemack} MD, {Nolta} MR, {Page} LA, {Reese} ED, {Sehgal} N,
  {Sievers} J, {Spergel} DN, {Staggs} ST, {Wollack} E (2012) {The Atacama
  Cosmology Telescope: ACT-CL J0102-4915 ``El Gordo,'' a Massive Merging
  Cluster at Redshift 0.87}. \apj 748(1):7. \doi{10.1088/0004-637X/748/1/7}.
  {\href{https://arxiv.org/abs/1109.0953}{{arXiv:1109.0953}}} {[astro-ph.CO]}

\bibitem[{{Meneghetti} et~al.(2013){Meneghetti}, {Bartelmann}, {Dahle}, and
  {Limousin}}]{men13}
{Meneghetti} M, {Bartelmann} M, {Dahle} H, {Limousin} M (2013) {Arc
  Statistics}. \ssr 177(1-4):31--74. \doi{10.1007/s11214-013-9981-x}.
  {\href{https://arxiv.org/abs/1303.3363}{{arXiv:1303.3363}}} {[astro-ph.CO]}

\bibitem[{{Meneghetti} et~al.(2017){Meneghetti}, {Natarajan}, {Coe}, {Contini},
  {De Lucia}, {Giocoli}, {Acebron}, {Borgani}, {Bradac}, {Diego}, {Hoag},
  {Ishigaki}, {Johnson}, {Jullo}, {Kawamata}, {Lam}, {Limousin}, {Liesenborgs},
  {Oguri}, {Sebesta}, {Sharon}, {Williams}, and {Zitrin}}]{men17}
{Meneghetti} M, {Natarajan} P, {Coe} D, {Contini} E, {De Lucia} G, {Giocoli} C,
  {Acebron} A, {Borgani} S, {Bradac} M, {Diego} JM, {Hoag} A, {Ishigaki} M,
  {Johnson} TL, {Jullo} E, {Kawamata} R, {Lam} D, {Limousin} M, {Liesenborgs}
  J, {Oguri} M, {Sebesta} K, {Sharon} K, {Williams} LLR, {Zitrin} A (2017) {The
  Frontier Fields lens modelling comparison project}. \mnras 472(3):3177--3216.
  \doi{10.1093/mnras/stx2064}.
  {\href{https://arxiv.org/abs/1606.04548}{{arXiv:1606.04548}}} {[astro-ph.CO]}

\bibitem[{{Meneghetti} et~al.(2020){Meneghetti}, {Davoli}, {Bergamini},
  {Rosati}, {Natarajan}, {Giocoli}, {Caminha}, {Metcalf}, {Rasia}, {Borgani},
  {Calura}, {Grillo}, {Mercurio}, and {Vanzella}}]{Meneghetti+2020}
{Meneghetti} M, {Davoli} G, {Bergamini} P, {Rosati} P, {Natarajan} P, {Giocoli}
  C, {Caminha} GB, {Metcalf} RB, {Rasia} E, {Borgani} S, {Calura} F, {Grillo}
  C, {Mercurio} A, {Vanzella} E (2020) {An excess of small-scale gravitational
  lenses observed in galaxy clusters}. Science 369(6509):1347--1351.
  \doi{10.1126/science.aax5164}.
  {\href{https://arxiv.org/abs/2009.04471}{{arXiv:2009.04471}}} {[astro-ph.GA]}

\bibitem[{{Meneghetti} et~al.(2022){Meneghetti}, {Ragagnin}, {Borgani},
  {Calura}, {Despali}, {Giocoli}, {Granato}, {Grillo}, {Moscardini}, {Rasia},
  {Rosati}, {Angora}, {Bassini}, {Bergamini}, {Caminha}, {Granata}, {Mercurio},
  {Metcalf}, {Natarajan}, {Nonino}, {Pignataro}, {Ragone-Figueroa}, {Vanzella},
  {Acebron}, {Dolag}, {Murante}, {Taffoni}, {Tornatore}, {Tortorelli}, and
  {Valentini}}]{Meneghetti+2022}
{Meneghetti} M, {Ragagnin} A, {Borgani} S, {Calura} F, {Despali} G, {Giocoli}
  C, {Granato} GL, {Grillo} C, {Moscardini} L, {Rasia} E, {Rosati} P, {Angora}
  G, {Bassini} L, {Bergamini} P, {Caminha} GB, {Granata} G, {Mercurio} A,
  {Metcalf} RB, {Natarajan} P, {Nonino} M, {Pignataro} GV, {Ragone-Figueroa} C,
  {Vanzella} E, {Acebron} A, {Dolag} K, {Murante} G, {Taffoni} G, {Tornatore}
  L, {Tortorelli} L, {Valentini} M (2022) {The probability of galaxy-galaxy
  strong lensing events in hydrodynamical simulations of galaxy clusters}. \aap
  668:A188. \doi{10.1051/0004-6361/202243779}.
  {\href{https://arxiv.org/abs/2204.09065}{{arXiv:2204.09065}}} {[astro-ph.CO]}

\bibitem[{{Merten}(2016)}]{merten16}
{Merten} J (2016) {Mesh-free free-form lensing - I. Methodology and application
  to mass reconstruction}. \mnras 461(3):2328--2345.
  \doi{10.1093/mnras/stw1413}.
  {\href{https://arxiv.org/abs/1412.5186}{{arXiv:1412.5186}}} {[astro-ph.CO]}

\bibitem[{{Merten} et~al.(2011){Merten}, {Coe}, {Dupke}, {Massey}, {Zitrin},
  {Cypriano}, {Okabe}, {Frye}, {Braglia}, {Jim{\'e}nez-Teja}, {Ben{\'\i}tez},
  {Broadhurst}, {Rhodes}, {Meneghetti}, {Moustakas}, {Sodr{\'e}}, {Krick}, and
  {Bregman}}]{merten11}
{Merten} J, {Coe} D, {Dupke} R, {Massey} R, {Zitrin} A, {Cypriano} ES, {Okabe}
  N, {Frye} B, {Braglia} FG, {Jim{\'e}nez-Teja} Y, {Ben{\'\i}tez} N,
  {Broadhurst} T, {Rhodes} J, {Meneghetti} M, {Moustakas} LA, {Sodr{\'e}} J L,
  {Krick} J, {Bregman} JN (2011) {Creation of cosmic structure in the complex
  galaxy cluster merger Abell 2744}. \mnras 417(1):333--347.
  \doi{10.1111/j.1365-2966.2011.19266.x}.
  {\href{https://arxiv.org/abs/1103.2772}{{arXiv:1103.2772}}} {[astro-ph.CO]}

\bibitem[{{Merten} et~al.(2015){Merten}, {Meneghetti}, {Postman}, {Umetsu},
  {Zitrin}, {Medezinski}, {Nonino}, {Koekemoer}, {Melchior}, {Gruen},
  {Moustakas}, {Bartelmann}, {Host}, {Donahue}, {Coe}, {Molino}, {Jouvel},
  {Monna}, {Seitz}, {Czakon}, {Lemze}, {Sayers}, {Balestra}, {Rosati},
  {Ben{\'\i}tez}, {Biviano}, {Bouwens}, {Bradley}, {Broadhurst}, {Carrasco},
  {Ford}, {Grillo}, {Infante}, {Kelson}, {Lahav}, {Massey}, {Moustakas},
  {Rasia}, {Rhodes}, {Vega}, and {Zheng}}]{Merten:2015}
{Merten} J, {Meneghetti} M, {Postman} M, {Umetsu} K, {Zitrin} A, {Medezinski}
  E, {Nonino} M, {Koekemoer} A, {Melchior} P, {Gruen} D, {Moustakas} LA,
  {Bartelmann} M, {Host} O, {Donahue} M, {Coe} D, {Molino} A, {Jouvel} S,
  {Monna} A, {Seitz} S, {Czakon} N, {Lemze} D, {Sayers} J, {Balestra} I,
  {Rosati} P, {Ben{\'\i}tez} N, {Biviano} A, {Bouwens} R, {Bradley} L,
  {Broadhurst} T, {Carrasco} M, {Ford} H, {Grillo} C, {Infante} L, {Kelson} D,
  {Lahav} O, {Massey} R, {Moustakas} J, {Rasia} E, {Rhodes} J, {Vega} J,
  {Zheng} W (2015) {CLASH: The Concentration-Mass Relation of Galaxy Clusters}.
  \apj 806(1):4. \doi{10.1088/0004-637X/806/1/4}.
  {\href{https://arxiv.org/abs/1404.1376}{{arXiv:1404.1376}}} {[astro-ph.CO]}

\bibitem[{{Miralda-Escude}(1991)}]{MiraldaEscude1991}
{Miralda-Escude} J (1991) {Gravitational Lensing by Clusters of Galaxies:
  Constraining the Mass Distribution}. \apj 370:1. \doi{10.1086/169789}

\bibitem[{{Miralda-Escud{\'e}}(1993)}]{mir93b}
{Miralda-Escud{\'e}} J (1993) {Statistics of Highly Magnified Gravitational
  Images in Clusters of Galaxies. II. Implications for the Sources}. \apj
  403:509. \doi{10.1086/172221}

\bibitem[{{Miralda-Escud{\'e}}(2002)}]{mir02}
{Miralda-Escud{\'e}} J (2002) {A Test of the Collisional Dark Matter Hypothesis
  from Cluster Lensing}. \apj 564(1):60--64. \doi{10.1086/324138}.
  {\href{https://arxiv.org/abs/astro-ph/0002050}{{arXiv:astro-ph/0002050}}}
  {[astro-ph]}

\bibitem[{{Mohammed} et~al.(2015){Mohammed}, {Saha}, and {Liesenborgs}}]{moh15}
{Mohammed} I, {Saha} P, {Liesenborgs} J (2015) {Lensing time delays as a
  substructure constraint: a case study with the cluster SDSS J1004+4112}.
  \pasj 67(2):21. \doi{10.1093/pasj/psu155}.
  {\href{https://arxiv.org/abs/1412.3464}{{arXiv:1412.3464}}} {[astro-ph.CO]}

\bibitem[{{Mohammed} et~al.(2016){Mohammed}, {Saha}, {Williams}, {Liesenborgs},
  and {Sebesta}}]{moh16}
{Mohammed} I, {Saha} P, {Williams} LLR, {Liesenborgs} J, {Sebesta} K (2016)
  {Quantifying substructures in Hubble Frontier Field clusters: comparison with
  {\ensuremath{\Lambda}}CDM simulations}. \mnras 459(2):1698--1709.
  \doi{10.1093/mnras/stw727}.
  {\href{https://arxiv.org/abs/1507.01532}{{arXiv:1507.01532}}} {[astro-ph.CO]}

\bibitem[{{Monna} et~al.(2014){Monna}, {Seitz}, {Greisel}, {Eichner}, {Drory},
  {Postman}, {Zitrin}, {Coe}, {Halkola}, {Suyu}, {Grillo}, {Rosati}, {Lemze},
  {Balestra}, {Snigula}, {Bradley}, {Umetsu}, {Koekemoer}, {Kuchner},
  {Moustakas}, {Bartelmann}, {Ben{\'\i}tez}, {Bouwens}, {Broadhurst},
  {Donahue}, {Ford}, {Host}, {Infante}, {Jimenez-Teja}, {Jouvel}, {Kelson},
  {Lahav}, {Medezinski}, {Melchior}, {Meneghetti}, {Merten}, {Molino},
  {Moustakas}, {Nonino}, and {Zheng}}]{Monna:2014}
{Monna} A, {Seitz} S, {Greisel} N, {Eichner} T, {Drory} N, {Postman} M,
  {Zitrin} A, {Coe} D, {Halkola} A, {Suyu} SH, {Grillo} C, {Rosati} P, {Lemze}
  D, {Balestra} I, {Snigula} J, {Bradley} L, {Umetsu} K, {Koekemoer} A,
  {Kuchner} U, {Moustakas} L, {Bartelmann} M, {Ben{\'\i}tez} N, {Bouwens} R,
  {Broadhurst} T, {Donahue} M, {Ford} H, {Host} O, {Infante} L, {Jimenez-Teja}
  Y, {Jouvel} S, {Kelson} D, {Lahav} O, {Medezinski} E, {Melchior} P,
  {Meneghetti} M, {Merten} J, {Molino} A, {Moustakas} J, {Nonino} M, {Zheng} W
  (2014) {CLASH: z {\ensuremath{\sim}} 6 young galaxy candidate quintuply
  lensed by the frontier field cluster RXC J2248.7-4431}. \mnras
  438(2):1417--1434. \doi{10.1093/mnras/stt2284}.
  {\href{https://arxiv.org/abs/1308.6280}{{arXiv:1308.6280}}} {[astro-ph.CO]}

\bibitem[{{Montes} and {Trujillo}(2019)}]{mon19}
{Montes} M, {Trujillo} I (2019) {Intracluster light: a luminous tracer for dark
  matter in clusters of galaxies}. \mnras 482(2):2838--2851.
  \doi{10.1093/mnras/sty2858}.
  {\href{https://arxiv.org/abs/1807.11488}{{arXiv:1807.11488}}} {[astro-ph.GA]}

\bibitem[{More et~al.(2012)More, Cabanac, More, Alard, Limousin, Kneib,
  Gavazzi, and Motta}]{More_2012}
More A, Cabanac R, More S, Alard C, Limousin M, Kneib JP, Gavazzi R, Motta V
  (2012) {The CFHTLS-Strong Lensing Legacy Survey (SL2S): Investigating the
  group-scale lenses with the SARCS sample}. \apj 749(1):38.
  \doi{10.1088/0004-637x/749/1/38}

\bibitem[{{Morioka} and {Futamase}(2015)}]{mor15}
{Morioka} M, {Futamase} T (2015) {Lens Statistics with Gravitationally Lensed
  yet Morphologically Regular Images}. \apj 805(2):184.
  \doi{10.1088/0004-637X/805/2/184}

\bibitem[{{Mu{\~n}oz} et~al.(2022){Mu{\~n}oz}, {Kochanek}, {Fohlmeister},
  {Wambsganss}, {Falco}, and {For{\'e}s-Toribio}}]{mun22}
{Mu{\~n}oz} JA, {Kochanek} CS, {Fohlmeister} J, {Wambsganss} J, {Falco} E,
  {For{\'e}s-Toribio} R (2022) {The Longest Delay: A 14.5 yr Campaign to
  Determine the Third Time Delay in the Lensing Cluster SDSS J1004+4112}. \apj
  937(1):34. \doi{10.3847/1538-4357/ac8877}.
  {\href{https://arxiv.org/abs/2206.08597}{{arXiv:2206.08597}}} {[astro-ph.GA]}

\bibitem[{{Napier} et~al.(2023){Napier}, {Sharon}, {Dahle}, {Bayliss},
  {Gladders}, {Mahler}, {Rigby}, and {Florian}}]{nap23}
{Napier} K, {Sharon} K, {Dahle} H, {Bayliss} M, {Gladders} MD, {Mahler} G,
  {Rigby} JR, {Florian} M (2023) {Hubble Constant Measurement from Three
  Large-separation Quasars Strongly Lensed by Galaxy Clusters}. \apj
  959(2):134. \doi{10.3847/1538-4357/ad045a}.
  {\href{https://arxiv.org/abs/2301.11240}{{arXiv:2301.11240}}} {[astro-ph.CO]}

\bibitem[{{Narasimha} and {Chitre}(1989)}]{nar89}
{Narasimha} D, {Chitre} SM (1989) {Lensing of Extended Sources by Dark Galactic
  Halos}. \aj 97:327. \doi{10.1086/114983}

\bibitem[{{Narayan} et~al.(1984){Narayan}, {Blandford}, and
  {Nityananda}}]{Narayan1984}
{Narayan} R, {Blandford} R, {Nityananda} R (1984) {Multiple imaging of quasars
  by galaxies and clusters}. \nat 310(5973):112--115. \doi{10.1038/310112a0}

\bibitem[{{Natarajan} and {Kneib}(1996)}]{natarajan+1996}
{Natarajan} P, {Kneib} JP (1996) {Probing the dynamics of cluster-lenses}.
  \mnras 283(3):1031--1046. \doi{10.1093/mnras/283.3.1031}.
  {\href{https://arxiv.org/abs/astro-ph/9602035}{{arXiv:astro-ph/9602035}}}
  {[astro-ph]}

\bibitem[{{Natarajan} and {Kneib}(1997)}]{Natarajan+1997}
{Natarajan} P, {Kneib} JP (1997) {Lensing by galaxy haloes in clusters of
  galaxies}. \mnras 287(4):833--847. \doi{10.1093/mnras/287.4.833}.
  {\href{https://arxiv.org/abs/astro-ph/9609008}{{arXiv:astro-ph/9609008}}}
  {[astro-ph]}

\bibitem[{{Natarajan} and {Springel}(2004)}]{Natarajan+2004}
{Natarajan} P, {Springel} V (2004) {Abundance of Substructure in Clusters of
  Galaxies}. \apjl 617(1):L13--L16. \doi{10.1086/427079}.
  {\href{https://arxiv.org/abs/astro-ph/0411515}{{arXiv:astro-ph/0411515}}}
  {[astro-ph]}

\bibitem[{{Natarajan} et~al.(1998){Natarajan}, {Kneib}, {Smail}, and
  {Ellis}}]{Natarajan+1998}
{Natarajan} P, {Kneib} JP, {Smail} I, {Ellis} RS (1998) {The Mass-to-Light
  Ratio of Early-Type Galaxies: Constraints from Gravitational Lensing in the
  Rich Cluster AC 114}. \apj 499(2):600--607. \doi{10.1086/305660}.
  {\href{https://arxiv.org/abs/astro-ph/9706129}{{arXiv:astro-ph/9706129}}}
  {[astro-ph]}

\bibitem[{{Natarajan} et~al.(2002){Natarajan}, {Kneib}, and
  {Smail}}]{PNtidal2002}
{Natarajan} P, {Kneib} JP, {Smail} I (2002) {Evidence for Tidal Stripping of
  Dark Matter Halos in Massive Cluster Lenses}. \apjl 580(1):L11--L15.
  \doi{10.1086/345399}.
  {\href{https://arxiv.org/abs/astro-ph/0207049}{{arXiv:astro-ph/0207049}}}
  {[astro-ph]}

\bibitem[{Natarajan et~al.(2002)Natarajan, Loeb, Kneib, and
  Smail}]{Natarajan_2002}
Natarajan P, Loeb A, Kneib JP, Smail I (2002) {Constraints on the Collisional
  Nature of the Dark Matter from Gravitational Lensing in the Cluster A2218}.
  \apj 580(1):L17. \doi{10.1086/345547}

\bibitem[{{Natarajan} et~al.(2007){Natarajan}, {De Lucia}, and
  {Springel}}]{Natarajan:2007}
{Natarajan} P, {De Lucia} G, {Springel} V (2007) {Substructure in lensing
  clusters and simulations}. \mnras 376(17):180--192.
  \doi{10.1111/j.1365-2966.2007.11399.x}.
  {\href{https://arxiv.org/abs/astro-ph/0604414}{{arXiv:astro-ph/0604414}}}
  {[astro-ph]}

\bibitem[{{Natarajan} et~al.(2009){Natarajan}, {Kneib}, {Smail}, {Treu},
  {Ellis}, {Moran}, {Limousin}, and {Czoske}}]{Natarajan:2009}
{Natarajan} P, {Kneib} JP, {Smail} I, {Treu} T, {Ellis} R, {Moran} S,
  {Limousin} M, {Czoske} O (2009) {The Survival of Dark Matter Halos in the
  Cluster Cl 0024+16}. \apj 693(1):970--983. \doi{10.1088/0004-637X/693/1/970}.
  {\href{https://arxiv.org/abs/0711.4587}{{arXiv:0711.4587}}} {[astro-ph]}

\bibitem[{{Natarajan} et~al.(2017){Natarajan}, {Chadayammuri}, {Jauzac},
  {Richard}, {Kneib}, {Ebeling}, {Jiang}, {van den Bosch}, {Limousin}, {Jullo},
  {Atek}, {Pillepich}, {Popa}, {Marinacci}, {Hernquist}, {Meneghetti}, and
  {Vogelsberger}}]{Natarajan:2017}
{Natarajan} P, {Chadayammuri} U, {Jauzac} M, {Richard} J, {Kneib} JP, {Ebeling}
  H, {Jiang} F, {van den Bosch} F, {Limousin} M, {Jullo} E, {Atek} H,
  {Pillepich} A, {Popa} C, {Marinacci} F, {Hernquist} L, {Meneghetti} M,
  {Vogelsberger} M (2017) {Mapping substructure in the HST Frontier Fields
  cluster lenses and in cosmological simulations}. \mnras 468(2):1962--1980.
  \doi{10.1093/mnras/stw3385}.
  {\href{https://arxiv.org/abs/1702.04348}{{arXiv:1702.04348}}} {[astro-ph.GA]}

\bibitem[{{Navarro} et~al.(1996){Navarro}, {Frenk}, and {White}}]{Navarro:1996}
{Navarro} JF, {Frenk} CS, {White} SDM (1996) {The Structure of Cold Dark Matter
  Halos}. \apj 462:563. \doi{10.1086/177173}.
  {\href{https://arxiv.org/abs/astro-ph/9508025}{{arXiv:astro-ph/9508025}}}
  {[astro-ph]}

\bibitem[{{Navarro} et~al.(1997){Navarro}, {Frenk}, and {White}}]{Navarro:1997}
{Navarro} JF, {Frenk} CS, {White} SDM (1997) {A Universal Density Profile from
  Hierarchical Clustering}. \apj 490(2):493--508. \doi{10.1086/304888}.
  {\href{https://arxiv.org/abs/astro-ph/9611107}{{arXiv:astro-ph/9611107}}}
  {[astro-ph]}

\bibitem[{{Navarro} et~al.(2004){Navarro}, {Hayashi}, {Power}, {Jenkins},
  {Frenk}, {White}, {Springel}, {Stadel}, and {Quinn}}]{nav04}
{Navarro} JF, {Hayashi} E, {Power} C, {Jenkins} AR, {Frenk} CS, {White} SDM,
  {Springel} V, {Stadel} J, {Quinn} TR (2004) {The inner structure of
  {\ensuremath{\Lambda}}CDM haloes - III. Universality and asymptotic slopes}.
  \mnras 349(3):1039--1051. \doi{10.1111/j.1365-2966.2004.07586.x}.
  {\href{https://arxiv.org/abs/astro-ph/0311231}{{arXiv:astro-ph/0311231}}}
  {[astro-ph]}

\bibitem[{{Newman} et~al.(2013){Newman}, {Treu}, {Ellis}, and
  {Sand}}]{newman2013b}
{Newman} AB, {Treu} T, {Ellis} RS, {Sand} DJ (2013) {The Density Profiles of
  Massive, Relaxed Galaxy Clusters. II. Separating Luminous and Dark Matter in
  Cluster Cores}. \apj 765(1):25. \doi{10.1088/0004-637X/765/1/25}.
  {\href{https://arxiv.org/abs/1209.1392}{{arXiv:1209.1392}}} {[astro-ph.CO]}

\bibitem[{{Newman} et~al.(2015{\natexlab{a}}){Newman}, {Ellis}, and
  {Treu}}]{newman2015}
{Newman} AB, {Ellis} RS, {Treu} T (2015{\natexlab{a}}) {Luminous and Dark
  Matter Profiles from Galaxies to Clusters: Bridging the Gap with Group-scale
  Lenses}. \apj 814(1):26. \doi{10.1088/0004-637X/814/1/26}.
  {\href{https://arxiv.org/abs/1503.05282}{{arXiv:1503.05282}}} {[astro-ph.GA]}

\bibitem[{{Newman} et~al.(2015{\natexlab{b}}){Newman}, {Ellis}, and
  {Treu}}]{Newman_2015}
{Newman} AB, {Ellis} RS, {Treu} T (2015{\natexlab{b}}) {Luminous and Dark
  Matter Profiles from Galaxies to Clusters: Bridging the Gap with Group-scale
  Lenses}. \apj 814(1):26. \doi{10.1088/0004-637X/814/1/26}.
  {\href{https://arxiv.org/abs/1503.05282}{{arXiv:1503.05282}}} {[astro-ph.GA]}

\bibitem[{{Niemiec} et~al.(2020){Niemiec}, {Jauzac}, {Jullo}, {Limousin},
  {Sharon}, {Kneib}, {Natarajan}, and {Richard}}]{Niemiec+2020}
{Niemiec} A, {Jauzac} M, {Jullo} E, {Limousin} M, {Sharon} K, {Kneib} JP,
  {Natarajan} P, {Richard} J (2020) {hybrid-LENSTOOL: a self-consistent
  algorithm to model galaxy clusters with strong- and weak-lensing
  simultaneously}. \mnras 493(3):3331--3340. \doi{10.1093/mnras/staa473}.
  {\href{https://arxiv.org/abs/2002.04635}{{arXiv:2002.04635}}} {[astro-ph.CO]}

\bibitem[{{Ogrean} et~al.(2015){Ogrean}, {van Weeren}, {Jones}, {Clarke},
  {Sayers}, {Mroczkowski}, {Nulsen}, {Forman}, {Murray}, {Pandey-Pommier},
  {Randall}, {Churazov}, {Bonafede}, {Kraft}, {David}, {Andrade-Santos},
  {Merten}, {Zitrin}, {Umetsu}, {Goulding}, {Roediger}, {Bagchi}, {Bulbul},
  {Donahue}, {Ebeling}, {Johnston-Hollitt}, {Mason}, {Rosati}, and
  {Vikhlinin}}]{Ogrean:2015}
{Ogrean} GA, {van Weeren} RJ, {Jones} C, {Clarke} TE, {Sayers} J, {Mroczkowski}
  T, {Nulsen} PEJ, {Forman} W, {Murray} SS, {Pandey-Pommier} M, {Randall} S,
  {Churazov} E, {Bonafede} A, {Kraft} R, {David} L, {Andrade-Santos} F,
  {Merten} J, {Zitrin} A, {Umetsu} K, {Goulding} A, {Roediger} E, {Bagchi} J,
  {Bulbul} E, {Donahue} M, {Ebeling} H, {Johnston-Hollitt} M, {Mason} B,
  {Rosati} P, {Vikhlinin} A (2015) {Frontier Fields Clusters: Chandra and JVLA
  View of the Pre-merging Cluster MACS J0416.1-2403}. \apj 812(2):153.
  \doi{10.1088/0004-637X/812/2/153}.
  {\href{https://arxiv.org/abs/1505.05560}{{arXiv:1505.05560}}} {[astro-ph.CO]}

\bibitem[{{Oguri}(2002)}]{ogu02}
{Oguri} M (2002) {Systematic Effects on Tangential and Radial Arc Statistics:
  The Finite Source Size and Ellipticities of the Lens and Source}. \apj
  573(1):51--59. \doi{10.1086/340594}.
  {\href{https://arxiv.org/abs/astro-ph/0203142}{{arXiv:astro-ph/0203142}}}
  {[astro-ph]}

\bibitem[{{Oguri}(2010)}]{oguri10}
{Oguri} M (2010) {The Mass Distribution of SDSS J1004+4112 Revisited}. \pasj
  62:1017. \doi{10.1093/pasj/62.4.1017}.
  {\href{https://arxiv.org/abs/1005.3103}{{arXiv:1005.3103}}} {[astro-ph.CO]}

\bibitem[{{Oguri}(2015)}]{Oguri2015}
{Oguri} M (2015) {Predicted properties of multiple images of the strongly
  lensed supernova SN Refsdal.} \mnras 449:L86--L89.
  \doi{10.1093/mnrasl/slv025}.
  {\href{https://arxiv.org/abs/1411.6443}{{arXiv:1411.6443}}} {[astro-ph.CO]}

\bibitem[{{Okura} et~al.(2008){Okura}, {Umetsu}, and {Futamase}}]{okura08}
{Okura} Y, {Umetsu} K, {Futamase} T (2008) {A Method for Weak-Lensing Flexion
  Analysis by the HOLICs Moment Approach}. \apj 680(1):1--16.
  \doi{10.1086/587676}.
  {\href{https://arxiv.org/abs/0710.2262}{{arXiv:0710.2262}}} {[astro-ph]}

\bibitem[{{Orban de Xivry} and {Marshall}(2009)}]{orb09}
{Orban de Xivry} G, {Marshall} P (2009) {An atlas of predicted exotic
  gravitational lenses}. \mnras 399(1):2--20.
  \doi{10.1111/j.1365-2966.2009.14925.x}.
  {\href{https://arxiv.org/abs/0904.1454}{{arXiv:0904.1454}}} {[astro-ph.CO]}

\bibitem[{{Paczynski}(1987)}]{Paczynski1987}
{Paczynski} B (1987) {Giant luminous arcs discovered in two clusters of
  galaxies}. \nat 325(6105):572--573. \doi{10.1038/325572a0}

\bibitem[{{Peter} et~al.(2013){Peter}, {Rocha}, {Bullock}, and
  {Kaplinghat}}]{pet13}
{Peter} AHG, {Rocha} M, {Bullock} JS, {Kaplinghat} M (2013) {Cosmological
  simulations with self-interacting dark matter - II. Halo shapes versus
  observations}. \mnras 430(1):105--120. \doi{10.1093/mnras/sts535}.
  {\href{https://arxiv.org/abs/1208.3026}{{arXiv:1208.3026}}} {[astro-ph.CO]}

\bibitem[{{Plazas Malag{\'o}n}(2020)}]{pla20}
{Plazas Malag{\'o}n} AA (2020) {Image Simulations for Strong and Weak
  Gravitational Lensing}. Symmetry 12(4):494. \doi{10.3390/sym12040494}.
  {\href{https://arxiv.org/abs/2003.06090}{{arXiv:2003.06090}}} {[astro-ph.CO]}

\bibitem[{{Postman} et~al.(2012){Postman}, {Coe}, {Ben{\'\i}tez}
  et~al.}]{Postman:2012}
{Postman} M, {Coe} D, {Ben{\'\i}tez} N, et~al. (2012) {The Cluster Lensing and
  Supernova Survey with Hubble: An Overview}. \apjs 199(2):25.
  \doi{10.1088/0067-0049/199/2/25}.
  {\href{https://arxiv.org/abs/1106.3328}{{arXiv:1106.3328}}} {[astro-ph.CO]}

\bibitem[{{Presotto} et~al.(2014){Presotto}, {Girardi}, {Nonino}, {Mercurio},
  {Grillo}, {Rosati}, {Biviano}, {Annunziatella}, {Balestra}, {Cui},
  {Sartoris}, {Lemze}, {Ascaso}, {Moustakas}, {Ford}, {Fritz}, {Czoske},
  {Ettori}, {Kuchner}, {Lombardi}, {Maier}, {Medezinski}, {Molino},
  {Scodeggio}, {Strazzullo}, {Tozzi}, {Ziegler}, {Bartelmann}, {Benitez},
  {Bradley}, {Brescia}, {Broadhurst}, {Coe}, {Donahue}, {Gobat}, {Graves},
  {Kelson}, {Koekemoer}, {Melchior}, {Meneghetti}, {Merten}, {Moustakas},
  {Munari}, {Postman}, {Reg{\H{o}}s}, {Seitz}, {Umetsu}, {Zheng}, and
  {Zitrin}}]{Presotto:2014}
{Presotto} V, {Girardi} M, {Nonino} M, {Mercurio} A, {Grillo} C, {Rosati} P,
  {Biviano} A, {Annunziatella} M, {Balestra} I, {Cui} W, {Sartoris} B, {Lemze}
  D, {Ascaso} B, {Moustakas} J, {Ford} H, {Fritz} A, {Czoske} O, {Ettori} S,
  {Kuchner} U, {Lombardi} M, {Maier} C, {Medezinski} E, {Molino} A, {Scodeggio}
  M, {Strazzullo} V, {Tozzi} P, {Ziegler} B, {Bartelmann} M, {Benitez} N,
  {Bradley} L, {Brescia} M, {Broadhurst} T, {Coe} D, {Donahue} M, {Gobat} R,
  {Graves} G, {Kelson} D, {Koekemoer} A, {Melchior} P, {Meneghetti} M, {Merten}
  J, {Moustakas} LA, {Munari} E, {Postman} M, {Reg{\H{o}}s} E, {Seitz} S,
  {Umetsu} K, {Zheng} W, {Zitrin} A (2014) {Intracluster light properties in
  the CLASH-VLT cluster MACS J1206.2-0847}. \aap 565:A126.
  \doi{10.1051/0004-6361/201323251}.
  {\href{https://arxiv.org/abs/1403.4979}{{arXiv:1403.4979}}} {[astro-ph.CO]}

\bibitem[{{Priewe} et~al.(2017){Priewe}, {Williams}, {Liesenborgs}, {Coe}, and
  {Rodney}}]{pri17}
{Priewe} J, {Williams} LLR, {Liesenborgs} J, {Coe} D, {Rodney} SA (2017) {Lens
  models under the microscope: comparison of Hubble Frontier Field cluster
  magnification maps}. \mnras 465(1):1030--1045. \doi{10.1093/mnras/stw2785}.
  {\href{https://arxiv.org/abs/1605.07621}{{arXiv:1605.07621}}} {[astro-ph.CO]}

\bibitem[{{Puchwein} and {Hilbert}(2009)}]{puc09}
{Puchwein} E, {Hilbert} S (2009) {Cluster strong lensing in the Millennium
  simulation: the effect of galaxies and structures along the line-of-sight}.
  \mnras 398(3):1298--1308. \doi{10.1111/j.1365-2966.2009.15227.x}.
  {\href{https://arxiv.org/abs/0904.0253}{{arXiv:0904.0253}}} {[astro-ph.CO]}

\bibitem[{{Ragagnin} et~al.(2022){Ragagnin}, {Meneghetti}, {Bassini},
  {Ragone-Figueroa}, {Granato}, {Despali}, {Giocoli}, {Granata}, {Moscardini},
  {Bergamini}, {Rasia}, {Valentini}, {Borgani}, {Calura}, {Dolag}, {Grillo},
  {Mercurio}, {Murante}, {Natarajan}, {Rosati}, {Taffoni}, {Tornatore}, and
  {Tortorelli}}]{Ragagnin+2022}
{Ragagnin} A, {Meneghetti} M, {Bassini} L, {Ragone-Figueroa} C, {Granato} GL,
  {Despali} G, {Giocoli} C, {Granata} G, {Moscardini} L, {Bergamini} P, {Rasia}
  E, {Valentini} M, {Borgani} S, {Calura} F, {Dolag} K, {Grillo} C, {Mercurio}
  A, {Murante} G, {Natarajan} P, {Rosati} P, {Taffoni} G, {Tornatore} L,
  {Tortorelli} L (2022) {Galaxies in the central regions of simulated galaxy
  clusters}. \aap 665:A16. \doi{10.1051/0004-6361/202243651}.
  {\href{https://arxiv.org/abs/2204.09067}{{arXiv:2204.09067}}} {[astro-ph.CO]}

\bibitem[{{Raghunathan} et~al.(2019){Raghunathan}, {Patil}, {Baxter}
  et~al.}]{raghu19}
{Raghunathan} S, {Patil} S, {Baxter} E, et~al. (2019) {Mass Calibration of
  Optically Selected DES Clusters Using a Measurement of CMB-cluster Lensing
  with SPTpol Data}. \apj 872(2):170. \doi{10.3847/1538-4357/ab01ca}.
  {\href{https://arxiv.org/abs/1810.10998}{{arXiv:1810.10998}}} {[astro-ph.CO]}

\bibitem[{{Randall} et~al.(2008){Randall}, {Markevitch}, {Clowe}, {Gonzalez},
  and {Brada{\v{c}}}}]{randall08}
{Randall} SW, {Markevitch} M, {Clowe} D, {Gonzalez} AH, {Brada{\v{c}}} M (2008)
  {Constraints on the Self-Interaction Cross Section of Dark Matter from
  Numerical Simulations of the Merging Galaxy Cluster 1E 0657-56}. \apj
  679(2):1173--1180. \doi{10.1086/587859}.
  {\href{https://arxiv.org/abs/0704.0261}{{arXiv:0704.0261}}} {[astro-ph]}

\bibitem[{{Raney} et~al.(2020{\natexlab{a}}){Raney}, {Keeton}, and
  {Brennan}}]{Raney:2020}
{Raney} CA, {Keeton} CR, {Brennan} S (2020{\natexlab{a}}) {Exploring Effects on
  Magnifications due to Line-of-Sight Galaxies in the Hubble Frontier Fields}.
  \mnras 492(1):503--527. \doi{10.1093/mnras/stz3116}.
  {\href{https://arxiv.org/abs/1911.02101}{{arXiv:1911.02101}}} {[astro-ph.CO]}

\bibitem[{{Raney} et~al.(2020{\natexlab{b}}){Raney}, {Keeton}, {Brennan}, and
  {Fan}}]{ran20}
{Raney} CA, {Keeton} CR, {Brennan} S, {Fan} H (2020{\natexlab{b}}) {Systematic
  versus statistical uncertainties in masses and magnifications of the Hubble
  Frontier Fields}. \mnras 494(4):4771--4793. \doi{10.1093/mnras/staa921}.
  {\href{https://arxiv.org/abs/2004.05952}{{arXiv:2004.05952}}} {[astro-ph.CO]}

\bibitem[{{Remolina Gonz{\'a}lez} et~al.(2018){Remolina Gonz{\'a}lez},
  {Sharon}, and {Mahler}}]{rem18}
{Remolina Gonz{\'a}lez} JD, {Sharon} K, {Mahler} G (2018) {An Evaluation of 10
  Lensing Models of the Frontier Fields Cluster MACS J0416.1-2403}. \apj
  863(1):60. \doi{10.3847/1538-4357/aacf8e}.
  {\href{https://arxiv.org/abs/1807.03291}{{arXiv:1807.03291}}} {[astro-ph.CO]}

\bibitem[{{Richard} et~al.(2010){Richard}, {Smith}, {Kneib}, {Ellis},
  {Sanderson}, {Pei}, {Targett}, {Sand}, {Swinbank}, {Dannerbauer}, {Mazzotta},
  {Limousin}, {Egami}, {Jullo}, {Hamilton-Morris}, and {Moran}}]{richard2010}
{Richard} J, {Smith} GP, {Kneib} JP, {Ellis} RS, {Sanderson} AJR, {Pei} L,
  {Targett} TA, {Sand} DJ, {Swinbank} AM, {Dannerbauer} H, {Mazzotta} P,
  {Limousin} M, {Egami} E, {Jullo} E, {Hamilton-Morris} V, {Moran} SM (2010)
  {LoCuSS: first results from strong-lensing analysis of 20 massive galaxy
  clusters at z = 0.2}. \mnras 404(1):325--349.
  \doi{10.1111/j.1365-2966.2009.16274.x}.
  {\href{https://arxiv.org/abs/0911.3302}{{arXiv:0911.3302}}} {[astro-ph.CO]}

\bibitem[{{Richard} et~al.(2015){Richard}, {Patricio}, {Martinez}, {Bacon},
  {Clement}, {Weilbacher}, {Soto}, {Wisotzki}, {Vernet}, {Pello}, {Schaye},
  {Turner}, and {Martinsson}}]{Richard:2015}
{Richard} J, {Patricio} V, {Martinez} J, {Bacon} R, {Clement} B, {Weilbacher}
  P, {Soto} K, {Wisotzki} L, {Vernet} J, {Pello} R, {Schaye} J, {Turner} M,
  {Martinsson} T (2015) {MUSE observations of the lensing cluster
  SMACSJ2031.8-4036: new constraints on the mass distribution in the cluster
  core.} \mnras 446:L16--L20. \doi{10.1093/mnrasl/slu150}.
  {\href{https://arxiv.org/abs/1409.2488}{{arXiv:1409.2488}}} {[astro-ph.CO]}

\bibitem[{{Rigby} et~al.(2018){Rigby}, {Bayliss}, {Sharon}, {Gladders},
  {Chisholm}, {Dahle}, {Johnson}, {Paterno-Mahler}, {Wuyts}, and
  {Kelson}}]{rigby18a}
{Rigby} JR, {Bayliss} MB, {Sharon} K, {Gladders} MD, {Chisholm} J, {Dahle} H,
  {Johnson} T, {Paterno-Mahler} R, {Wuyts} E, {Kelson} DD (2018) {The Magellan
  Evolution of Galaxies Spectroscopic and Ultraviolet Reference Atlas
  (MegaSaura). I. The Sample and the Spectra}. \aj 155(3):104.
  \doi{10.3847/1538-3881/aaa2ff}.
  {\href{https://arxiv.org/abs/1710.07294}{{arXiv:1710.07294}}} {[astro-ph.GA]}

\bibitem[{{Robertson}(2021)}]{rob21}
{Robertson} A (2021) {The galaxy-galaxy strong lensing cross-sections of
  simulated {\ensuremath{\Lambda}}CDM galaxy clusters}. \mnras 504(1):L7--L11.
  \doi{10.1093/mnrasl/slab028}.
  {\href{https://arxiv.org/abs/2101.12067}{{arXiv:2101.12067}}} {[astro-ph.GA]}

\bibitem[{{Robertson} et~al.(2017){Robertson}, {Massey}, and
  {Eke}}]{bib2:robertson2017}
{Robertson} A, {Massey} R, {Eke} V (2017) {What does the Bullet Cluster tell us
  about self-interacting dark matter?} \mnras 465(1):569--587.
  \doi{10.1093/mnras/stw2670}.
  {\href{https://arxiv.org/abs/1605.04307}{{arXiv:1605.04307}}} {[astro-ph.CO]}

\bibitem[{{Robertson} et~al.(2019){Robertson}, {Harvey}, {Massey}, {Eke},
  {McCarthy}, {Jauzac}, {Li}, and {Schaye}}]{rob19}
{Robertson} A, {Harvey} D, {Massey} R, {Eke} V, {McCarthy} IG, {Jauzac} M, {Li}
  B, {Schaye} J (2019) {Observable tests of self-interacting dark matter in
  galaxy clusters: cosmological simulations with SIDM and baryons}. \mnras
  488(3):3646--3662. \doi{10.1093/mnras/stz1815}.
  {\href{https://arxiv.org/abs/1810.05649}{{arXiv:1810.05649}}} {[astro-ph.CO]}

\bibitem[{{Robertson} et~al.(2020){Robertson}, {Smith}, {Massey}, {Eke},
  {Jauzac}, {Bianconi}, and {Ryczanowski}}]{robertson2020}
{Robertson} A, {Smith} GP, {Massey} R, {Eke} V, {Jauzac} M, {Bianconi} M,
  {Ryczanowski} D (2020) {What does strong gravitational lensing? The mass and
  redshift distribution of high-magnification lenses}. \mnras
  495(4):3727--3739. \doi{10.1093/mnras/staa1429}.
  {\href{https://arxiv.org/abs/2002.01479}{{arXiv:2002.01479}}} {[astro-ph.CO]}

\bibitem[{{Rodney} et~al.(2015){Rodney}, {Patel}, {Scolnic}, {Foley}, {Molino},
  {Brammer}, {Jauzac}, {Brada{\v{c}}}, {Broadhurst}, {Coe}, {Diego}, {Graur},
  {Hjorth}, {Hoag}, {Jha}, {Johnson}, {Kelly}, {Lam}, {McCully}, {Medezinski},
  {Meneghetti}, {Merten}, {Richard}, {Riess}, {Sharon}, {Strolger}, {Treu},
  {Wang}, {Williams}, and {Zitrin}}]{rod15}
{Rodney} SA, {Patel} B, {Scolnic} D, {Foley} RJ, {Molino} A, {Brammer} G,
  {Jauzac} M, {Brada{\v{c}}} M, {Broadhurst} T, {Coe} D, {Diego} JM, {Graur} O,
  {Hjorth} J, {Hoag} A, {Jha} SW, {Johnson} TL, {Kelly} P, {Lam} D, {McCully}
  C, {Medezinski} E, {Meneghetti} M, {Merten} J, {Richard} J, {Riess} A,
  {Sharon} K, {Strolger} LG, {Treu} T, {Wang} X, {Williams} LLR, {Zitrin} A
  (2015) {Illuminating a Dark Lens : A Type Ia Supernova Magnified by the
  Frontier Fields Galaxy Cluster Abell 2744}. \apj 811(1):70.
  \doi{10.1088/0004-637X/811/1/70}.
  {\href{https://arxiv.org/abs/1505.06211}{{arXiv:1505.06211}}} {[astro-ph.CO]}

\bibitem[{{Rosati} et~al.(2014){Rosati}, {Balestra}, {Grillo}, {Mercurio},
  {Nonino}, {Biviano}, {Girardi}, {Vanzella}, and {Clash-VLT
  Team}}]{Rosati:2014}
{Rosati} P, {Balestra} I, {Grillo} C, {Mercurio} A, {Nonino} M, {Biviano} A,
  {Girardi} M, {Vanzella} E, {Clash-VLT Team} (2014) {CLASH-VLT: A VIMOS Large
  Programme to Map the Dark Matter Mass Distribution in Galaxy Clusters and
  Probe Distant Lensed Galaxies}. The Messenger 158:48--53

\bibitem[{{Rossi}(2020)}]{ros20}
{Rossi} G (2020) {The Sejong Suite: Cosmological Hydrodynamical Simulations
  with Massive Neutrinos, Dark Radiation, and Warm Dark Matter}. \apjs
  249(2):19. \doi{10.3847/1538-4365/ab9d1e}.
  {\href{https://arxiv.org/abs/2007.15279}{{arXiv:2007.15279}}} {[astro-ph.CO]}

\bibitem[{{Rykoff} et~al.(2016){Rykoff}, {Rozo}, {Hollowood}, {et~al.}, and
  {[DES Collaboration]}}]{ryk16}
{Rykoff} ES, {Rozo} E, {Hollowood} D, {et~al}, {[DES Collaboration]} (2016)
  {The RedMaPPer Galaxy Cluster Catalog From DES Science Verification Data}.
  \apjs 224(1):1. \doi{10.3847/0067-0049/224/1/1}.
  {\href{https://arxiv.org/abs/1601.00621}{{arXiv:1601.00621}}} {[astro-ph.CO]}

\bibitem[{{Saha} et~al.(2006){Saha}, {Read}, and {Williams}}]{sah06}
{Saha} P, {Read} JI, {Williams} LLR (2006) {Two Strong-Lensing Clusters
  Confront Universal Dark Matter Profiles}. \apjl 652(1):L5--L8.
  \doi{10.1086/509782}.
  {\href{https://arxiv.org/abs/astro-ph/0610011}{{arXiv:astro-ph/0610011}}}
  {[astro-ph]}

\bibitem[{{Sand} et~al.(2002){Sand}, {Treu}, and {Ellis}}]{Sand:2002}
{Sand} DJ, {Treu} T, {Ellis} RS (2002) {The Dark Matter Density Profile of the
  Lensing Cluster MS 2137-23: A Test of the Cold Dark Matter Paradigm}. \apjl
  574(2):L129--L133. \doi{10.1086/342530}.
  {\href{https://arxiv.org/abs/astro-ph/0207048}{{arXiv:astro-ph/0207048}}}
  {[astro-ph]}

\bibitem[{{Sand} et~al.(2004){Sand}, {Treu}, {Smith}, and {Ellis}}]{Sand:2004}
{Sand} DJ, {Treu} T, {Smith} GP, {Ellis} RS (2004) {The Dark Matter
  Distribution in the Central Regions of Galaxy Clusters: Implications for Cold
  Dark Matter}. \apj 604(1):88--107. \doi{10.1086/382146}.
  {\href{https://arxiv.org/abs/astro-ph/0309465}{{arXiv:astro-ph/0309465}}}
  {[astro-ph]}

\bibitem[{{Sand} et~al.(2008){Sand}, {Treu}, {Ellis}, {Smith}, and
  {Kneib}}]{sand2008}
{Sand} DJ, {Treu} T, {Ellis} RS, {Smith} GP, {Kneib} JP (2008) {Separating
  Baryons and Dark Matter in Cluster Cores: A Full Two-dimensional Lensing and
  Dynamic Analysis of Abell 383 and MS 2137-23}. \apj 674(2):711--727.
  \doi{10.1086/524652}.
  {\href{https://arxiv.org/abs/0710.1069}{{arXiv:0710.1069}}} {[astro-ph]}

\bibitem[{{Sartoris} et~al.(2014){Sartoris}, {Biviano}, {Rosati}, {Borgani},
  {Umetsu}, {Bartelmann}, {Girardi}, {Grillo}, {Lemze}, {Zitrin}, {Balestra},
  {Mercurio}, {Nonino}, {Postman}, {Czakon}, {Bradley}, {Broadhurst}, {Coe},
  {Medezinski}, {Melchior}, {Meneghetti}, {Merten}, {Annunziatella}, {Benitez},
  {Czoske}, {Donahue}, {Ettori}, {Ford}, {Fritz}, {Kelson}, {Koekemoer},
  {Kuchner}, {Lombardi}, {Maier}, {Moustakas}, {Munari}, {Presotto},
  {Scodeggio}, {Seitz}, {Tozzi}, {Zheng}, and {Ziegler}}]{Sartoris:2014}
{Sartoris} B, {Biviano} A, {Rosati} P, {Borgani} S, {Umetsu} K, {Bartelmann} M,
  {Girardi} M, {Grillo} C, {Lemze} D, {Zitrin} A, {Balestra} I, {Mercurio} A,
  {Nonino} M, {Postman} M, {Czakon} N, {Bradley} L, {Broadhurst} T, {Coe} D,
  {Medezinski} E, {Melchior} P, {Meneghetti} M, {Merten} J, {Annunziatella} M,
  {Benitez} N, {Czoske} O, {Donahue} M, {Ettori} S, {Ford} H, {Fritz} A,
  {Kelson} D, {Koekemoer} A, {Kuchner} U, {Lombardi} M, {Maier} C, {Moustakas}
  LA, {Munari} E, {Presotto} V, {Scodeggio} M, {Seitz} S, {Tozzi} P, {Zheng} W,
  {Ziegler} B (2014) {CLASH-VLT: Constraints on the Dark Matter Equation of
  State from Accurate Measurements of Galaxy Cluster Mass Profiles}. \apjl
  783(1):L11. \doi{10.1088/2041-8205/783/1/L11}.
  {\href{https://arxiv.org/abs/1401.5800}{{arXiv:1401.5800}}} {[astro-ph.CO]}

\bibitem[{{Sartoris} et~al.(2016){Sartoris}, {Biviano}, {Fedeli}, {Bartlett},
  {Borgani}, {Costanzi}, {Giocoli}, {Moscardini}, {Weller}, {Ascaso},
  {Bardelli}, {Maurogordato}, and {Viana}}]{Sartoris:2016}
{Sartoris} B, {Biviano} A, {Fedeli} C, {Bartlett} JG, {Borgani} S, {Costanzi}
  M, {Giocoli} C, {Moscardini} L, {Weller} J, {Ascaso} B, {Bardelli} S,
  {Maurogordato} S, {Viana} PTP (2016) {Next generation cosmology: constraints
  from the Euclid galaxy cluster survey}. \mnras 459(2):1764--1780.
  \doi{10.1093/mnras/stw630}.
  {\href{https://arxiv.org/abs/1505.02165}{{arXiv:1505.02165}}} {[astro-ph.CO]}

\bibitem[{{Sch{\"a}fer} et~al.(2020){Sch{\"a}fer}, {Fourestey}, and
  {Kneib}}]{sch20}
{Sch{\"a}fer} C, {Fourestey} G, {Kneib} JP (2020) {Lenstool-HPC: A High
  Performance Computing based mass modelling tool for cluster-scale
  gravitational lenses}. Astronomy and Computing 30:100360.
  \doi{10.1016/j.ascom.2019.100360}.
  {\href{https://arxiv.org/abs/2004.06352}{{arXiv:2004.06352}}} {[astro-ph.IM]}

\bibitem[{{Schneider}(1984)}]{Schneider1984}
{Schneider} P (1984) {The amplification caused by gravitational bending of
  light}. \aap 140(1):119--124

\bibitem[{{Schneider}(2014)}]{Schneider:2014}
{Schneider} P (2014) {Can one determine cosmological parameters from
  multi-plane strong lens systems?} \aap 568:L2.
  \doi{10.1051/0004-6361/201424450}.
  {\href{https://arxiv.org/abs/1406.6152}{{arXiv:1406.6152}}} {[astro-ph.CO]}

\bibitem[{{Schneider} and {Seitz}(1995)}]{schneiderseitz95}
{Schneider} P, {Seitz} C (1995) {Steps towards nonlinear cluster inversion
  through gravitational distortions. I. Basic considerations and circular
  clusters.} \aap 294:411--431.
  {\href{https://arxiv.org/abs/astro-ph/9407032}{{arXiv:astro-ph/9407032}}}
  {[astro-ph]}

\bibitem[{{Schombert}(1986)}]{Schombert1986}
{Schombert} JM (1986) {The Structure of Brightest Cluster Members. I. Surface
  Photometry}. \apjs 60:603. \doi{10.1086/191100}

\bibitem[{{Schrabback} et~al.(2021){Schrabback}, {Bocquet}, {Sommer}, {Zohren},
  {van den Busch}, {Hern{\'a}ndez-Mart{\'\i}n}, {Hoekstra}, {Raihan},
  {Schirmer}, {Applegate}, {Bayliss}, {Benson}, {Bleem}, {Dietrich}, {Floyd},
  {Hilbert}, {Hlavacek-Larrondo}, {McDonald}, {Saro}, {Stark}, and
  {Weissgerber}}]{schrabback21}
{Schrabback} T, {Bocquet} S, {Sommer} M, {Zohren} H, {van den Busch} JL,
  {Hern{\'a}ndez-Mart{\'\i}n} B, {Hoekstra} H, {Raihan} SF, {Schirmer} M,
  {Applegate} D, {Bayliss} M, {Benson} BA, {Bleem} LE, {Dietrich} JP, {Floyd}
  B, {Hilbert} S, {Hlavacek-Larrondo} J, {McDonald} M, {Saro} A, {Stark} AA,
  {Weissgerber} N (2021) {Mass calibration of distant SPT galaxy clusters
  through expanded weak-lensing follow-up observations with HST, VLT, \&
  Gemini-South}. \mnras 505(3):3923--3943. \doi{10.1093/mnras/stab1386}.
  {\href{https://arxiv.org/abs/2009.07591}{{arXiv:2009.07591}}} {[astro-ph.CO]}

\bibitem[{{Schwinn} et~al.(2017){Schwinn}, {Jauzac}, {Baugh}, {Bartelmann},
  {Eckert}, {Harvey}, {Natarajan}, and {Massey}}]{sch17}
{Schwinn} J, {Jauzac} M, {Baugh} CM, {Bartelmann} M, {Eckert} D, {Harvey} D,
  {Natarajan} P, {Massey} R (2017) {Abell 2744: too much substructure for
  {\ensuremath{\Lambda}}CDM?} \mnras 467(3):2913--2923.
  \doi{10.1093/mnras/stx277}.
  {\href{https://arxiv.org/abs/1611.02790}{{arXiv:1611.02790}}} {[astro-ph.CO]}

\bibitem[{{Schwinn} et~al.(2018){Schwinn}, {Baugh}, {Jauzac}, {Bartelmann}, and
  {Eckert}}]{sch18}
{Schwinn} J, {Baugh} CM, {Jauzac} M, {Bartelmann} M, {Eckert} D (2018)
  {Uncovering substructure with wavelets: proof of concept using Abell 2744}.
  \mnras 481(4):4300--4310. \doi{10.1093/mnras/sty2566}.
  {\href{https://arxiv.org/abs/1804.07401}{{arXiv:1804.07401}}} {[astro-ph.CO]}

\bibitem[{{Sebesta} et~al.(2016){Sebesta}, {Williams}, {Mohammed}, {Saha}, and
  {Liesenborgs}}]{seb16}
{Sebesta} K, {Williams} LLR, {Mohammed} I, {Saha} P, {Liesenborgs} J (2016)
  {Testing light-traces-mass in Hubble Frontier Fields Cluster
  MACS-J0416.1-2403}. \mnras 461(2):2126--2134. \doi{10.1093/mnras/stw1433}.
  {\href{https://arxiv.org/abs/1507.08960}{{arXiv:1507.08960}}} {[astro-ph.CO]}

\bibitem[{{Sereno} et~al.(2017){Sereno}, {Covone}, {Izzo}, {Ettori}, {Coupon},
  and {Lieu}}]{sereno17}
{Sereno} M, {Covone} G, {Izzo} L, {Ettori} S, {Coupon} J, {Lieu} M (2017)
  {PSZ2LenS. Weak lensing analysis of the Planck clusters in the CFHTLenS and
  in the RCSLenS}. \mnras 472(2):1946--1971. \doi{10.1093/mnras/stx2085}.
  {\href{https://arxiv.org/abs/1703.06886}{{arXiv:1703.06886}}} {[astro-ph.CO]}

\bibitem[{{Sharon} and {Johnson}(2015)}]{SharonTraci2015}
{Sharon} K, {Johnson} TL (2015) {Revised Lens Model for the Multiply Imaged
  Lensed Supernova, {\textquotedblleft}SN Refsdal{\textquotedblright} in MACS
  J1149+2223}. \apjl 800(2):L26. \doi{10.1088/2041-8205/800/2/L26}.
  {\href{https://arxiv.org/abs/1411.6933}{{arXiv:1411.6933}}} {[astro-ph.CO]}

\bibitem[{{Sharon} et~al.(2020){Sharon}, {Bayliss}, {Dahle}, {Dunham},
  {Florian}, {Gladders}, {Johnson}, {Mahler}, {Paterno-Mahler}, {Rigby},
  {Whitaker}, {Akhshik}, {Koester}, {Murray}, {Remolina Gonz{\'a}lez}, and
  {Wuyts}}]{sharon2020}
{Sharon} K, {Bayliss} MB, {Dahle} H, {Dunham} SJ, {Florian} MK, {Gladders} MD,
  {Johnson} TL, {Mahler} G, {Paterno-Mahler} R, {Rigby} JR, {Whitaker} KE,
  {Akhshik} M, {Koester} BP, {Murray} K, {Remolina Gonz{\'a}lez} JD, {Wuyts} E
  (2020) {Strong Lens Models for 37 Clusters of Galaxies from the SDSS Giant
  Arcs Survey}. \apjs 247(1):12. \doi{10.3847/1538-4365/ab5f13}.
  {\href{https://arxiv.org/abs/1904.05940}{{arXiv:1904.05940}}} {[astro-ph.GA]}

\bibitem[{{Sharon} et~al.(2022{\natexlab{a}}){Sharon}, {Mahler},
  {Rivera-Thorsen}, {Dahle}, {Gladders}, {Bayliss}, {Florian}, {Kim},
  {Khullar}, {Mainali}, {Napier}, {Navarre}, {Rigby}, {Remolina Gonz{\'a}lez},
  and {Sharma}}]{sharon2022sunburst}
{Sharon} K, {Mahler} G, {Rivera-Thorsen} TE, {Dahle} H, {Gladders} MD,
  {Bayliss} MB, {Florian} MK, {Kim} KJ, {Khullar} G, {Mainali} R, {Napier} KA,
  {Navarre} A, {Rigby} JR, {Remolina Gonz{\'a}lez} JD, {Sharma} S
  (2022{\natexlab{a}}) {The Cosmic Telescope That Lenses the Sunburst Arc, PSZ1
  G311.65-18.48: Strong Gravitational Lensing Model and Source Plane Analysis}.
  \apj 941(2):203. \doi{10.3847/1538-4357/ac927a}.
  {\href{https://arxiv.org/abs/2209.03417}{{arXiv:2209.03417}}} {[astro-ph.GA]}

\bibitem[{{Sharon} et~al.(2022{\natexlab{b}}){Sharon}, {Mahler},
  {Rivera-Thorsen}, {Dahle}, {Gladders}, {Bayliss}, {Florian}, {Kim},
  {Khullar}, {Mainali}, {Napier}, {Navarre}, {Rigby}, {Remolina Gonz{\'a}lez},
  and {Sharma}}]{Sharon2022}
{Sharon} K, {Mahler} G, {Rivera-Thorsen} TE, {Dahle} H, {Gladders} MD,
  {Bayliss} MB, {Florian} MK, {Kim} KJ, {Khullar} G, {Mainali} R, {Napier} KA,
  {Navarre} A, {Rigby} JR, {Remolina Gonz{\'a}lez} JD, {Sharma} S
  (2022{\natexlab{b}}) {The Cosmic Telescope That Lenses the Sunburst Arc, PSZ1
  G311.65-18.48: Strong Gravitational Lensing Model and Source Plane Analysis}.
  \apj 941(2):203. \doi{10.3847/1538-4357/ac927a}.
  {\href{https://arxiv.org/abs/2209.03417}{{arXiv:2209.03417}}} {[astro-ph.GA]}

\bibitem[{{Smail} and {Dickinson}(1995)}]{Smail1995}
{Smail} I, {Dickinson} M (1995) {Lensing by Distant Clusters: HST Observations
  of Weak Shear in the Field of 3C 324}. \apjl 455:L99. \doi{10.1086/309842}.
  {\href{https://arxiv.org/abs/astro-ph/9510050}{{arXiv:astro-ph/9510050}}}
  {[astro-ph]}

\bibitem[{{Soucail} et~al.(1987){Soucail}, {Fort}, {Mellier}, and
  {Picat}}]{sou87}
{Soucail} G, {Fort} B, {Mellier} Y, {Picat} JP (1987) {A blue ring-like
  structure in the center of the A 370 cluster of galaxies.} \aap 172:L14--L16

\bibitem[{{Soucail} et~al.(1988){Soucail}, {Mellier}, {Fort}, and
  {Cailloux}}]{Soucail1988}
{Soucail} G, {Mellier} Y, {Fort} B, {Cailloux} M (1988) {Spectroscopic
  observations of the distant cluster of galaxies Abell 370 : a catalogue of 84
  spectra.} \aap 73:471--514

\bibitem[{{Soucail} et~al.(2004){Soucail}, {Kneib}, and {Golse}}]{Soucail:2004}
{Soucail} G, {Kneib} JP, {Golse} G (2004) {Multiple-images in the cluster lens
  Abell 2218: Constraining the geometry of the Universe?} \aap 417:L33--L37.
  \doi{10.1051/0004-6361:20040077}.
  {\href{https://arxiv.org/abs/astro-ph/0402658}{{arXiv:astro-ph/0402658}}}
  {[astro-ph]}

\bibitem[{{Spergel} and {Steinhardt}(2000)}]{spe00}
{Spergel} DN, {Steinhardt} PJ (2000) {Observational Evidence for
  Self-Interacting Cold Dark Matter}. \prl 84(17):3760--3763.
  \doi{10.1103/PhysRevLett.84.3760}.
  {\href{https://arxiv.org/abs/astro-ph/9909386}{{arXiv:astro-ph/9909386}}}
  {[astro-ph]}

\bibitem[{{Springel} and {Farrar}(2007)}]{SpringelFarrar2007}
{Springel} V, {Farrar} GR (2007) {The speed of the `bullet' in the merging
  galaxy cluster 1E0657-56}. \mnras 380(3):911--925.
  \doi{10.1111/j.1365-2966.2007.12159.x}.
  {\href{https://arxiv.org/abs/astro-ph/0703232}{{arXiv:astro-ph/0703232}}}
  {[astro-ph]}

\bibitem[{{Springel} et~al.(2008){Springel}, {Wang}, {Vogelsberger}, {Ludlow},
  {Jenkins}, {Helmi}, {Navarro}, {Frenk}, and {White}}]{spr08}
{Springel} V, {Wang} J, {Vogelsberger} M, {Ludlow} A, {Jenkins} A, {Helmi} A,
  {Navarro} JF, {Frenk} CS, {White} SDM (2008) {The Aquarius Project: the
  subhaloes of galactic haloes}. \mnras 391(4):1685--1711.
  \doi{10.1111/j.1365-2966.2008.14066.x}.
  {\href{https://arxiv.org/abs/0809.0898}{{arXiv:0809.0898}}} {[astro-ph]}

\bibitem[{{Steinhardt} et~al.(2020){Steinhardt}, {Jauzac}, {Acebron}
  et~al.}]{sec2:steinhardt20}
{Steinhardt} CL, {Jauzac} M, {Acebron} A, et~al. (2020) {The BUFFALO HST
  Survey}. \apjs 247(2):64. \doi{10.3847/1538-4365/ab75ed}.
  {\href{https://arxiv.org/abs/2001.09999}{{arXiv:2001.09999}}} {[astro-ph.GA]}

\bibitem[{{Strait} et~al.(2018){Strait}, {Brada{\v c}}, {Hoag}, {Huang},
  {Treu}, {Wang}, {Amorin}, {Castellano}, {Fontana}, {Lemaux}, {Merlin},
  {Schmidt}, {Schrabback}, {Tomczack}, {Trenti}, and {Vulcani}}]{strait18}
{Strait} V, {Brada{\v c}} M, {Hoag} A, {Huang} KH, {Treu} T, {Wang} X, {Amorin}
  R, {Castellano} M, {Fontana} A, {Lemaux} BC, {Merlin} E, {Schmidt} KB,
  {Schrabback} T, {Tomczack} A, {Trenti} M, {Vulcani} B (2018) {Mass and Light
  of Abell 370: A Strong and Weak Lensing Analysis}. \apj 868:129.
  \doi{10.3847/1538-4357/aae834}.
  {\href{https://arxiv.org/abs/1805.08789}{{arXiv:1805.08789}}}

\bibitem[{{Suyu} and {Halkola}(2010)}]{Suyu:2010}
{Suyu} SH, {Halkola} A (2010) {The halos of satellite galaxies: the companion
  of the massive elliptical lens SL2S J08544-0121}. \aap 524:A94.
  \doi{10.1051/0004-6361/201015481}.
  {\href{https://arxiv.org/abs/1007.4815}{{arXiv:1007.4815}}} {[astro-ph.CO]}

\bibitem[{{Suyu} et~al.(2012){Suyu}, {Hensel}, {McKean}, {Fassnacht}, {Treu},
  {Halkola}, {Norbury}, {Jackson}, {Schneider}, {Thompson}, {Auger},
  {Koopmans}, and {Matthews}}]{Suyu:2012}
{Suyu} SH, {Hensel} SW, {McKean} JP, {Fassnacht} CD, {Treu} T, {Halkola} A,
  {Norbury} M, {Jackson} N, {Schneider} P, {Thompson} D, {Auger} MW, {Koopmans}
  LVE, {Matthews} K (2012) {Disentangling Baryons and Dark Matter in the Spiral
  Gravitational Lens B1933+503}. \apj 750(1):10.
  \doi{10.1088/0004-637X/750/1/10}.
  {\href{https://arxiv.org/abs/1110.2536}{{arXiv:1110.2536}}} {[astro-ph.CO]}

\bibitem[{{Torres-Ballesteros} and {Casta{\~n}eda}(2023)}]{tor22}
{Torres-Ballesteros} DA, {Casta{\~n}eda} L (2023) {RELENSING: Reconstructing
  the mass profile of galaxy clusters from gravitational lensing}. \mnras
  518(3):4494--4516. \doi{10.1093/mnras/stac3253}.
  {\href{https://arxiv.org/abs/2201.10076}{{arXiv:2201.10076}}} {[astro-ph.CO]}

\bibitem[{{Treu} et~al.(2016){Treu}, {Brammer}, {Diego}, {Grillo}, {Kelly},
  {Oguri}, {Rodney}, {Rosati}, {Sharon}, {Zitrin}, {Balestra}, {Brada{\v{c}}},
  {Broadhurst}, {Caminha}, {Halkola}, {Hoag}, {Ishigaki}, {Johnson}, {Karman},
  {Kawamata}, {Mercurio}, {Schmidt}, {Strolger}, {Suyu}, {Filippenko}, {Foley},
  {Jha}, and {Patel}}]{Treu+2016}
{Treu} T, {Brammer} G, {Diego} JM, {Grillo} C, {Kelly} PL, {Oguri} M, {Rodney}
  SA, {Rosati} P, {Sharon} K, {Zitrin} A, {Balestra} I, {Brada{\v{c}}} M,
  {Broadhurst} T, {Caminha} GB, {Halkola} A, {Hoag} A, {Ishigaki} M, {Johnson}
  TL, {Karman} W, {Kawamata} R, {Mercurio} A, {Schmidt} KB, {Strolger} LG,
  {Suyu} SH, {Filippenko} AV, {Foley} RJ, {Jha} SW, {Patel} B (2016)
  {``Refsdal'' Meets Popper: Comparing Predictions of the Re-appearance of the
  Multiply Imaged Supernova Behind MACSJ1149.5+2223}. \apj 817(1):60.
  \doi{10.3847/0004-637X/817/1/60}.
  {\href{https://arxiv.org/abs/1510.05750}{{arXiv:1510.05750}}} {[astro-ph.CO]}

\bibitem[{{Tutusaus} et~al.(2020){Tutusaus}, {Martinelli}, {Cardone}, {Camera}
  et~al.}]{Tutusaus:2020}
{Tutusaus} I, {Martinelli} M, {Cardone} VF, {Camera} S, et~al. (2020) {Euclid:
  The importance of galaxy clustering and weak lensing cross-correlations
  within the photometric Euclid survey}. \aap 643:A70.
  \doi{10.1051/0004-6361/202038313}.
  {\href{https://arxiv.org/abs/2005.00055}{{arXiv:2005.00055}}} {[astro-ph.CO]}

\bibitem[{{Tyson} et~al.(1990){Tyson}, {Valdes}, and {Wenk}}]{Tyson1990}
{Tyson} JA, {Valdes} F, {Wenk} RA (1990) {Detection of Systematic Gravitational
  Lens Galaxy Image Alignments: Mapping Dark Matter in Galaxy Clusters}. \apjl
  349:L1. \doi{10.1086/185636}

\bibitem[{{Umetsu}(2020)}]{Umetsu2020}
{Umetsu} K (2020) {Cluster-galaxy weak lensing}. \aapr 28(1):7.
  \doi{10.1007/s00159-020-00129-w}.
  {\href{https://arxiv.org/abs/2007.00506}{{arXiv:2007.00506}}} {[astro-ph.CO]}

\bibitem[{{Umetsu} et~al.(2012){Umetsu}, {Medezinski}, {Nonino}, {Merten},
  {Zitrin}, {Molino}, {Grillo}, {Carrasco}, {Donahue}, {Mahdavi}, {Coe},
  {Postman}, {Koekemoer}, {Czakon}, {Sayers}, {Mroczkowski}, {Golwala}, {Koch},
  {Lin}, {Molnar}, {Rosati}, {Balestra}, {Mercurio}, {Scodeggio}, {Biviano},
  {Anguita}, {Infante}, {Seidel}, {Sendra}, {Jouvel}, {Host}, {Lemze},
  {Broadhurst}, {Meneghetti}, {Moustakas}, {Bartelmann}, {Ben{\'\i}tez},
  {Bouwens}, {Bradley}, {Ford}, {Jim{\'e}nez-Teja}, {Kelson}, {Lahav},
  {Melchior}, {Moustakas}, {Ogaz}, {Seitz}, and {Zheng}}]{Umetsu:2012}
{Umetsu} K, {Medezinski} E, {Nonino} M, {Merten} J, {Zitrin} A, {Molino} A,
  {Grillo} C, {Carrasco} M, {Donahue} M, {Mahdavi} A, {Coe} D, {Postman} M,
  {Koekemoer} A, {Czakon} N, {Sayers} J, {Mroczkowski} T, {Golwala} S, {Koch}
  PM, {Lin} KY, {Molnar} SM, {Rosati} P, {Balestra} I, {Mercurio} A,
  {Scodeggio} M, {Biviano} A, {Anguita} T, {Infante} L, {Seidel} G, {Sendra} I,
  {Jouvel} S, {Host} O, {Lemze} D, {Broadhurst} T, {Meneghetti} M, {Moustakas}
  L, {Bartelmann} M, {Ben{\'\i}tez} N, {Bouwens} R, {Bradley} L, {Ford} H,
  {Jim{\'e}nez-Teja} Y, {Kelson} D, {Lahav} O, {Melchior} P, {Moustakas} J,
  {Ogaz} S, {Seitz} S, {Zheng} W (2012) {CLASH: Mass Distribution in and around
  MACS J1206.2-0847 from a Full Cluster Lensing Analysis}. \apj 755(1):56.
  \doi{10.1088/0004-637X/755/1/56}.
  {\href{https://arxiv.org/abs/1204.3630}{{arXiv:1204.3630}}} {[astro-ph.CO]}

\bibitem[{{Umetsu} et~al.(2014){Umetsu}, {Medezinski}, {Nonino}, {Merten},
  {Postman}, {Meneghetti}, {Donahue}, {Czakon}, {Molino}, {Seitz}, {Gruen},
  {Lemze}, {Balestra}, {Ben{\'\i}tez}, {Biviano}, {Broadhurst}, {Ford},
  {Grillo}, {Koekemoer}, {Melchior}, {Mercurio}, {Moustakas}, {Rosati}, and
  {Zitrin}}]{Umetsu:2014}
{Umetsu} K, {Medezinski} E, {Nonino} M, {Merten} J, {Postman} M, {Meneghetti}
  M, {Donahue} M, {Czakon} N, {Molino} A, {Seitz} S, {Gruen} D, {Lemze} D,
  {Balestra} I, {Ben{\'\i}tez} N, {Biviano} A, {Broadhurst} T, {Ford} H,
  {Grillo} C, {Koekemoer} A, {Melchior} P, {Mercurio} A, {Moustakas} J,
  {Rosati} P, {Zitrin} A (2014) {CLASH: Weak-lensing Shear-and-magnification
  Analysis of 20 Galaxy Clusters}. \apj 795(2):163.
  \doi{10.1088/0004-637X/795/2/163}.
  {\href{https://arxiv.org/abs/1404.1375}{{arXiv:1404.1375}}} {[astro-ph.CO]}

\bibitem[{{Umetsu} et~al.(2016){Umetsu}, {Zitrin}, {Gruen}, {Merten},
  {Donahue}, and {Postman}}]{ume16}
{Umetsu} K, {Zitrin} A, {Gruen} D, {Merten} J, {Donahue} M, {Postman} M (2016)
  {CLASH: Joint Analysis of Strong-lensing, Weak-lensing Shear, and
  Magnification Data for 20 Galaxy Clusters}. \apj 821(2):116.
  \doi{10.3847/0004-637X/821/2/116}.
  {\href{https://arxiv.org/abs/1507.04385}{{arXiv:1507.04385}}} {[astro-ph.CO]}

\bibitem[{{Umetsu} et~al.(2018){Umetsu}, {Sereno}, {Tam}, {Chiu}, {Fan},
  {Ettori}, {Gruen}, {Okumura}, {Medezinski}, {Donahue}, {Meneghetti}, {Frye},
  {Koekemoer}, {Broadhurst}, {Zitrin}, {Balestra}, {Ben{\'\i}tez}, {Higuchi},
  {Melchior}, {Mercurio}, {Merten}, {Molino}, {Nonino}, {Postman}, {Rosati},
  {Sayers}, and {Seitz}}]{umetsu2018}
{Umetsu} K, {Sereno} M, {Tam} SI, {Chiu} IN, {Fan} Z, {Ettori} S, {Gruen} D,
  {Okumura} T, {Medezinski} E, {Donahue} M, {Meneghetti} M, {Frye} B,
  {Koekemoer} A, {Broadhurst} T, {Zitrin} A, {Balestra} I, {Ben{\'\i}tez} N,
  {Higuchi} Y, {Melchior} P, {Mercurio} A, {Merten} J, {Molino} A, {Nonino} M,
  {Postman} M, {Rosati} P, {Sayers} J, {Seitz} S (2018) {The Projected Dark and
  Baryonic Ellipsoidal Structure of 20 CLASH Galaxy Clusters}. \apj 860(2):104.
  \doi{10.3847/1538-4357/aac3d9}.
  {\href{https://arxiv.org/abs/1804.00664}{{arXiv:1804.00664}}} {[astro-ph.CO]}

\bibitem[{{Uson} et~al.(1991){Uson}, {Boughn}, and {Kuhn}}]{Uson+1991}
{Uson} JM, {Boughn} SP, {Kuhn} JR (1991) {Diffuse Light in Dense Clusters of
  Galaxies. I. R-Band Observations of Abell 2029}. \apj 369:46.
  \doi{10.1086/169737}

\bibitem[{{van Weeren} et~al.(2017){van Weeren}, {Ogrean}, {Jones}, {Forman},
  {Andrade-Santos}, {Pearce}, {Bonafede}, {Br{\"u}ggen}, {Bulbul}, {Clarke},
  {Churazov}, {David}, {Dawson}, {Donahue}, {Goulding}, {Kraft}, {Mason},
  {Merten}, {Mroczkowski}, {Nulsen}, {Rosati}, {Roediger}, {Randall}, {Sayers},
  {Umetsu}, {Vikhlinin}, and {Zitrin}}]{vanWeeran+2017}
{van Weeren} RJ, {Ogrean} GA, {Jones} C, {Forman} WR, {Andrade-Santos} F,
  {Pearce} CJJ, {Bonafede} A, {Br{\"u}ggen} M, {Bulbul} E, {Clarke} TE,
  {Churazov} E, {David} L, {Dawson} WA, {Donahue} M, {Goulding} A, {Kraft} RP,
  {Mason} B, {Merten} J, {Mroczkowski} T, {Nulsen} PEJ, {Rosati} P, {Roediger}
  E, {Randall} SW, {Sayers} J, {Umetsu} K, {Vikhlinin} A, {Zitrin} A (2017)
  {Chandra and JVLA Observations of HST Frontier Fields Cluster MACS
  J0717.5+3745}. \apj 835(2):197. \doi{10.3847/1538-4357/835/2/197}.
  {\href{https://arxiv.org/abs/1701.04096}{{arXiv:1701.04096}}} {[astro-ph.CO]}

\bibitem[{{Vega-Ferrero} et~al.(2018){Vega-Ferrero}, {Diego}, {Miranda}, and
  {Bernstein}}]{HubbleRefsdal2018}
{Vega-Ferrero} J, {Diego} JM, {Miranda} V, {Bernstein} GM (2018) {The Hubble
  Constant from SN Refsdal}. \apjl 853(2):L31. \doi{10.3847/2041-8213/aaa95f}.
  {\href{https://arxiv.org/abs/1712.05800}{{arXiv:1712.05800}}} {[astro-ph.CO]}

\bibitem[{{Verdugo} et~al.(2011){Verdugo}, {Motta}, {Mu{\~n}oz}, {Limousin},
  {Cabanac}, and {Richard}}]{Verdugo_2011}
{Verdugo} T, {Motta} V, {Mu{\~n}oz} RP, {Limousin} M, {Cabanac} R, {Richard} J
  (2011) {Gravitational lensing and dynamics in SL2S J02140-0535: probing the
  mass out to large radius}. \aap 527:A124. \doi{10.1051/0004-6361/201014965}.
  {\href{https://arxiv.org/abs/1005.1566}{{arXiv:1005.1566}}} {[astro-ph.CO]}

\bibitem[{{Verdugo} et~al.(2014){Verdugo}, {Motta}, {Fo{\"e}x},
  {Forero-Romero}, {Mu{\~n}oz}, {Pello}, {Limousin}, {More}, {Cabanac},
  {Soucail}, {Blakeslee}, {Mej{\'\i}a-Narv{\'a}ez}, {Magris}, and
  {Fern{\'a}ndez-Trincado}}]{Verdugo_2014}
{Verdugo} T, {Motta} V, {Fo{\"e}x} G, {Forero-Romero} JE, {Mu{\~n}oz} RP,
  {Pello} R, {Limousin} M, {More} A, {Cabanac} R, {Soucail} G, {Blakeslee} JP,
  {Mej{\'\i}a-Narv{\'a}ez} AJ, {Magris} G, {Fern{\'a}ndez-Trincado} JG (2014)
  {Characterizing SL2S galaxy groups using the Einstein radius}. \aap 571:A65.
  \doi{10.1051/0004-6361/201423696}.
  {\href{https://arxiv.org/abs/1409.2900}{{arXiv:1409.2900}}} {[astro-ph.CO]}

\bibitem[{{Wagner}(2019)}]{wag19}
{Wagner} J (2019) {A Model-Independent Characterisation of Strong Gravitational
  Lensing by Observables}. Universe 5(7):177. \doi{10.3390/universe5070177}.
  {\href{https://arxiv.org/abs/1906.05285}{{arXiv:1906.05285}}} {[astro-ph.CO]}

\bibitem[{{Wagner}(2020)}]{wag20b}
{Wagner} J (2020) {Cosmic structures from a mathematical perspective 1: dark
  matter halo mass density profiles}. General Relativity and Gravitation
  52(6):61. \doi{10.1007/s10714-020-02715-w}.
  {\href{https://arxiv.org/abs/2002.00960}{{arXiv:2002.00960}}} {[astro-ph.CO]}

\bibitem[{{Wagner}(2022)}]{wag22}
{Wagner} J (2022) Generalised model-independent characterisation of strong
  gravitational lenses - vii. impact of source properties and higher-order lens
  properties on the local lens reconstruction. A\&A 663:A157.
  \doi{10.1051/0004-6361/202243562},
  \urlprefix\url{https://doi.org/10.1051/0004-6361/202243562}

\bibitem[{{Wagner} and {Tessore}(2018)}]{wag18b}
{Wagner} J, {Tessore} N (2018) {Generalised model-independent characterisation
  of strong gravitational lenses. II. Transformation matrix between multiple
  images}. \aap 613:A6. \doi{10.1051/0004-6361/201730947}.
  {\href{https://arxiv.org/abs/1704.01822}{{arXiv:1704.01822}}} {[astro-ph.CO]}

\bibitem[{{Wagner} and {Williams}(2020)}]{wag20}
{Wagner} J, {Williams} LLR (2020) {Model-independent and model-based local
  lensing properties of B0128+437 from resolved quasar images}. \aap 635:A86.
  \doi{10.1051/0004-6361/201936628}.
  {\href{https://arxiv.org/abs/1909.01349}{{arXiv:1909.01349}}} {[astro-ph.GA]}

\bibitem[{{Wagner} et~al.(2018){Wagner}, {Liesenborgs}, and {Tessore}}]{wag18}
{Wagner} J, {Liesenborgs} J, {Tessore} N (2018) {Model-independent and
  model-based local lensing properties of CL0024+1654 from multiply imaged
  galaxies}. \aap 612:A17. \doi{10.1051/0004-6361/201731932}.
  {\href{https://arxiv.org/abs/1709.03531}{{arXiv:1709.03531}}}

\bibitem[{{Walsh} et~al.(1979){Walsh}, {Carswell}, and {Weymann}}]{wal79}
{Walsh} D, {Carswell} RF, {Weymann} RJ (1979) {0957+561 A, B: twin quasistellar
  objects or gravitational lens?} \nat 279:381--384. \doi{10.1038/279381a0}

\bibitem[{{Wambsganss} et~al.(2004){Wambsganss}, {Bode}, and
  {Ostriker}}]{Wambsganss+2004}
{Wambsganss} J, {Bode} P, {Ostriker} JP (2004) {Giant Arc Statistics in Concord
  with a Concordance Lambda Cold Dark Matter Universe}. \apjl 606(2):L93--L96.
  \doi{10.1086/421459}.
  {\href{https://arxiv.org/abs/astro-ph/0306088}{{arXiv:astro-ph/0306088}}}
  {[astro-ph]}

\bibitem[{{Wang} et~al.(2022){Wang}, {Ca{\~n}ameras}, {Caminha}, {Suyu},
  {Y{\i}ld{\i}r{\i}m}, {Chiriv{\`\i}}, {Christensen}, {Grillo}, and
  {Schuldt}}]{sec2:wang22}
{Wang} H, {Ca{\~n}ameras} R, {Caminha} GB, {Suyu} SH, {Y{\i}ld{\i}r{\i}m} A,
  {Chiriv{\`\i}} G, {Christensen} L, {Grillo} C, {Schuldt} S (2022)
  {Constraining the multi-scale dark-matter distribution in CASSOWARY 31 with
  strong gravitational lensing and stellar dynamics}. \aap 668:A162.
  \doi{10.1051/0004-6361/202243600}.
  {\href{https://arxiv.org/abs/2203.13759}{{arXiv:2203.13759}}} {[astro-ph.GA]}

\bibitem[{{Wang} et~al.(2015){Wang}, {Hoag}, {Huang}, {Treu}, {Brada{\v c}},
  {Schmidt}, {Brammer}, {Vulcani}, {Jones}, {Ryan}, {Amor{\'{\i}}n},
  {Castellano}, {Fontana}, {Merlin}, and {Trenti}}]{wang15}
{Wang} X, {Hoag} A, {Huang} KH, {Treu} T, {Brada{\v c}} M, {Schmidt} KB,
  {Brammer} GB, {Vulcani} B, {Jones} TA, {Ryan} RE Jr, {Amor{\'{\i}}n} R,
  {Castellano} M, {Fontana} A, {Merlin} E, {Trenti} M (2015) {The Grism
  Lens-amplified Survey from Space (GLASS). IV. Mass Reconstruction of the
  Lensing Cluster Abell 2744 from Frontier Field Imaging and GLASS
  Spectroscopy}. \apj 811:29. \doi{10.1088/0004-637X/811/1/29}.
  {\href{https://arxiv.org/abs/1504.02405}{{arXiv:1504.02405}}}

\bibitem[{{Wechsler} et~al.(2002){Wechsler}, {Bullock}, {Primack}, {Kravtsov},
  and {Dekel}}]{Wechsler:2002}
{Wechsler} RH, {Bullock} JS, {Primack} JR, {Kravtsov} AV, {Dekel} A (2002)
  {Concentrations of Dark Halos from Their Assembly Histories}. \apj
  568(1):52--70. \doi{10.1086/338765}.
  {\href{https://arxiv.org/abs/astro-ph/0108151}{{arXiv:astro-ph/0108151}}}
  {[astro-ph]}

\bibitem[{{Williams} and {Lewis}(1998)}]{wil98}
{Williams} LLR, {Lewis} GF (1998) {Undistorted lensed images in galaxy
  clusters.} \mnras 294:299--306. \doi{10.1046/j.1365-8711.1998.01179.x}.
  {\href{https://arxiv.org/abs/astro-ph/9710042}{{arXiv:astro-ph/9710042}}}
  {[astro-ph]}

\bibitem[{{Williams} and {Liesenborgs}(2019)}]{wil19}
{Williams} LLR, {Liesenborgs} J (2019) {The role of multiple images and model
  priors in measuring H$_{0}$ from supernova Refsdal in galaxy cluster MACS
  J1149.5+2223}. \mnras 482(4):5666--5677. \doi{10.1093/mnras/sty3113}.
  {\href{https://arxiv.org/abs/1806.11113}{{arXiv:1806.11113}}} {[astro-ph.CO]}

\bibitem[{{Williams} and {Saha}(2004)}]{wil04}
{Williams} LLR, {Saha} P (2004) {Models of the Giant Quadruple Quasar SDSS
  J1004+4112}. \aj 128(6):2631--2641. \doi{10.1086/426007}.
  {\href{https://arxiv.org/abs/astro-ph/0409418}{{arXiv:astro-ph/0409418}}}
  {[astro-ph]}

\bibitem[{{Williams} et~al.(1999){Williams}, {Navarro}, and
  {Bartelmann}}]{wil99}
{Williams} LLR, {Navarro} JF, {Bartelmann} M (1999) {The Core Structure of
  Galaxy Clusters from Gravitational Lensing}. \apj 527(2):535--544.
  \doi{10.1086/308127}.
  {\href{https://arxiv.org/abs/astro-ph/9905134}{{arXiv:astro-ph/9905134}}}
  {[astro-ph]}

\bibitem[{{Williams} et~al.(2010){Williams}, {Hjorth}, and {Wojtak}}]{wil10}
{Williams} LLR, {Hjorth} J, {Wojtak} R (2010) {Statistical Mechanics of
  Collisionless Orbits. III. Comparison with N-body Simulations}. \apj
  725(1):282--287. \doi{10.1088/0004-637X/725/1/282}.
  {\href{https://arxiv.org/abs/1010.0267}{{arXiv:1010.0267}}} {[astro-ph.CO]}

\bibitem[{{Williams} et~al.(2018){Williams}, {Sebesta}, and
  {Liesenborgs}}]{wil18}
{Williams} LLR, {Sebesta} K, {Liesenborgs} J (2018) {Evidence for the
  line-of-sight structure in the Hubble Frontier Field cluster,
  MACSJ0717.5+3745}. \mnras 480(3):3140--3151. \doi{10.1093/mnras/sty2088}.
  {\href{https://arxiv.org/abs/1711.05265}{{arXiv:1711.05265}}} {[astro-ph.CO]}

\bibitem[{{Windhorst} et~al.(2023){Windhorst}, {Cohen}, {Jansen}, {Summers},
  {Tompkins}, {Conselice}, {Driver}, {Yan}, {Coe}, {Frye}, {Grogin},
  {Koekemoer}, {Marshall}, {O'Brien}, {Pirzkal}, {Robotham}, {Ryan}, {Willmer},
  {Carleton}, {Diego}, {Keel}, {Porto}, {Redshaw}, {Scheller}, {Wilkins},
  {Willner}, {Zitrin}, {Adams}, {Austin}, {Arendt}, {Beacom}, {Bhatawdekar},
  {Bradley}, {Broadhurst}, {Cheng}, {Civano}, {Dai}, {Dole}, {D'Silva},
  {Duncan}, {Fazio}, {Ferrami}, {Ferreira}, {Finkelstein}, {Furtak}, {Gim},
  {Griffiths}, {Hammel}, {Harrington}, {Hathi}, {Holwerda}, {Honor}, {Huang},
  {Hyun}, {Im}, {Joshi}, {Kamieneski}, {Kelly}, {Larson}, {Li}, {Lim}, {Ma},
  {Maksym}, {Manzoni}, {Meena}, {Milam}, {Nonino}, {Pascale}, {Petric},
  {Pierel}, {del Carmen Polletta}, {R{\"o}ttgering}, {Rutkowski}, {Smail},
  {Straughn}, {Strolger}, {Swirbul}, {Trussler}, {Wang}, {Welch}, {B. Wyithe},
  {Yun}, {Zackrisson}, {Zhang}, and {Zhao}}]{Windhorst2022}
{Windhorst} RA, {Cohen} SH, {Jansen} RA, {Summers} J, {Tompkins} S, {Conselice}
  CJ, {Driver} SP, {Yan} H, {Coe} D, {Frye} B, {Grogin} N, {Koekemoer} A,
  {Marshall} MA, {O'Brien} R, {Pirzkal} N, {Robotham} A, {Ryan} RE, {Willmer}
  CNA, {Carleton} T, {Diego} JM, {Keel} WC, {Porto} P, {Redshaw} C, {Scheller}
  S, {Wilkins} SM, {Willner} SP, {Zitrin} A, {Adams} NJ, {Austin} D, {Arendt}
  RG, {Beacom} JF, {Bhatawdekar} RA, {Bradley} LD, {Broadhurst} T, {Cheng} C,
  {Civano} F, {Dai} L, {Dole} H, {D'Silva} JCJ, {Duncan} KJ, {Fazio} GG,
  {Ferrami} G, {Ferreira} L, {Finkelstein} SL, {Furtak} LJ, {Gim} HB,
  {Griffiths} A, {Hammel} HB, {Harrington} KC, {Hathi} NP, {Holwerda} BW,
  {Honor} R, {Huang} JS, {Hyun} M, {Im} M, {Joshi} BA, {Kamieneski} PS, {Kelly}
  P, {Larson} RL, {Li} J, {Lim} J, {Ma} Z, {Maksym} P, {Manzoni} G, {Meena} AK,
  {Milam} SN, {Nonino} M, {Pascale} M, {Petric} A, {Pierel} JDR, {del Carmen
  Polletta} M, {R{\"o}ttgering} HJA, {Rutkowski} MJ, {Smail} I, {Straughn} AN,
  {Strolger} LG, {Swirbul} A, {Trussler} JAA, {Wang} L, {Welch} B, {B Wyithe}
  JS, {Yun} M, {Zackrisson} E, {Zhang} J, {Zhao} X (2023) {JWST PEARLS. Prime
  Extragalactic Areas for Reionization and Lensing Science: Project Overview
  and First Results}. \aj 165(1):13. \doi{10.3847/1538-3881/aca163}.
  {\href{https://arxiv.org/abs/2209.04119}{{arXiv:2209.04119}}} {[astro-ph.CO]}

\bibitem[{Wu et~al.(2021)Wu, Weinberg, Salcedo, and Wibking}]{Wu+2021}
Wu HY, Weinberg DH, Salcedo AN, Wibking BD (2021) Cosmology with galaxy cluster
  weak lensing: Statistical limits and experimental design. \apj 910(1):28.
  \doi{10.3847/1538-4357/abdc23},
  \urlprefix\url{https://dx.doi.org/10.3847/1538-4357/abdc23}

\bibitem[{{Yang} and {Yu}(2021)}]{yan21}
{Yang} D, {Yu} HB (2021) {Self-interacting dark matter and small-scale
  gravitational lenses in galaxy clusters}. \prd 104(10):103031.
  \doi{10.1103/PhysRevD.104.103031}.
  {\href{https://arxiv.org/abs/2102.02375}{{arXiv:2102.02375}}} {[astro-ph.GA]}

\bibitem[{{Young} et~al.(1980){Young}, {Gunn}, {Kristian}, {Oke}, and
  {Westphal}}]{Young1980}
{Young} P, {Gunn} JE, {Kristian} J, {Oke} JB, {Westphal} JA (1980) {The double
  quasar Q0957+561 A, B: a gravitational lens image formed by a galaxy at
  z=0.39.} \apj 241:507--520. \doi{10.1086/158365}

\bibitem[{Yue et~al.(2018)Yue, Castellano, Ferrara, Fontana, Merlin,
  Amor{\'{\i}}n, Grazian, M{\'a}rmol-Queralto, Michalowski, Mortlock, Paris,
  Parsa, Pilo, Santini, and Criscienzo}]{Yue+2018}
Yue B, Castellano M, Ferrara A, Fontana A, Merlin E, Amor{\'{\i}}n R, Grazian
  A, M{\'a}rmol-Queralto E, Michalowski MJ, Mortlock A, Paris D, Parsa S, Pilo
  S, Santini P, Criscienzo MD (2018) On the faint end of the galaxy luminosity
  function in the epoch of reionization: Updated constraints from the hst
  frontier fields. \apj 868(2):115. \doi{10.3847/1538-4357/aae77f}

\bibitem[{{Zhao} et~al.(2003){Zhao}, {Mo}, {Jing}, and
  {B{\"o}rner}}]{Zhao:2003}
{Zhao} DH, {Mo} HJ, {Jing} YP, {B{\"o}rner} G (2003) {The growth and structure
  of dark matter haloes}. \mnras 339(1):12--24.
  \doi{10.1046/j.1365-8711.2003.06135.x}.
  {\href{https://arxiv.org/abs/astro-ph/0204108}{{arXiv:astro-ph/0204108}}}
  {[astro-ph]}

\bibitem[{{Zitrin} and {Broadhurst}(2009)}]{zit09b}
{Zitrin} A, {Broadhurst} T (2009) {Discovery of the Largest Known Lensed Images
  Formed by a Critically Convergent Lensing Cluster}. \apjl 703(2):L132--L136.
  \doi{10.1088/0004-637X/703/2/L132}.
  {\href{https://arxiv.org/abs/0906.5079}{{arXiv:0906.5079}}} {[astro-ph.CO]}

\bibitem[{{Zitrin} et~al.(2009{\natexlab{a}}){Zitrin}, {Broadhurst}, {Umetsu},
  {Coe}, {Ben{\'\i}tez}, {Ascaso}, {Bradley}, {Ford}, {Jee}, {Medezinski},
  {Rephaeli}, and {Zheng}}]{Zitrin:09}
{Zitrin} A, {Broadhurst} T, {Umetsu} K, {Coe} D, {Ben{\'\i}tez} N, {Ascaso} B,
  {Bradley} L, {Ford} H, {Jee} J, {Medezinski} E, {Rephaeli} Y, {Zheng} W
  (2009{\natexlab{a}}) {New multiply-lensed galaxies identified in ACS/NIC3
  observations of Cl0024+1654 using an improved mass model}. \mnras
  396(4):1985--2002. \doi{10.1111/j.1365-2966.2009.14899.x}.
  {\href{https://arxiv.org/abs/0902.3971}{{arXiv:0902.3971}}} {[astro-ph.CO]}

\bibitem[{{Zitrin} et~al.(2009{\natexlab{b}}){Zitrin}, {Broadhurst}, {Umetsu},
  {Coe}, {Ben{\'\i}tez}, {Ascaso}, {Bradley}, {Ford}, {Jee}, {Medezinski},
  {Rephaeli}, and {Zheng}}]{zit09}
{Zitrin} A, {Broadhurst} T, {Umetsu} K, {Coe} D, {Ben{\'\i}tez} N, {Ascaso} B,
  {Bradley} L, {Ford} H, {Jee} J, {Medezinski} E, {Rephaeli} Y, {Zheng} W
  (2009{\natexlab{b}}) {New multiply-lensed galaxies identified in ACS/NIC3
  observations of Cl0024+1654 using an improved mass model}. \mnras
  396(4):1985--2002. \doi{10.1111/j.1365-2966.2009.14899.x}.
  {\href{https://arxiv.org/abs/0902.3971}{{arXiv:0902.3971}}} {[astro-ph.CO]}

\bibitem[{{Zitrin} et~al.(2011){Zitrin}, {Broadhurst}, {Barkana}, {Rephaeli},
  and {Ben{\'\i}tez}}]{zitrin2011}
{Zitrin} A, {Broadhurst} T, {Barkana} R, {Rephaeli} Y, {Ben{\'\i}tez} N (2011)
  {Strong-lensing analysis of a complete sample of 12 MACS clusters at z > 0.5:
  mass models and Einstein radii}. \mnras 410(3):1939--1956.
  \doi{10.1111/j.1365-2966.2010.17574.x}.
  {\href{https://arxiv.org/abs/1002.0521}{{arXiv:1002.0521}}} {[astro-ph.CO]}

\bibitem[{{Zitrin} et~al.(2013){Zitrin}, {Menanteau}, {Hughes}, {Coe},
  {Barrientos}, {Infante}, and {Mandelbaum}}]{Zitrin:2013}
{Zitrin} A, {Menanteau} F, {Hughes} JP, {Coe} D, {Barrientos} LF, {Infante} L,
  {Mandelbaum} R (2013) {A Highly Elongated Prominent Lens at z = 0.87: First
  Strong-lensing Analysis of El Gordo}. \apjl 770(1):L15.
  \doi{10.1088/2041-8205/770/1/L15}.
  {\href{https://arxiv.org/abs/1304.0455}{{arXiv:1304.0455}}} {[astro-ph.CO]}

\bibitem[{{Zitrin} et~al.(2015{\natexlab{a}}){Zitrin}, {Fabris}, {Merten},
  {Melchior}, {Meneghetti}, {Koekemoer}, {Coe}, {Maturi}, {Bartelmann},
  {Postman}, {Umetsu}, {Seidel}, {Sendra}, {Broadhurst}, {Balestra}, {Biviano},
  {Grillo}, {Mercurio}, {Nonino}, {Rosati}, {Bradley}, {Carrasco}, {Donahue},
  {Ford}, {Frye}, and {Moustakas}}]{Zitrin:15}
{Zitrin} A, {Fabris} A, {Merten} J, {Melchior} P, {Meneghetti} M, {Koekemoer}
  A, {Coe} D, {Maturi} M, {Bartelmann} M, {Postman} M, {Umetsu} K, {Seidel} G,
  {Sendra} I, {Broadhurst} T, {Balestra} I, {Biviano} A, {Grillo} C, {Mercurio}
  A, {Nonino} M, {Rosati} P, {Bradley} L, {Carrasco} M, {Donahue} M, {Ford} H,
  {Frye} BL, {Moustakas} J (2015{\natexlab{a}}) {Hubble Space Telescope
  Combined Strong and Weak Lensing Analysis of the CLASH Sample: Mass and
  Magnification Models and Systematic Uncertainties}. \apj 801(1):44.
  \doi{10.1088/0004-637X/801/1/44}.
  {\href{https://arxiv.org/abs/1411.1414}{{arXiv:1411.1414}}} {[astro-ph.CO]}

\bibitem[{{Zitrin} et~al.(2015{\natexlab{b}}){Zitrin}, {Fabris}, {Merten},
  {Melchior}, {Meneghetti}, {Koekemoer}, {Coe}, {Maturi}, {Bartelmann},
  {Postman}, {Umetsu}, {Seidel}, {Sendra}, {Broadhurst}, {Balestra}, {Biviano},
  {Grillo}, {Mercurio}, {Nonino}, {Rosati}, {Bradley}, {Carrasco}, {Donahue},
  {Ford}, {Frye}, and {Moustakas}}]{Zitrin:2015}
{Zitrin} A, {Fabris} A, {Merten} J, {Melchior} P, {Meneghetti} M, {Koekemoer}
  A, {Coe} D, {Maturi} M, {Bartelmann} M, {Postman} M, {Umetsu} K, {Seidel} G,
  {Sendra} I, {Broadhurst} T, {Balestra} I, {Biviano} A, {Grillo} C, {Mercurio}
  A, {Nonino} M, {Rosati} P, {Bradley} L, {Carrasco} M, {Donahue} M, {Ford} H,
  {Frye} BL, {Moustakas} J (2015{\natexlab{b}}) {Hubble Space Telescope
  Combined Strong and Weak Lensing Analysis of the CLASH Sample: Mass and
  Magnification Models and Systematic Uncertainties}. \apj 801(1):44.
  \doi{10.1088/0004-637X/801/1/44}.
  {\href{https://arxiv.org/abs/1411.1414}{{arXiv:1411.1414}}} {[astro-ph.CO]}

\bibitem[{{Zwicky}(1937)}]{Zwicky1937}
{Zwicky} F (1937) {On the Masses of Nebulae and of Clusters of Nebulae}. \apj
  86:217. \doi{10.1086/143864}

\end{thebibliography}
\nocite{*}

\end{document}